\documentclass[12pt]{article}
\usepackage[a4paper,
 left=2.5cm,right=2.5cm,top=2.5cm,bottom=2.3cm]{geometry}
\usepackage[utf8]{inputenc}
\usepackage{graphicx,subfigure}
\usepackage{epsfig}
\usepackage{mathrsfs}
\usepackage{amsmath}
\usepackage{amsfonts}
\usepackage{amssymb}
\usepackage{hyperref}
\allowdisplaybreaks
\usepackage[dvipsnames]{xcolor}

\numberwithin{equation}{section}

\newcommand{\nl}{\nonumber \\}

\def\m{\mu}
\def\n{\nu}

\begin{document}

\title{{\bf Rotating black holes in Einstein-aether theory
}  }
\author{Alexander Adam,$^{\text{\ae}}$ Pau Figueras,$^{\text{\oe}}$ Ted Jacobson$^{\text{\AE}}$ and Toby Wiseman$^{\text{\ae}}$ }
\date{}
\maketitle
\thispagestyle{empty}

\begin{center}
\textit{$^{\text{\ae}}$Theoretical Physics Group, Blackett Laboratory, Imperial College London}\\ \textit{London SW7 2AZ, UK}
\vspace{0.5cm}

\textit{$^{\text{\oe}}$School of Mathematical Sciences, Queen Mary University of London}\\ \textit{ Mile End Road, London, E1 4NS, UK}
\vspace{1cm}

\textit{$^{\text{\AE}}$Maryland Center for Fundamental Physics, University of Maryland,}\\ \textit{College Park, MD 20742, USA}
\vspace{0.5cm}

{\small
\texttt{alexander.g.adam@gmail.com}, \texttt{p.figueras@qmul.ac.uk}, \texttt{jacobson@umd.edu}, 
\texttt{t.wiseman@imperial.ac.uk}
}
\end{center}

\vspace{1cm}
\begin{abstract}

We introduce new methods to numerically construct for the first time stationary axisymmetric black hole solutions in 
Einstein-aether theory and study their properties. The key technical challenge is 
to impose regularity at the spin-2, 1, and 0 wave mode horizons. Interestingly we find the metric horizon, and various wave mode horizons, are not Killing horizons, having null generators 
to which no linear combination of Killing vectors is tangent,
and which spiral from pole to equator or vice versa.
Existing phenomenological constraints result in two regions of coupling parameters where the theory is viable and some couplings are large; region I with a large twist coupling and region II with also a 
(somewhat) large expansion coupling. Currently these constraints do not include tests from strong field dynamics, such as observations of black holes and their mergers.
Given the large aether coupling(s) one might 
expect such dynamics to deviate significantly from general relativity, 
and hence to further constrain the theory.
Here we argue this is not the case, 
since for these parameter regions solutions exist where the 
aether is ``painted'' onto a metric background 
that is very close to that of general relativity.
This painting for region I is approximately independent of the large twist coupling,
and for region II is also approximately independent of the large expansion coupling
and normal to a maximal foliation of the spacetime. 
We support this picture analytically for weak fields,
and numerically for rotating black hole solutions, which 
closely approximate the Kerr metric.

\end{abstract}

\newpage

\setcounter{page}{1}

\tableofcontents

\section{Introduction}
\label{sec:intro}

With the new and remarkable ability to measure gravitational waves given off from strongly gravitating binary systems~\cite{LIGOScientific:2016aoc}, together with the recent imaging by the EHT~\cite{EventHorizonTelescope:2019dse}, black holes have come to the fore in testing Einstein's General Relativity (GR). Considerable effort is now focused on comparing the predictions of modified theories of gravity to these and proposed future experiments with the hope of better understanding to what extent we believe GR to be the correct description of our dynamical spacetime in the strong field regime.\footnote{Here we mean `strong field regime' in the classical sense, so spacetimes where the curvature radius is comparable or smaller than the relevant dynamical scales or other length scales in the matter system, for example a star's size or orbit radius.}
At the heart of such endeavours is understanding how black holes behave in these theories, and particularly rotating stationary black holes.
The focus of this paper will be the Einstein-aether theory~\cite{Jacobson:2004ts}.
LIGO has given a remarkable constraint on modified theories of gravity, namely that the spin-2 graviton speed is constrained to be equal to the speed of electromagnetic waves to within one part in $10^{15}$ \cite{TheLIGOScientific:2017qsa,Creminelli:2017sry}.\footnote{Generally in a modified theory of gravity wavespeeds may be scale dependent in such a way that these two speeds deviate significantly on scales other than that probed by LIGO\cite{deRham:2018red}. However in classical Einstein-aether theory which we focus on here wave speeds are not scale dependent.} We therefore restrict attention here to the case where these speeds are exactly equal.
For the Einstein-aether theory there are then two natural regions of parameter space where some couplings are large and yet the tight bounds from Solar system tests of gravity, together with recent constraints from binary pulsars are satisfied~\cite{Oost:2018tcv,Gupta:2021vdj}. We denote these regions I and II in parameter space, as these have one or two couplings allowed to be large respectively, so $\sim O(1)$, while remaining coupling combinations are constrained to be very small. The naive expectation is that with such large couplings then in the strong field regime 
 the behaviour would differ significantly from GR. For stationary black holes this would mean the spacetime would deviate significantly from Kerr, potentially allowing us to discriminate between GR and Einstein-aether theory using observations of black holes.

Whilst it is possible there exist such exotic spacetime solutions to the Einstein-aether for strong fields, we present new arguments that in these allowed regions of parameter space there may still be solutions which are very close to those of conventional GR with a suitable aether field `painted on'.
 In these parameter regions the aether action is dominated by specific quadratic kinetic terms associated to the large couplings.
 For region I the single dominant term is given by the aether's twist, and for region II the aether's twist and expansion form the two large terms.
 We show that the aether may take approximately twist and expansion free configurations where these terms vanish. This allows the aether to have minimal backreaction on the geometry despite the large couplings. 
We show that this behaviour occurs in the weak field regime at leading order in the Post-Newtonian expansion, and is behind the consistency of the theory with Solar system constraints. While we give the equations that the aether will approximately satisfy for the two regions I and II, we cannot analytically prove existence of solutions in strong field settings.
We therefore turn to the study of black holes in the Einstein-aether theory to understand whether they exhibit approximate Kerr-like behaviour with a `painted on' aether.

Black hole solutions are considerably more complicated for modified gravity theories than for GR.
We later give an argument that even in theories where different degrees of freedom have different propagation speeds, stationary black holes may exist with a common Killing horizon for all the degrees of freedom. 
It is perhaps counter-intuitive that outside the horizon speeds differ, and yet they arrange themselves to agree at the horizon. 
We will find a sufficient condition is that the metric has a smooth bifurcate Killing horizon and that all the remaining fields in the theory are smooth on the spacetime. 
In such cases the exterior spacetime to this horizon can be approached numerically using existing formulations such as~\cite{Kleihaus:2000kg, Headrick:2009pv} where the problem is phrased as an elliptic boundary value problem
 with the smooth horizon forming one part of the boundary and the asymptotic region the other. 
However in Einstein-aether theory the aether cannot be smooth at a bifurcation surface, and consequently each wave mode in the theory possesses a different future horizon compatible with the stationary symmetry. 
In fact cosmic rays provide the strong phenomenological constraint that the wave speeds of degrees of freedom should be equal or greater than that of matter, otherwise ultra-high energy cosmic rays would quickly decay to gravi-Cherenkov radiation. Hence the metric horizon, which (with our parameter restriction) coincides with the spin-2 graviton horizon, must be the outermost horizon, and the spin-1 and spin-0 horizons potentially lie inside this. 
Furthermore we will see that that the rotating Einstein-aether black hole spacetimes do not have the usual $t$-$\phi$ orthogonality property, and the horizons are not Killing horizons.
In order to numerically construct such stationary rotating black holes 
smoothness must be imposed at all these  wave mode horizons to obtain a well posed problem. Thus at least some of the spacetime inside the metric horizon must be constructed, making this a novel problem that has not been addressed before beyond the static spherically symmetric case.\footnote{One might wonder whether such a smoothness condition is the correct boundary condition for black holes which are asymptotically stationary having formed from dynamical collapse of matter. Further one might question whether collapse may form singular end states that have no horizons, such as the Einstein-aether solutions in~\cite{Eling:2006df,Oost:2021tqi}.
This is morally the question of weak cosmic censorship, the general expectation being that well posed propagation of degrees of freedom will act to smooth and disperse localized curvature and field gradients, leaving regular horizons shielding any central singularity. While the status of weak cosmic censorship in modified gravity theories is far from certain, we will assume 
in 
this work that static 
and stationary black holes formed from collapse have smooth horizons and are the relevant end states for gravitational collapse. 
In the context of Einstein-aether theory this has been borne out by spherically symmetric collapse simulations~\cite{Garfinkle:2007bk}.
Note Ref.~\cite{Garfinkle:2007bk} also found that a finite area naked singularity formed when some aether coupling parameters
were large,  similar to those couplings for which no static solutions with regular spin-0 horizon were found in \cite{Eling:2006ec}.
On the other hand, later work \cite{Barausse:2011pu} at higher numerical resolution found static solutions with regular spin-0 horizons and significantly larger aether coupling, which suggests that the resolution used in the time dependent collapse calculation of \cite{Garfinkle:2007bk} may have been insufficient to resolve the evolution around the spin-0 horizon. 
}

In the simpler static spherically symmetric case some analytic black hole solutions exist
for special values of the coupling parameters~\cite{Berglund:2012bu}. More generally static spherically  symmetric  solutions  may be found by numerically solving a coupled ode system, whose variable is a radial coordinate, subject to the correct asymptotic behaviour and smooth behaviour at horizons.  
This is relatively straightforward, and typically a ‘shooting’ method can be employed as in~\cite{Eling:2006ec,Barausse:2011pu} where static black holes were numerically constructed. In~\cite{Zhang:2020too}  some static spherically symmetric black holes were constructed in parameter region I and it was observed that Schwarzschild behaviour did indeed emerge. However for static spherical symmetry the aether is automatically twist free~\cite{Eling:2003rd}, and so is it clear that one recovers GR-like behaviour for region I with such solutions. Adding rotation perturbatively about static solutions has been studied in~\cite{Barausse:2015frm} where it was seen that the rotation creates twist in the aether. Thus in the phenomenologically more relevant case of rotating black holes it is much less obvious that one could recover GR-like behaviour with an approximately twist free aether `painted on'.

In order to address the question of recovering GR-like behaviour we develop novel numerical tools to directly construct rotating stationary black hole solutions with multiple wave mode horizons. 
We do this employing the harmonic formulation~\cite{Headrick:2009pv} together with horizon penetrating ingoing coordinates that extend within this innermost future horizon.  
We validate our numerical scheme by adding rotation to the static spherically symmetric solutions previously constructed in~\cite{Barausse:2011pu}. Then we focus on the phenomenologically allowed parameter regions I and II. As we tune the couplings towards these regions indeed we find that the spacetime and aether tend to Kerr with an aether `painted on' that is twist free for region I, and twist and expansion free for region II. Furthermore we find that as the small couplings tend to zero, the limiting aether depends only on 
the Kerr mass and angular momentum parameters and on whether the $O(1)$ aether couplings are in region I or II, but not on the value of those couplings.
These stationary black holes confirm our picture that GR behaviour may be recovered in the strong field regime for these viable Einstein-aether parameter regimes, despite some aether couplings being large. 
Combined with our weak field analysis, this suggests that in typical strong field astrophysical settings the theory will exhibit approximate GR behaviour. Initial data for collapse which at early times is in the weak field regime will have a small aether twist and behave as in GR. We argue the dynamics leading to the strong field regime will closely approximate that of GR with an aether with small twist `painted on', resulting in a rotating black hole that is approximately Kerr such as the solutions we find numerically here.

The structure of the paper is as follows. 
We begin by reviewing the Einstein-aether theory in Section~\ref{sec:aether}, and detail the two  phenomenologically allowed regions with one or two large couplings. 
Next in Section~\ref{sec:painting} we give the argument that in these two regimes solutions may exist where the metric is very close to that of GR solutions, with a twist free aether `painted' on top. We discuss how this occurs in the weak field regime relevant for Solar system constraints, and our attention then turns to the strong field regime and rotating stationary black holes.
In Section~\ref{sec:horizons} we discuss black holes in theories with multiple wave speeds, arguing that
despite modified gravity theories generally having different wavespeeds, in some theories black holes may still exist with a single Killing horizon for all propagating degrees of freedom. We also discuss how in Einstein-aether this is not possible, and black holes cannot have a single smooth Killing horizon, and instead  have multiple horizons.
In the following Section~\ref{sec:numericalconstruction} we introduce the numerical methods to find stationary black holes in theories with multiple wave mode horizons. 
We demonstrate the method in the simple case of Einstein gravity and Kerr, before applying it to the Einstein-aether theory in Section~\ref{sec:results}. We firstly add rotation to previously found static solutions.
We discuss the novel non-Killing nature of the multiple wave mode horizons of these rotating black holes. Then we focus on rotating black holes in the phenomenologically allowed regions I and II which have some large couplings. 
Having found that these solutions do approximate GR behaviour with a `painted on' twist free aether, we conclude with a discussion focused on how this GR-like behaviour may be recovered more generally in astrophysical settings.

\section{Einstein-aether theory}
\label{sec:aether}

We now review the Einstein-aether theory~\cite{Jacobson:2004ts}. The theory contains the metric and the aether vector $u$ which is constrained to have unit timelike norm, so $u^2 = -1$. 
Rather than use the usual parameterization of the theory, here we employ the irreducible couplings
introduced in~\cite{Jacobson:2013xta}. 
Defining the acceleration of the aether, $a^\mu = u \cdot \nabla u^\mu$, then we may decompose the covariant derivative of the aether as,
\begin{eqnarray}\label{decomp}
\nabla_\mu u_\nu  = \frac{1}{3} \theta h_{\mu\nu} + \sigma_{\mu\nu} + \omega_{\mu\nu} - u_\mu a_\nu
\end{eqnarray}
where the expansion $\theta = \nabla \cdot u$, and the shear $\sigma_{\mu\nu} = \sigma_{(\mu\nu)}$ with $\sigma^\mu_\mu = 0$ is orthogonal to the aether, as is the twist $\omega_{\mu\nu} = \omega_{[\mu\nu]}$. Here $h_{\mu\nu}$ is the metric orthogonal to the aether, $h_{\mu\nu} = g_{\mu\nu} + u_\mu u_\nu$. 
The terms in the decomposition \eqref{decomp} are irreducible with respect to the spatial rotation group  
in the local rest frame of the aether.
A virtue of this irreducible
parameterization is that the gravitational part of the action, 
\begin{equation}
\label{eq:action}
I_{grav} = \frac{1}{16 \pi G}\int d^4x\sqrt{-g}\bigg[R - \frac{1}{3} c_\theta \theta^2 - c_\sigma \sigma^2 - c_\omega \omega^2 + c_a a^2 + {\lambda} (g^{\mu\nu}u_\mu\,u_\nu+1)\bigg]
\end{equation}
can be seen to be a sum of squares of these quantities and this will play an important role later. 
The field $\lambda$ is a Lagrange multiplier than enforces the timelike constraint on the vector $u^2 = -1$. The aether coupling constants
$c_A$, $A = \{ \theta, \sigma, \omega, a \}$ are dimensionless and determine the aether dynamics. For the convenience of the reader we have included Appendix~\ref{app:aetherTheory} where the theory is described in the usual variables and couplings $c_{1,2,3,4}$, and we detail the translation between this and the irreducible parameterization.

In the limit $c_A \to 0$ the aether field decouples leaving the metric to be governed by Einstein GR. 
The dynamical wave modes (in the absence of other matter) are the spin-0, spin-1 and spin-2 combinations of the metric and aether field. The effective metrics for their propagation are;
\begin{eqnarray}
\label{eq:effmetric}
g^{\rm eff}_{\mu\nu} = g_{\mu\nu} + (1 - s_{(a)}^2) u_\mu u_\nu
\end{eqnarray}
with $s_{(a)}$ the wavespeeds in the rest frame defined by the aether.
The theory has been highly constrained by the LIGO neutron star merger observation~\cite{TheLIGOScientific:2017qsa,Creminelli:2017sry} which imposes to high precision (one part in $\sim 10^{15}$) that the spin-2 gravitational wave speed is that of light. Since,
\begin{eqnarray}
\mathrm{spin-2}: && s_{(2)}^2 = \frac{1}{1 - c_{\sigma}}
\end{eqnarray}
from now on, unless otherwise stated, we assume that $c_\sigma = 0$ \emph{exactly} and so there is no shear term in the action. Thus the metric and spin-2 effective metric are the same, and for a black hole the horizon for matter is the same as that for the spin-2 graviton. Given that $c_\sigma = 0$ then the remaining wavespeeds are,\footnote{The full wavespeeds without taking $c_\sigma = 0$ are reviewed in the Appendix~\ref{app:aetherTheory}.}
\begin{eqnarray}
\mathrm{spin-0}: && s_{(0)}^2 = \frac{ c_\theta \left( 2 - c_a \right) }{ 3 c_a \left( 2 + c_\theta \right) } \nl
\mathrm{spin-1}: && s_{(1)}^2 = \frac{ c_\omega }{2 c_a} 
\end{eqnarray}
The theory has been shown to be well-posed for the parameter ranges we will be interested in~\cite{Sarbach:2019yso}. For black holes we will see later that these different effective metrics will generally have different positions of their future horizons. 

Written in the variables of the irreducible parameterization, once one has eliminated the Lagrange multiplier and imposed $c_\sigma = 0$, the aether equation is $A_{\text{\AE}}^\mu = 0$ where,
\begin{eqnarray}
\label{eq:aethereq}
&& A_{\text{\AE}}^\mu \equiv \frac{c_\theta}{3} \left( \nabla^\mu \theta +  u^\mu u \cdot \nabla \theta \right)
+ c_\omega \left( \nabla_\nu \omega^{\nu\mu} -  \omega^{\mu\nu} a_\nu -  u^\mu \omega^2 \right)
    \nl
 && \qquad \qquad
 - c_a \left( \frac{2}{3} \theta a^\mu + u \cdot \nabla a^\mu - \left( \sigma^{\mu\nu} + \omega^{\mu\nu} \right) a _\nu - u^\mu a^2  \right) 
\end{eqnarray}
and given the aether vector's timelike norm constraint, it is orthogonal to the aether field, so $u \cdot A_{\text{\AE}} = 0$. The Einstein equation is $G_{\mu\nu} = T^{\text{\AE}}_{\mu\nu} + 8 \pi G T^{\mathrm{matter}}_{\mu\nu}$ where,
\begin{eqnarray}
\label{eq:Einsteineq}
T^{\text{\AE}}_{\mu\nu} & \equiv & \frac{c_\theta}{3} \left( - 2 u_{(\mu} \nabla_{\nu)} \theta +g_{\mu\nu} \left( \frac{1}{2} \theta^2 + u \cdot \nabla \theta \right) -  u_\mu u_\nu ( u \cdot \nabla \theta ) \right) \nl
&& \qquad + c_\omega \left( 2 \omega_{\mu\alpha} \omega_{\nu}^{~~\alpha} - \frac{1}{2} g_{\mu\nu} \omega^2 +  u_\mu u_\nu \omega^2 \right) \nl
&& \qquad - c_a \left( a_\mu a_\nu + 4 u_{(\mu} \omega_{\nu) \alpha} a^\alpha - \frac{1}{2} g_{\mu\nu} a^2 -  u_\mu u_\nu \nabla \cdot a \right)
\end{eqnarray}
We see the structure of the action being a sum of squares manifests itself in the aether equation and Einstein equation. If one of the quantities $\theta$, $\omega^{\mu\nu}$ or $a^\mu$ vanishes, then the corresponding terms multiplied by $c_\theta$,  $c_\omega$ or $c_a$ respectively, vanish in both the aether and Einstein equations, and this will play an important role later.

\subsection{Phenomenological Regimes}
\label{sec:phenoregimes}

While the aether has four parameters, treating the theory as a low energy effective description of gravity~\cite{Withers:2009qg}, these are constrained in order to have realistic phenomenology. 
Taking $c_A$ very close to zero, the theory obviously behaves metrically as GR, and so passes phenomenological tests, provided the aether obeys basic constraints (for example, having a stable evolution, and $s_{0,1}^2 \ge 1$ to avoid matter energy loss via Cherenkov radiation into aether modes), but does not behave in an interesting manner (unless one can observe the aether vector directly).

However, having imposed the LIGO constraint, so the spin-2 wavespeed is that of light, substantially changes the original observational constraints computed in~\cite{Foster:2005dk}.
There remain two phenomenologically viable parameter regions, one where the twist coupling may be large, so $\sim O(1)$, while the other couplings are very small in magnitude, and the other where both the twist and expansion couplings are large~\cite{Oost:2018tcv}.
An important point is that these constraints do not involve strong field dynamics (apart from considering nucleosynthesis) and given the large couplings, this raises the fascinating possibility that observations of the strong field regime could differ from GR. For example, black holes in these parameter regions may strongly deviate from the Kerr black hole of GR, in the sense that geometric invariants of the black holes deviate from their Kerr counterparts by $O(1)$ amounts for fixed mass and angular momentum. This in principle could allow the theory to be distinguished from GR on the basis of its strong field behaviour, such as the properties of black holes. Whether this indeed occurs is the key focus of this paper.

We now discuss these two allowed regions. The 
absence of decay of cosmic rays to Cherenkov radiation in the spin-0 and spin-1 modes implies that $s_{0,1}^2 \ge 1$ to high precision.
Further constraints come from agreeing with weak field tests of gravity~\cite{Foster:2005dk}.  The two PPN parameters $\alpha_{1,2}$ are constrained to be  small, with Solar system observations 
requiring $|\alpha_1| \lesssim 10^{-4}$ and $|\alpha_2| \lesssim 10^{-7}$. Written in terms of the couplings $c_{\omega, \theta,a}$ together with the condition that $c_\sigma = 0$, \footnote{The  general expressions are given in Appendix~\ref{app:aetherTheory}. }
\begin{eqnarray}
\label{alphaconstraints}
\alpha_1 & = & - 4 c_a \nl
\alpha_2 & \simeq & - \frac{c_a}{2} \left(  1 -    \frac{3 c_a}{c_\theta}   \right)
\end{eqnarray}
where we have used that $|\alpha_1| \lesssim 10^{-4}$ implies $| c_a | \ll 1 $.
Nucleosynthesis constrains $|c_\theta| \lesssim 0.3$ (assuming standard matter content).\footnote{Recently it was argued~\cite{Frusciante:2020gkx} that in the related Ho\v{r}ava gravity theory that the CMB combined with other cosmological observations provides substantially stronger constraints than those just from nucleosynthesis. It would be interesting to see if such detailed cosmology analysis might lead to improved bounds in the Einstein-aether theory.
}
Positivity of the kinetic terms implies,
\begin{eqnarray}
0 \le c_a, c_\theta \, .
\end{eqnarray}
Dissipative dynamics of binary (and ternary) pulsar systems has recently been found to 
improve the constraint on $\alpha_1$ by a factor of 10, i.e.\,  $|\alpha_1| \lesssim 10^{-5}$ \cite{Gupta:2021vdj}. 
Another strong constraint $|\hat\alpha_2|< 2\times10^{-9}$ comes from spin precession of a millisecond pulsar~\cite{Shao:2013wga,Will:2014kxa},
where $\hat\alpha_2$ is a strong field analog of $\alpha_2$. However the translation of the constraint on $\hat{\alpha}_2$ to one on the coupling parameters is complicated as it involves the neutron star sensitivities to higher order in velocity relative to the aether than they are currently known
(see also~\cite{Oost:2018tcv} for a discussion of this point).

\subsubsection*{One large parameter: region I}

Firstly we assume that $c_\omega \sim O(1)$ and not small in magnitude -- otherwise we will find ourselves in the trivial regime, $|c_A | \ll 1$ where the aether decouples from the theory.\footnote{In what follows when we say a quantity is $\sim O(1)$ we mean to imply that it is also not small in magnitude.}
Then taking the remaining parameters as,
\begin{eqnarray}
c_\omega = O(1) \; , \quad 10^{-7} \ll c_{a} \lesssim 10^{-5}  \; , \quad \frac{c_\theta}{3} =  c_a  + O(10^{-7}) 
\end{eqnarray}
satisfies the phenomenological constraints.
Hence in this region $c_\omega$ is the only large $O(1)$ parameter. All the other parameters are of order $O(10^{-5})$ or smaller, 
so the aether action is
dominated  by the twist term.
We restrict this region to refer to the parameter range where $c_a \gg 10^{-7}$ which implies $c_\theta \simeq 3 c_a$.
We will call this parameter region I.
We see the spin-0 wavespeed is close to that of light, $s_{(0)}^2 \simeq 1$, while the spin-1 wavespeed diverges as $s_{(1)}^2 \sim 1/c_a = O(10^5)$. 

\subsubsection*{Two large parameters: region II}

Taking both $|\alpha_1|, |\alpha_2| \sim 10^{-7}$, so that $\alpha_1$ is smaller than necessary to satisfy current bounds, allows for two independent parameters to be large.  
Again taking $c_\omega\sim O(1)$ and not small in magnitude, and further allowing $c_\theta$ to be an independent large $O(1)$ parameter, these bounds are satisfied by,
\begin{eqnarray}
c_\omega  = O(1)\, , \; 10^{-7} \ll c_\theta \lesssim 0.3 \; , \quad c_{a} = O(10^{-7}) 
\end{eqnarray}
The aether action is then dominated by the twist and expansion terms.
We term this parameter region II.
Now both the spin-0 and spin-1 speeds become very large as $s_{(0)}^2, s_{(1)}^2 \sim  1/c_{a} = O(10^7)$. 

\subsubsection*{The transition between regions I and II}

\begin{figure}
  \centerline{\includegraphics[height=10cm]{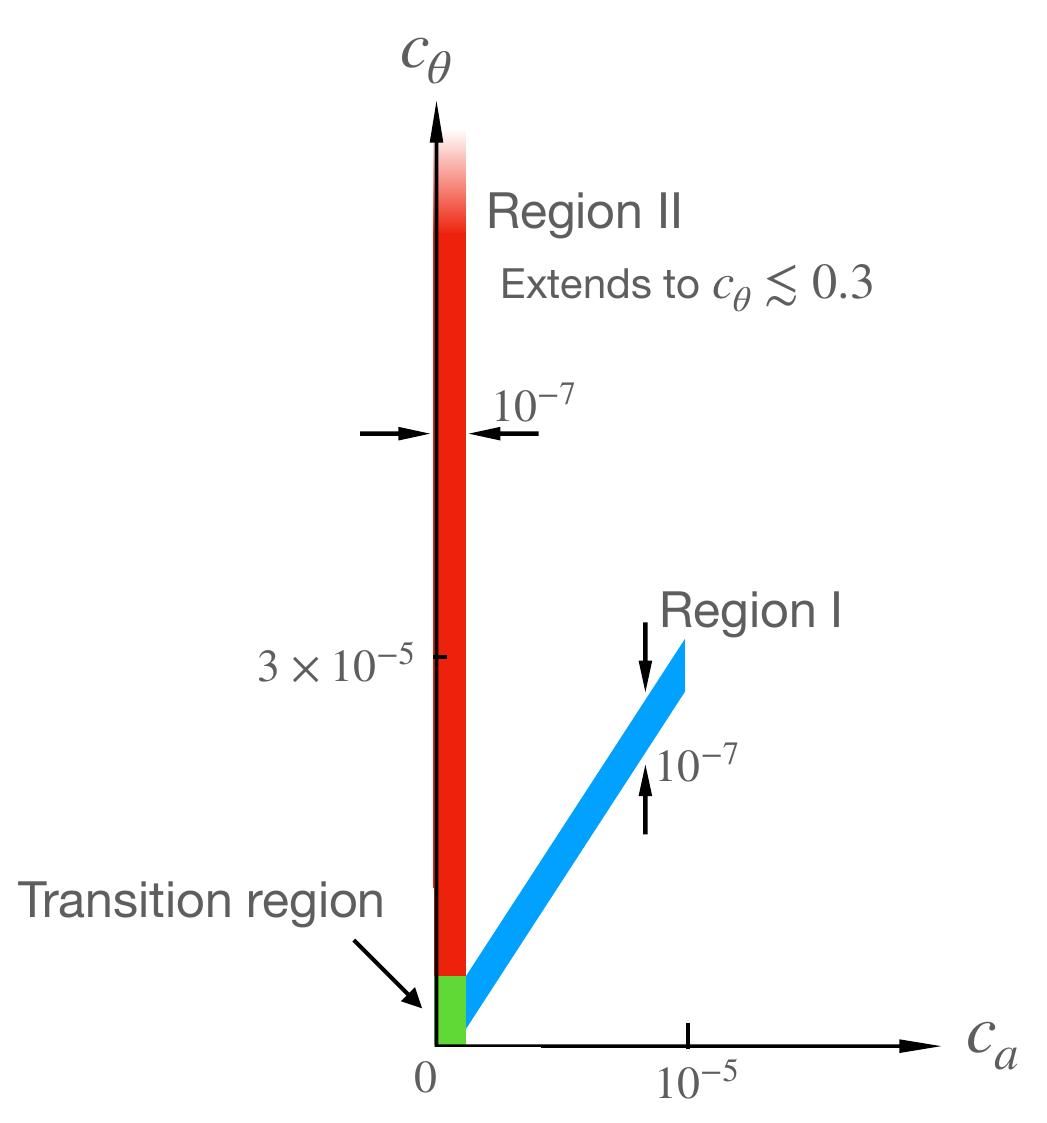}}
  \caption{\label{fig:regions}Figure showing the two phenomenologically allowed regions, region I and II, in the aether coupling space. The coupling $c_\omega$ is not shown as it is unconstrained for both these regions.}
\end{figure}

In figure~\ref{fig:regions} we depict these two regions.
The transition region joining these two regions I and II in  parameter space is given by,
\begin{eqnarray}
c_\omega = O(1) \; , \quad c_{a} = O(10^{-7}) \; , \quad c_\theta = O(10^{-7}) 
\end{eqnarray}
This joins to the region II taking $c_\theta$ very small rather than $O(1)$. On the other hand this space joins with region I when $c_{a} = O(10^{-7})$. 
A subtlety of this region is that if $c_\theta \to 0$ for fixed $c_a$ then it is unclear the PPN analysis holds as the spin-0 wavespeed tends to zero.
As this transition region is much smaller than the parameter regions I and II implied by the observational constraints we will not consider it further in this work. 

\section{Recovering GR in phenomenological large parameter regions}
\label{sec:painting}

Both the one and two large parameter regions (I and II) have $O(1)$ aether parameters, so at least some $c_A = O(1)$. Hence one would naively expect strong field solutions, such as black holes, to deviate from those of GR by $O(1)$ amounts. The aether with its unit timelike norm will vary on scales comparable to the curvature scale of the metric, and one might then expect the terms in the aether action with large couplings  to give rise to significant backreaction in the Einstein equation of the theory, deforming the metric response to matter from that of GR.

However we now discuss how GR may be recovered in these phenomenologically allowed regions. While some terms in the aether action may have $O(1)$ couplings, nonetheless they may be dynamically suppressed. For our two regimes above the action is dominated by the twist term, and for regime II also the expansion. 
By Frobenius's theorem a twist free aether field is hypersurface orthogonal,
and can then be written locally as $u_\mu = k\, \partial_\mu f$, for functions $k$ and $f$, where level sets of $f$ define these hypersurfaces~\cite{Jacobson:2013xta}.
Due to the aether norm constraint, 
$k^2 (\partial f)^2 = -1$,
which can be thought of as fixing $k$ in terms of $f$. Hence a twist free aether is 
determined by a single scalar potential function, $f$. The important point is that the aether has sufficient degrees of freedom, after imposing the timelike unit norm constraint, to still be twist free. Furthermore it also has sufficient freedom left, the scalar function $f$, to potentially satisfy the further scalar constraint coming from requiring it to be expansion free. It is this ability for the aether to be twist free, and also expansion free, that lies behind the potential recovery of GR, as it allows the terms in the aether action with large couplings to still be small.

\subsection{GR behaviour for region II}
\label{sec:paintingII}

We begin by considering region II as it is simpler to analyse than region I. Here the aether action is dominated by the quadratic twist and expansion terms. 
When the shear and acceleration couplings are set to zero in the 
aether Lagrangian~\eqref{eq:action}, leaving only the twist and expansion couplings,
any solution to the usual GR Einstein equation admitting a maximal foliation
is also an exact Einstein-aether solution, with the aether set
equal to the unit normal to the foliation. The reason is that
the twist and expansion of this aether vanish, 
and the action is quadratic in those quantities, 
so its variation 
away from such a configuration vanishes when the Lagrange multiplier for the unit norm constraint vanishes.
If the acceleration 
coupling is instead small, rather than
strictly zero, one may thus expect there to exist a solution that is a small perturbation of that in which it does vanish. 
We will now develop this more explicitly.

We begin by taking $\epsilon \equiv c_a = O(10^{-7})$ which we consider to be a small parameter.
For simplicity let us consider a vacuum solution, and hence no matter. 
Let us attempt to write a consistent vacuum solution 
as an expansion in $\epsilon$ about a solution of vacuum GR and a twist free aether configuration, given in terms of a potential function $f$, as,
\begin{eqnarray}
\label{eq:GRbehaviour}
g_{\mu\nu} & = & \bar{g}_{\mu\nu} + \epsilon \,  g^{(1)}_{\mu\nu} + O(\epsilon^2) \nl
u_\mu & = & \bar{u}_\mu + \epsilon \, u^{(1)}_{\mu} + O(\epsilon^2) \; , \quad \bar{u}_\mu = k \partial_\mu f
\end{eqnarray}
where $\bar{g}_{\mu\nu}$ is a vacuum GR solution, so Ricci flat. Now at order $O(\epsilon^0)$ the aether equation receives a contribution from the twist and expansion terms in the action. The twist term vanishes as our leading aether configuration is twist free, and since the term is  quadratic, its contribution to the equations of motion and the aether stress tensor vanishes too. However from the aether equation~\eqref{eq:aethereq} we see the expansion term leads to the condition;
\begin{eqnarray}
c_\theta \bar{h}^{\mu\nu} \partial_\nu \bar{\theta}  = 0
\end{eqnarray}
where indices are raised/lowered with the leading metric $\bar{g}_{\mu\nu}$ and 
where $\bar{\theta}$ is the expansion of the leading order twist-free aether,
\begin{eqnarray}
\bar{\theta} = k \bar{h}^{\mu\nu} \bar{\nabla}_\mu \partial_\nu f 
\end{eqnarray}
where $\bar{h}_{\mu\nu} = \bar{g}_{\mu\nu} + \bar{u}_\mu \bar{u}_\nu$ is the projection of the metric onto the constant $f$ hypersurfaces. Thus the expansion $\bar{\theta}$ must be constant on the hypersurfaces of constant $f$. 

Having taken a vacuum GR solution, then the Einstein tensor $\bar{G}_{\mu\nu}$ of the metric $\bar{g}_{\mu\nu}$ vanishes, and we must also check consistency with the Einstein equation. From \eqref{eq:Einsteineq} at leading $O(\epsilon^0)$ order we find the expansion term backreacts as,
\begin{eqnarray}
0 = \bar{G}_{\mu\nu} = \frac{c_\theta}{3} \left( - 2 \bar{u}_{(\mu} \bar{\nabla}_{\nu)} \bar{\theta} + \bar{g}_{\mu\nu} \left( \frac{1}{2} \bar{\theta}^2 + \bar{u} \cdot \bar{\nabla} \bar{\theta} \right) -  \bar{u}_\mu \bar{u}_\nu ( \bar{u} \cdot \bar{\nabla} \bar{\theta} ) \right)
\end{eqnarray}
and again indices are raised/lowered with the leading metric $\bar{g}_{\mu\nu}$.
Since the twist vanishes and the twist term in the action is quadratic it does not backreact. Now contracting with $u^\mu u^\nu$ yields the condition,
\begin{eqnarray}
\bar{\theta}^2 = 0
\end{eqnarray}
which in fact forces the expansion to vanish for consistency. 

Thus we see that we may consistently take the spacetime to be that of vacuum GR at leading order $O(\epsilon^0)$ with a `painted on' aether vector that is twist and expansion free, provided we can find a potential function $f$ which obeys the vanishing expansion condition on this spacetime, given by the non-linear p.d.e.,
\begin{eqnarray}
\label{eq:aetherRegion2}
\left( \bar{g}^{\mu\nu} - \frac{1}{({\partial} f)^2} \partial^\mu f  {\partial}^\nu f \right) \bar{\nabla}_\mu \partial_\nu f  = 0 \; .
\end{eqnarray}
We note the two derivative terms in $f$ are contracted by the induced inverse metric on a constant $f$ surface. This induced inverse metric must be Riemannian, since it is orthogonal to the aether which is constrained to be timelike, and hence this p.d.e. has elliptic character.  
We may view the aether as locally defining a foliation of the spacetime by the constant $f$ hypersurfaces, with the unit timelike norm and expansion free condition implying that it is a maximal slicing.
This leading order Ricci flat spacetime is then perturbed at order $O(\epsilon^1)$ by $g^{(1)}_{\mu\nu}$, and the aether by $u^{(1)}_\mu$. However 
for region II we have $\epsilon = O(10^{-7})$ so 
to one part in $10^7$ the spacetime geometry is simply that of vacuum GR. Thus to a great accuracy the vacuum behaviour of GR can potentially be recovered even though both $c_\omega$ and $c_\theta$ are $O(1)$. 

While for convenience we have excluded matter, including it
does not change the picture. With matter, $\bar{g}_{\mu\nu}$ and the leading matter behaviour will be a solution to the usual GR Einstein-matter equations. The aether will again be twist and expansion free at leading order, and to high accuracy the usual Einstein-matter dynamics will be recovered for such solutions. Furthermore these conditions on the aether are independent of the values of the large aether couplings $c_{\omega}$ and $c_\theta$. Thus the 
aether behaviour for these near GR solutions is approximately (ie. to one part in $10^7$) universal within region II.

An important point to emphasize is that there is no guarantee that solutions to the above twist and expansion free  equation~\eqref{eq:aetherRegion2} exist for all GR spacetimes. For example, in a cosmological setting the asymptotic expansion of the aether will not vanish and hence one could not hope to find a global solution to the above conditions. 
While it was long
ago proved that vacuum spacetimes ``close'' to Minkowski spacetime
 admit a maximal foliation, to our knowledge there is no stronger result that could guarantee such a foliation more generally, such as for black hole spacetimes that we will later be interested in.\footnote{For a discussion and references see~\cite{Eardley:1978tr}. Interestingly, that paper includes an example of a spherical, pressureless dust collapse solution
 to GR for which there is no maximal foliation that covers the entire exterior of the black hole. 
 } 
For the case of Kerr, maximal spacelike foliations have been constructed numerically~\cite{PhysRevD.31.1267}.\footnote{
The constant Boyer-Lindquist time slices of Kerr spacetime provide a
maximal foliation~\cite{Gomes:2013bbl,Bergamini:2003ch}. However, this
foliation is not
suitable for the aether construction since it is not spacelike inside the horizon.
}
We will shortly give numerical evidence that a twist and expansion free aether, associated to such a foliation, indeed describes the behaviour of black holes that we are able to construct associated to region II.

If solutions do exist another question is whether these are unique, and whether there might be a moduli space of such solutions. While the above elliptic p.d.e. for the potential $f$ above locally determines $f$, such a moduli space may arise if there is global data for solutions once boundary conditions are prescribed.
A key point is that if a solution or moduli space of solutions for $f$ exist, they are univeral to region II in the sense that they are independent of the $O(1)$ couplings $c_\omega$ and $c_\theta$, as we see explicitly from the coupling independence of equation~\eqref{eq:aetherRegion2}

\subsection{GR behaviour for region I}
\label{sec:paintingI}

Now let us consider region I which has one large parameter, the twist coupling $c_\omega$. 
The logic is similar to that above for region II. If the expansion and acceleration couplings vanish exactly, then any solution to the usual GR Einstein equation that admits a twist free aether will be an exact solution to the Einstein-aether equations. Now if the expansion and acceleration couplings
are not zero, but are small, we then expect a solution to exist which is a small perturbation of this solution where it does vanish.  

We again take the small parameter to perform our expansion to be $\epsilon \equiv c_a = O(10^{-5})$. 
The aether action is now dominated only by the twist term. The expansion and acceleration terms have small coefficients obeying the phenomenological constraint above, $\frac{c_\theta}{3} = c_a  + O(10^{-7})$. 
Since we have restricted region I to have $10^{-7} \ll \epsilon \lesssim 10^{-5}$ then we can write,
\begin{eqnarray}
\frac{c_\theta}{3} = \epsilon + k_\theta \epsilon^2
\end{eqnarray}
so that $| k_\theta \epsilon^2 | \ll \epsilon$.
Again for simplicity let us consider the spacetime at leading order to be a GR vacuum solution, so $\bar{g}_{\mu\nu}$ is Ricci flat and there is no matter.
Then `painting on' a twist free aether, given by the potential function $f$, we again expand the metric and aether as above in equation~\eqref{eq:GRbehaviour}.

Now since at leading order $O(\epsilon^0)$ the aether action comprises only the quadratic twist term, and the leading aether is twist free, both the aether and Einstein equations are trivially satisfied at this order. One might then wonder whether the aether potential $f$ can be arbitrary? In fact it cannot, as it is constrained by consistency of the $O(\epsilon^1)$ equations. At this order the aether equation gives,
\begin{eqnarray}
\label{eq:deltaomega}
c_\omega \left( \bar{\nabla}_\nu \omega^{(1)\nu\mu} +  \bar{a}_\nu \omega^{(1)\nu\mu}   \right) & = & \bar{j}^\mu
\end{eqnarray}
where $\omega_{\mu\nu} = \epsilon\,  \omega^{(1)}_{\mu\nu} + O(\epsilon^2)$ and we note that $\bar{u}_\mu  \omega^{(1)\mu\nu} = 0$, and consequently $\bar{u}_\mu \bar{j}^\mu = 0$ and,
\begin{eqnarray}
\label{eq:fEqOneParam}
 \bar{j}^\mu & = &
 -
\left( \bar{\nabla}^\mu \bar{\theta} +  \bar{u}^\mu \bar{u} \cdot \bar{\nabla} \bar{\theta} \right) +  \left( \bar{u} \cdot \bar{\nabla} \bar{a}^\mu - \bar{u}^\mu \bar{a}^2 + \frac{2}{3} \bar{\theta} \bar{a}^\mu  -  \bar{\sigma}^{\mu\nu} \bar{a}_\nu  \right)   
\end{eqnarray}
which we emphasize is constructed only from the leading metric and aether, $\bar{g}_{\mu\nu}$ and $\bar{u}_\mu$.
We may regard equation~\eqref{eq:deltaomega} as determining the aether perturbation $\omega^{(1)}_{\mu\nu}$. However, somewhat analogously to source conservation in the Maxwell equation, one also finds the constraint,
\begin{eqnarray}
\label{eq:aetherRegion1}
 \bar{\nabla} \cdot \bar{j} + \bar{a} \cdot \bar{j} = 0 
\end{eqnarray}
a key point being that this involves only the leading order aether and metric.
This appears to be a complicated differential equation for the potential $f$ which is fourth order in derivatives
and which we may think of as its equation of motion.
Another important point is that  solutions for $f$ of this p.d.e. are universal for region I in the sense that they do not depend on the $O(1)$ aether coupling $c_\omega$, as we explicitly see from the above equation.

The origin of this condition may be understood by considering the aether action under an $O(\epsilon)$ variation of  the aether, ${u}^{(1)}_\mu \to {u}^{(1)}_\mu + \delta{u}^{(1)}_\mu$; the variation of the action at order $O(\epsilon^2)$ is,
\begin{eqnarray}
\delta S^{(2)} = \frac{1}{8\pi G} \int d^4x \sqrt{-\bar{g}} \left( - c_\omega {\omega}^{(1)\mu\nu} \delta{\omega}^{(1)}_{\mu\nu} - \delta{u}^{(1)\mu} \bar{j}_\mu \right)
\end{eqnarray}
which gives rise to the equation~\eqref{eq:deltaomega} for ${\omega}^{(1)}$ above. Now consider the variation generated by changing the foliation at order $O(\epsilon)$;
\begin{eqnarray}
f \to f + \epsilon \, \delta f^{(1)} \quad \implies \quad \delta u^{(1)}_\mu = k \, \bar{h}_\mu^{~\nu} \bar{\nabla}_\nu \delta f^{(1)} \,
\end{eqnarray}
In analogy to a gauge transformation for Maxwell leaving the field strength invariant, this variation leaves the twist $\omega^{(1)}$ invariant, so $\delta{\omega}^{(1)}_{\mu\nu} = 0$. Then taking the remaining variation and integrating by parts,
\begin{eqnarray}
\delta S^{(2)} =  \frac{1}{8\pi G} \int d^4x \sqrt{-\bar{g}} \, \delta f^{(1)} \bar{\nabla}^\mu \left( k \bar{j}_\mu \right) 
\end{eqnarray}
and since the action should be stationary for a solution, $\bar{\nabla}^\mu \left( k \bar{j}_\mu \right)  = 0$. Then using $\bar{a}_\mu = \bar{h}_\mu^{~~\nu} \partial_\nu(\ln{k})$ implies equation~\eqref{eq:aetherRegion1} above.

As for the region II discussion, we have no argument that guarantees solutions to~\eqref{eq:aetherRegion1} for $f$ must generally exist given an underlying GR spacetime. If they do, given that it is a fourth order differential equation in $f$, it is also unclear how much global data would characterize them.
However, we will shortly give numerical evidence that for stationary black holes a solution does exist and furthermore it appears to be unique given the boundary condition that the aether is asymptotically at rest with respect to the black hole.

A possible solution to~\eqref{eq:aetherRegion1} is $\bar{j} = 0$, which is a simpler condition  involving only three derivatives of $f$. For example, if one considers static spherically symmetric black holes where the aether is twist free by symmetry, then indeed $\bar{j}$ must vanish (as $\omega^{(1)}$ vanishes in equation~\eqref{eq:deltaomega}). 
Could it be the case that while~\eqref{eq:aetherRegion1} holds true, it only does so because of the more fundamental condition $\bar{j} = 0$ which really plays the role of the equation governing the potential $f$?
We later show that this is not the case. For stationary rotating black holes our numerical solutions will show that in region I these take the approximate form of Kerr with the aether `painted on', and further that equation~\eqref{eq:aetherRegion1} is satisfied by an aether where $\bar{j} \ne 0$. Hence it appears that~\eqref{eq:aetherRegion1} is indeed the true condition governing the leading aether potential.
Moreover, we find that $\bar{\theta}$, $\bar{a}^\mu$, and $\bar{\sigma}^{\mu\nu}$ are
all nonvanishing, so there is no obvious simplification of the
form of $\bar{j}^\mu$ in~\eqref{eq:fEqOneParam}.

Subject to the existence of solutions for $f$, again we have solutions in region I that are very close to GR solutions, 
now to one part in $\sim 10^{5}$, and where the aether is close to being twist free and determined by a single function $f$ obeying a somewhat complicated four derivative condition. Interestingly this condition for $f$, and hence the leading aether solution, is independent of where the theory is in region I, given by the large parameter $c_\omega$.
Thus such GR-like solutions will have an approximately universal aether behaviour, to one part in $\sim 10^{5}$, 
in this phenomenological parameter region.\footnote{
In the transition region discussed in section~\ref{sec:phenoregimes} where both $c_a, c_\theta \sim 10^{-7}$, then the function $f$ is determined in the same way as for region I, except that now the analog of~\eqref{eq:fEqOneParam} will explicitly depend on the ratio $c_\theta/c_{a} \sim O(1)$.
}

\subsection{The weak field limit for regions I and II}

The conditions~\eqref{alphaconstraints} define regions I and II precisely so that the deviation of the metric from a GR response to matter in the weak field regime is small at leading order in a PPN expansion. However our discussion above suggests that not only should the metric behaviour in region I and II be close to GR, but the aether should also be approximately twist free and in the case of region II additionally expansion free. It is instructive to see this emerge in the weak field PPN calculation.  In Appendix~\ref{app:PPN} we review the relevant results from the original Einstein-aether PPN calculation~\cite{Foster:2005dk} and use them to compute the twist and expansion of the aether. 

Interestingly we find that the condition $c_\sigma = 0$ is sufficient at leading non-trivial PPN order to give a twist free aether. 
Thus the aether being twist free in weak field at leading PPN order is generic for $c_\sigma=0$, even if the metric response does not behave as GR (as seen through the preferred frame parameters $\alpha_{1,2}$). Indeed this is precisely the reason that $c_\omega$ does not enter the expressions for $\alpha_{1,2}$ in equation~\eqref{alphaconstraints}.

Further restricting to regions I and II the aether takes the expected form $u_\mu = \bar{u}_\mu + c_a \, u^{(1)}_{\mu} + O(c_a^2)$ with universal $\bar{u}_\mu = k \partial_\mu f$. As shown in Appendix~\ref{app:PPN}, in region II one finds the potential function simply goes as time, $f = t$, at leading non-trivial PPN order, which indeed results in $\bar{u}_\mu$ being expansion free. For region I the potential involves a contribution from the matter, and consequently  the expansion does not vanish, again as expected.
Since $c_\sigma = 0$ is sufficient to give a twist free aether at leading PPN order, in this weak field limit the leading correction to $\bar{u}_\mu$, given by $u^{(1)}_\mu$, and all subsequent terms in the $c_a$ expansion of $u_\mu$ must also be twist free. More generally for strong fields the aether will only be twist free in the $c_a \to 0$ limit, so these corrections will have non-vanishing twist as we explicitly see later for stationary black holes.

\subsection{Region II static black holes in the near GR limit}

An example of this GR limit for black holes in region II can be given analytically in the static case via the exact solutions  in~\cite{Berglund:2012bu} which were found for $c_a = 0$. Taking $c_{\sigma} = 0$ as we do here, then for any $c_{\omega}$ and $c_{\theta}$, an exact solution for the metric and aether is simply given by Schwarzschild,
\begin{eqnarray}
\label{eq:exact1}
ds^2 = - \left(1 - \frac{r_0}{r}\right) dv^2 + 2 dv dr + r^2 d\Omega^2 
\end{eqnarray}
with the aether taking the form,
\begin{eqnarray}
\label{eq:exact2}
u = \alpha(r) \frac{\partial}{\partial v} + \beta(r) \frac{\partial}{\partial r} 
\; , \quad 
\beta(r) = - \frac{r_{\text{\ae}}^2}{r^2}
\end{eqnarray}
with $\alpha(r)$ determined from $\beta(r)$ by the norm condition, and $r_{\text{\ae}}$ being a parameter. In spherical symmetry any aether is twist free. Writing the aether as $u_\mu = k \partial_\mu f$ we see the potential takes the form,
$f = v + \phi(r)$. One can verify that this aether has zero expansion. Hence this represents the leading aether behaviour as $c_a \to 0$ for the parameter region II. From the discussion above, a static black hole in region II would approximate Schwarzschild with this aether to an accuracy of order $O(10^{-7})$.

The parameter $r_{\text{\ae}}$ above reflects the fact that the expansion free condition is a differential equation with global data. However, as shown in~\cite{Berglund:2012bu}, if one requires the existence of a universal horizon, then $r_{\text{\ae}} = \frac{3^{3/4}}{4} r_0$.
For region II in the limit $c_a \to 0$ both the spin-0 and spin-1 wavespeeds diverge, and hence existence of a smooth universal horizon is the same statement as existence of 
smooth horizons for these degrees of freedom.
Thus in the static spherically symmetric case for region II it seems that, for a given mass, determined here by $r_0$, there is a unique aether independent of the aether coupling
parameters 
in the limit $c_a \to 0$.

The universal horizon is located at $r = \frac{3}{4} r_0$.
While the aether vector field is smooth there,
the function $\phi$, and hence $f$, diverges as $\sim -\log|r- \frac34 r_0|$, being explicitly given as,
\begin{eqnarray}
\phi(r) = -r - r_0 \log\left[ \left( \sqrt{3} ( 20 r + 7 r_0) + 9 \Delta \right) \left( \frac{\left| r - \frac{3}{4} r_0 \right|}{4 r + \frac{3}{2} r_0 + \frac{3 \sqrt{2}}{4} \Delta} \right)^{\sqrt{\frac{27}{32}}} \right] 
\end{eqnarray}
where we have defined $\Delta = \sqrt{16 r^2 + 8 r r_0 + 3 r_0^2}$.
Thus the twist potential $f$ is not globally smooth, but is smooth separately in the interior and exterior of the universal horizon where it defines maximal foliations. The exterior foliation is one discussed in~\cite{Estabrook:1973ue,Beig:1997fp}. 
In the case of Kerr again one expects that maximal foliations with the correct asymptotics  only penetrate a certain distance inside the Kerr horizon~\cite{PhysRevD.31.1267}.

For region I we expect an analogous solution, but with a different function $\beta$ corresponding to an aether potential function that satisfies the condition $\bar{j} = 0$. (As discussed above, since any static spherically symmetric aether is twist free, the only solution to~\eqref{eq:aetherRegion1} is that with vanishing $\bar{j}$.) 
We note this solution for the aether is implicit in numerical solutions in~\cite{Zhang:2020too} where static spherical black holes in region I were found, and indeed seen to be approximately Schwarzschild.

\section{Constructing stationary black holes with multiple wave mode horizons}
\label{sec:numericalconstruction}

We have argued that GR may emerge dynamically in the strong field regime for Einstein-aether theory in the phenomenologically allowed regions I and II to a good approximation. Even though various couplings are large, provided that suitable solutions  for the twist free aether potential, $f$, exist to equations~\eqref{eq:aetherRegion1} and~\eqref{eq:aetherRegion2} for regions I and II respectively, the solution closely approximates an aether `painted on' to a usual GR spacetime solution.
We have shown the solution for weak fields indeed takes this form at leading order in the PPN expansion. The key question is now whether solutions to these equations for the aether potential exist in the strong field regime. We thus consider numerical construction of stationary black holes to deduce how solutions behave in parameter regions I and II. From this point onwards, for convenience we choose units such that $8 \pi G = 1$.

While static spherically symmetric black holes are numerically straightforward to construct, since they trivially have a twist free aether due to the symmetry, they are not a good testing ground for studying this behaviour. For example, as discussed above, in region I it is obvious that GR behaviour will emerge if the aether is automatically twist free, as the only large coupling is that associated to the twist term in the action. The key question is then does this persist when rotation is added, as then there is no symmetry reason to protect the aether from having a twist. 
We ultimately will give evidence that `nearly Kerr' black holes with a `painted on' aether indeed exist for both regions I and II.  

Black holes in Einstein-aether are subtle due to having multiple horizons associated to the wave modes propagating at different speeds. This makes the problem novel and we therefore develop new numerical methods to tackle it. 
While it is generic in modified gravity theories to have multiple wave mode speeds, this by no means implies multiple horizons. For many such theories more conventional numerical methods may be used. Thus before we 
outline the new numerical methods for Einstein-aether, we briefly review some general facts about horizons in 
theories with more than one mode speed.
Focusing on theories with multiple effective metrics governing wavemode propagation,
we pinpoint the key difference 
between scalar-tensor theories 
and those with a timelike vector field, 
such as Einstein-aether theory, which makes this problem more subtle.

\subsection{Black hole horizons in theories with different wavespeeds}
\label{sec:horizons}

The setting under discussion is a spacetime with various tensor fields, 
including a metric, which are all invariant under the flow of 
a Killing vector $\chi$ that possesses a Killing horizon, i.e.\ 
a null surface generated by $\chi$. 
We divide the discussion into two 
parts. In the first part, we suppose that the Killing horizon 
has a bifurcation surface 
where $\chi$ vanishes and all the fields
are regular, and explain,
following an argument in~\cite{Jacobson:1993vj},
why this implies that the Killing horizon is a 
Killing horizon with respect to the effective metrics for all wave modes.
We also highlight why the Einstein-aether theory cannot have such solutions.
In the second part we review the results of \cite{Racz:1995nh} that establish conditions under which the existence of such a bifurcation surface 
is guaranteed.

We label the different linearized modes in the theory by an index $i$, 
and  we suppose that the mode equations are all hyperbolic, with local
characteristic surfaces (a.k.a.\ ``causal cones") determined by metric tensors
$g^i_{\m\n}$ that are constructed from the fields of the theory. 
Since by assumption all the fields of the configuration are invariant under the flow
of $\chi$, we have ${\cal L}_\chi g^i_{\m\n}=0$. That is, $\chi$ is a
Killing vector for all of the metrics. Moreover, this implies that 
the scalar quantities $g^i_{\m\n}\chi^\m\chi^\n$ are constant on the flow lines
of $\chi$, and in particular on the null curves that generate the Killing horizon. 
Since these generators all pass through the bifurcation surface 
where $\chi=0$, it follows that these scalars 
all vanish everywhere on the Killing horizon, which means that 
it is a Killing horizon
for all of the metrics $g^i_{\m\n}$.

This is a very powerful argument, but it relies on a strong
assumption: the existence of a bifurcation surface at which all 
of the fields are regular. 
The Einstein-aether theory cannot have such solutions,
since a regular unit timelike vector field cannot be  invariant under the Killing flow at the bifurcation surface, because the flow acts there as a boost in the tangent space at each point and a nonzero 
timelike vector is not invariant under any boost. 
Instead the Einstein-aether theory has black hole solutions which have different future horizons for each wave mode, as one can see for the earlier example in equations~\eqref{eq:exact1} and~\eqref{eq:exact2}, but these cannot be smoothly extended back to a bifurcation surface where the aether is regular.
Furthermore for rotating black holes we will later see these horizons are not even Killing horizons.

Let us briefly return to theories that admit such black holes with a common Killing horizon for all wave modes. 
 The above argument for a common 
Killing horizon
requires a bifurcation surface and fields to be smooth there.
In a maximally extended, analytic solution 
that requirement might be manifestly met, but this is not an
adequate criterion in the more phenomenological setting in
which a black hole forms via collapse of matter, and 
asymptotically approaches a stationary black hole solution in the future. 
What, if anything, can be said in that case? 
Some strong results due to R\'{a}cz and 
Wald~\cite{Racz:1995nh}
are useful here.
What they showed (among other things) is, roughly speaking,
that if  1) a spacetime possesses a 
neighborhood of a future portion of a Killing horizon; 
2) any fields on the spacetime other than the metric are also 
invariant under the Killing flow; 3) the metric and other fields
are either static
(and therefore time-reflection symmetric)
or stationary-axisymmetric 
and invariant under a 
$t$-$\phi$ reflection isometry, 
then
the spacetime metric 
and other fields are all extendable to a neighborhood
of a regular bifurcation surface. 
Field equations play no role in the argument. 
What this means is that while a physical spacetime 
in which a black hole forms by collapse has no real bifurcation surface, 
it has what might be called a ``virtual bifurcation surface". The existence
of such a virtual bifurcation surface suffices for running the argument
given above, so we may conclude that, under the hypotheses of the R\'{a}cz-Wald theorem,
a Killing horizon is a causal horizon for all of the fields in a 
gravity theory.

The first two assumptions are routine, but that
of the reflection symmetry is more subtle. 
For a nonrotating black hole, it seems plausible that
the metric would generically admit a static Killing vector,
whose time reflection isometry would be shared by any scalar
fields. 
But a unit timelike vector, as occurs in Einstein-aether theory,
would reverse its time orientation under the reflection, and hence would
not satisfy the requirement. 
Furthermore we will later show that rotating Einstein-aether black holes do not possess
 the $t$-$\phi$ orthogonality property, 
which presumably means they do not 
admit a $t$-$\phi$
reflection isometry.

Above we have assumed the local characteristic surfaces are governed by metric tensors. However as recently discussed in~\cite{Reall:2021voz} one may have more complicated situations where the principle symbol may not be decomposed in such a simple manner. Interestingly in the explicit scalar-tensor theories considered there, which all had second order equations of motion, it was shown by direct analysis of the principle symbol, that a metric Killing horizon is a characteristic surface for all degrees of freedom in the theory, without direct  reference to the bifurcation surface of the metric horizon.

\subsection{Harmonic formulation}

In order to numerically construct the stationary black hole spacetimes we will employ the harmonic method outlined in~\cite{Headrick:2009pv,Adam:2011dn,Figueras:2012rb}.
The harmonic Einstein equation 
takes the form (in four dimensions),
\begin{eqnarray}
\label{eq:harmonic}
R^H_{\mu\nu}[g] \equiv R_{\mu\nu} + \nabla_{(\mu} \xi_{\nu)} = T_{\mu\nu} - \frac{1}{2} T g_{\mu\nu}
\end{eqnarray}
where $T_{\mu\nu}$ is the stress tensor, and the vector $\xi^\mu = g^{\alpha\beta} \left( \Gamma^\mu_{~\alpha\beta} - \bar{\Gamma}^\mu_{~\alpha\beta} \right)$, with $\bar{\Gamma}$ a smooth reference connection on the manifold which we take to be the Levi-Civita connection of a smooth reference metric $\bar{g}$. 
In the context of Riemannian geometry this modification of the Ricci tensor was due to DeTurck who used it in the analysis of Ricci flow \cite{DeTurck}.
The harmonic Einstein equation is then solved in conjunction with the matter equations of motion.
This formulation removes the coordinate invariance of the equations, and in the vacuum gravity case (ie. $T_{\mu\nu} = 0$) gives a principle symbol controlled by the metric itself. 
The vector $\xi$ should vanish if we wish to find a solution to the Einstein equation, rather than a solution with non-vanishing $\xi$, which we term a `soliton' as in vacuum  equation~\eqref{eq:harmonic}  is known as the Ricci soliton equation.
We may view the vanishing of $\xi$ as a gauge condition, with the resulting coordinates determined by the prescribed reference metric $\bar{g}$.
This takes the form of a generalized harmonic gauge condition~\cite{Friedrich,Garfinkle:2001ni} with the coordinate functions locally obeying $\nabla_S^2 x^\mu = - g^{\alpha\beta} \bar{\Gamma}^\mu_{~\alpha\beta}$, where $\nabla^2_S$ is the scalar Laplacian.
Clearly in order to find solutions of the Einstein equation we must ensure boundary/asymptotic conditions such that $\xi$ may vanish there.

The existence of solutions with non-vanishing $\xi$ is constrained. We may understand this as the vector satisfies the linear equation
\begin{eqnarray}
\label{eq:Oop}
O \xi_\mu =  \nabla^2 \xi_{\mu} + R_{\mu}^{~~\nu} \xi_{\nu}= 0 \, .
\end{eqnarray}
Clearly $\xi = 0$ is a solution. However for $\xi \ne 0$ to be a solution, the linear operator $O$ must have a non-trivial kernel. If  boundary/asymptotic conditions ensure $\xi \to 0$, then the kernel may be forced to be trivial. This has been proven to occur in the vacuum GR setting for stationary black holes in~\cite{Figueras:2011va,Figueras:2016nmo}. In practice one may simply check that solutions obtained by solving the harmonic Einstein equation are not solitons.

\subsection{Review of method for stationary black holes with a Killing horizon}

We now briefly review the use of this harmonic formulation in the conventional stationary black hole setting where there is a single Killing horizon. In the rotating case we assume rigidity holds so that the black hole rotates in the direction of an asymptotically spatial Killing vector (eg. $\partial_\phi$ for Kerr). Then the problem of finding the exterior to the horizon may be phrased as an elliptic boundary value problem on a spatial slice that extends from infinity and terminates at the bifurcation surface~\cite{Adam:2011dn}. The horizon and asymptotic regions form the boundaries, and the lack of dependence of the metric components on time or the rotation direction implies that the principle symbol is elliptic. The boundary conditions at the horizon ensure its regularity, and also fix the physical data of the black hole; its surface gravity and the velocities of the horizon.

In this stationary setting, with a reasonable initial guess, one can hope to solve the resulting elliptic p.d.e.'s to find the desired $\xi = 0$ solution. This is typically achieved either as a flow (such as DeTurck flow in the vacuum case) or more directly as we will do here, using a Newton method. 
While one usually thinks of using the Newton method after finite differencing in order to solve the resulting coupled non-linear algebraic equations, we may formally think of it prior to this in the continuum. In the case of the vacuum equations it is simply,
\begin{eqnarray}
g_i = g_{i-1} - \Delta[g_{i-1}]^{-1} R^H[g_{i-1}]
\end{eqnarray}
where $g_i$ are metrics with $g_0$ an initial guess and $\Delta$ is the linearization, $R^H[g + \epsilon h] = R^H[g] + \epsilon \Delta[g] h + O(\epsilon^2)$, and the aim is that given a good initial guess $g_0$ then $g_i$ tends to a solution as $i \to \infty$.
Considering the Newton method in the continuum we see that provided $g_{i-1}$ is a smooth metric, and the problem is well posed so that the inverse $\Delta^{-1}$ exists, then we expect the new metric $g_i$ to remain smooth.

\subsection{An in-going approach}

 The approach described above is based  implicitly on the black hole of interest having a smooth Killing horizon, and hence a bifurcation surface that forms a boundary of the domain.
This will not work if the theory does not admit black holes with a common horizon for all wave modes, as for the Einstein-aether theory. If there are other fields with effective metrics which have horizons inside the metric one, as is the situation for Einstein-aether due to the Cherenkov constraints $s_{0,1}^2 \ge 1$, such an approach cannot correctly impose regularity of these horizons and the problem will be ill-posed.

Instead here we will introduce a new approach using in-going coordinates,
taking the coordinate domain to pierce inside the horizon of the problem, or horizons if there are multiple effective metrics.
This in-going approach has previously been used in~\cite{Figueras:2012rb,Sonner:2017jcf} where certain exotic black holes with `non-Killing horizons' were found that naturally occur in the holographic context with AdS asymptotics. Here we will use a similar approach, although the reason is somewhat different. In that holographic setting there was a unique horizon, but it failed to be Killing. Here the issue we must address is the presence of multiple effective horizons for different degrees of freedom inside the metric horizon. Our aim is to extend the domain far enough inside the metric horizon to capture all these effective horizons.
An early version of this method applied to Einstein-aether black holes can be found in~\cite{Adam:2013aa}.

\begin{figure}
\centerline{
  \includegraphics[width=10cm]{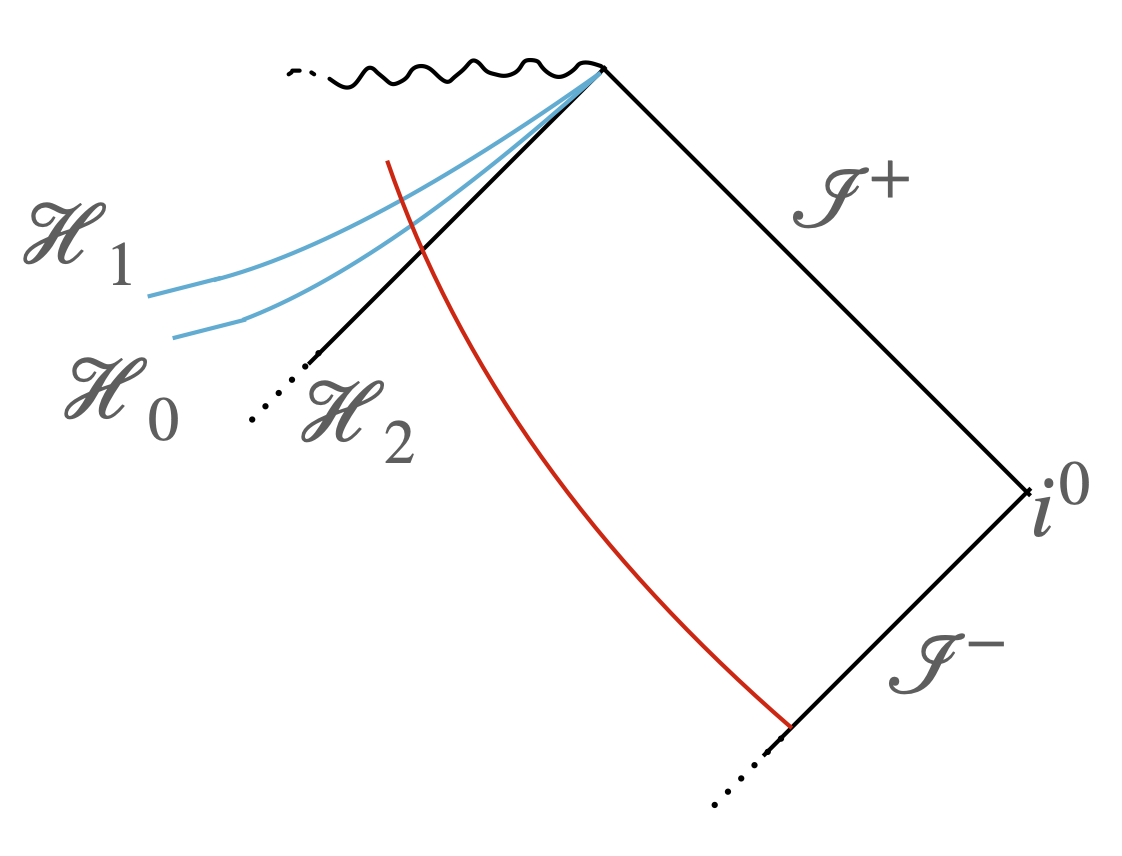}
  }
  \caption{\label{fig:ingoing}
  Figure schematically showing the computational domain embedded in a stationary spacetime with multiple wave mode horizons $\mathcal{H}_{0,1,2}$. The conformal diagram is drawn with respect to the effective metric with the outermost horizon, $\mathcal{H}_{2}$. Outside the horizons the harmonic Einstein equation has an elliptic character, while inside all the horizons it will be hyperbolic in character. Consequently there is no boundary condition to impose at the innermost points of the coordinate domain. 
Although a rotating black hole does not actually admit a planar Penrose diagram like this one, the relation between the computational domain and the horizons is faithful, and each point on the red line has the topology of an angular two-sphere. The computational 
  domain is therefore closed except at its inner and outer ends.}
\end{figure}

Let us firstly consider the case where the metric controls all wave speeds using such in-going horizon piercing coordinates. The problem no longer has an elliptic character and becomes hyperbolic inside the Killing horizon. Smoothness of the horizon is imposed simply by the metric functions being smooth in the interior of the domain where the horizon lies. 
As discussed above, considering the Newton method in the continuum, then if the metric at some step is smooth we expect the updated metric will also be smooth, and hence provided the method converges and one starts with a smooth initial guess, the resulting solution should have a smooth horizon.
Inside the horizon the problem is hyperbolic with the solution determined by the initial data set by the metric functions at the horizon. Thus the innermost boundary of the domain should be regarded as the last slice of this hyperbolic evolution, and no boundary condition should be imposed there. 
Now consider the case where there are multiple wave modes with different effective metrics controlling their propagation. We may apply the method in the same way, solving the Einstein and matter equations on an in-going slice that pierces all the horizons associated to these wave modes. Provided the metric functions are smooth, regularity of all the wave mode horizons will be ensured. Starting with a smooth initial guess close enough to a solution we expect the Newton method to maintain smoothness as it iterates to the solution. Outside all the horizons we expect the problem to be elliptic in character, and inside all the horizons we expect it to be hyperbolic. It will have a mixed character in the region between the various wave mode horizons. Again since the problem should be hyperbolic inside all the horizons, no boundary condition should be imposed on the innermost points of the coordinate domain, and instead the equations of motion should be solved there.
This is schematically illustrated in figure~\ref{fig:ingoing}.

In the previously described elliptic setting we expect a solution given the boundary data which fixes the particular black hole of interest by specifying the moduli of the solution, the surface gravity and angular velocities. In our ingoing setting we are imposing only smoothness at the horizon. Thus there is no unique solution to the problem, but rather any stationary black hole will be a solution. In order to have a tractable numerical problem we must have a unique solution (or at least a discrete set of solutions) and thus must fix the surface gravity, angular velocities and any other moduli. 
Without fixing these physical data a method such as the Newton method will fail as the linearized operator $\Delta$ above will not be invertible. An iterative method will also typically fail to produce a solution, and will likely drift around in the space of solutions not settling on any one, or if it does settle it will be arbitrary which one it has chosen.
Here we will employ a simple and numerically stable way to fix the physical data. Suppose we have $n$ moduli that should be fixed, we fix the values of $n$ metric functions at some specific point in the domain. 
Thus the values of certain metric functions are mapped to physical data, assuming solutions exist with those values.

While this in-going approach is certainly more complicated than the usual method outlined above for a single Killing horizon, its strength is that the generalization to multiple wave mode horizons is straightforward. One must simply ensure the coordinate domain pierces sufficiently into the interior of the black hole to capture all horizons of the effective propagation metrics of the various wave modes. Before applying this approach to black holes in Einstein-aether we firstly illustrate this method explicitly with the toy example of recovering the Kerr solution in the simplest setting of vacuum Einstein gravity.

\subsection{Example: GR and the Kerr black hole}
\label{sec:KerrExample}

Suppose we wish to numerically `find' the Kerr solution in vacuum gravity using this ingoing approach. Then we wish to solve the harmonic Einstein equation~\eqref{eq:harmonic} with no stress tensor. We begin by taking an in-going coordinate chart that will cover the exterior of the black hole and penetrate the future horizon so that smoothness is imposed there. We do this by taking the most general metric ansatz compatible with stationarity and axisymmetry generated by $\partial_v$ and $\partial_\phi$ respectively; 
\begin{eqnarray}
\label{eq:metricansatz}
g = - \left( T_0 + T \right) dv^2 - \frac{2 \left( V_0 + V \right) }{z^2} dv dz + \frac{2 U}{z}\, \sin\theta\, \cos\theta \, dv d\theta &-&
 2 \left( W_0 + W \right)  \sin^2\theta dv d\phi 
 \nonumber \\
 + \frac{A}{z^4} dz^2  \qquad \quad + \frac{2 F}{z^2}\, \sin\theta\, \cos\theta \, dz d\theta
  \; &+&  \frac{2 \left( P_0 + P \right)}{z^2} \sin^2\theta dz d\phi   \nonumber \\
 + \frac{B_0}{z^2} e^{S + \sin^2\theta B} d\theta^2    &+&  \frac{2 Q}{z^2} \sin\theta \,\cos\theta\,  d\theta d\phi   \nonumber \\
 &+& \frac{S_0}{z^2} \sin^2\theta e^{ S  } d\phi^2 
\end{eqnarray}
where $\mathcal{F}_0 = \{T_0, V_0, W_0, P_0, S_0, B_0\}$ are fixed functions of $z$ and $\theta$ and the unknown functions to be solved for are then the 10 functions $\mathcal{F} = \{ T, V, W, U, A, P, F, S, Q, B \}$, again depending on $z$ and $\theta$. 
We choose the reference metric $\bar{g}$ to be that above, $g$, with the unknown functions $\mathcal{F}$ vanishing, so that it is given by the fixed functions $\mathcal{F}_0$.
Here we will restrict our solutions to have a reflection symmetry in the equatorial plane of the black holes. 
We emphasize that the form of the metric above does not impose the `$t$-$\phi$' orthogonality property, ie. that the 2 planes orthogonal to the Killing vectors $\partial_v$ and $\partial_\phi$ are integrable. As we discuss later, this implies that the above ansatz does not assume horizons will take the form of a Killing horizon.

Here we think of $v$, $\theta$ and $\phi$ as analogous to those in ingoing Eddington-Finklestein coordinates for Schwarzschild, with $M/z$ being analogous to the usual radial coordinate $r$ in that coordinate system, where $M$ is the black hole mass. Hence we take $z = 0$ to be the asymptotic infinity $\mathcal{I}^-$, and the computation domain (given reflection symmetry in the equatorial plane) is then rectangular with $z \in [0, z_{max}]$ and $\theta \in [ 0, \pi/2 ]$ with $\theta = 0$ the rotation axis and $\theta = \pi/2$ the equatorial plane. 
The value $z_{max}$ should be large enough that this boundary of the domain is entirely contained within the horizon so that smoothness is imposed there provided the functions $\mathcal{F}$ are smooth. If the value $z_{max}$ is too small so that the boundary is outside (any) horizon then the problem will not be well-posed -- it will be analogous to an elliptic problem lacking data on one boundary. 

The various powers of $z$ and factors of $\sin\theta$ and $\cos\theta$ are included in \eqref{eq:metricansatz} to simplify the various boundary and asymptotic conditions. Regularity of the axis of symmetry and equatorial plane requires the functions $\mathcal{F}_0$ and $\mathcal{F}$ to be even in $\theta$ and in $( \pi/2 - \theta)$. Taking the $\mathcal{F}_0$ to be smooth functions at the boundary $z=0$ with Dirichlet boundary conditions,
\begin{eqnarray}
T_0 \to 1 \; , \quad
W_0 \to 0  \; , \quad
V_0 \to \mu  \; , \quad
P_0 \to \mu a  \; , \quad
B_0 \, , \; S_0 \to \mu^2  
\end{eqnarray}
where $\mu$ and $a$ are constants then imposes the requirement that the reference metric 
be asymptotically flat. 
We then impose analogous conditions on the metric $g$ by taking $\mathcal{F}$ to be smooth functions at $z = 0$ that all vanish there. 

We now must make a choice for $\mathcal{F}_0$, which determines the metric ansatz and the reference metric, subject to these boundary conditions. Later we will take the reference metric to be the Kerr solution in appropriate ingoing coordinates by choosing,
\begin{eqnarray}
\label{eq:KerrRef}
T_0 &=& 1 - \frac{2 \mu^2 z}{\Sigma} \; , \quad W_0 = \frac{2 a \mu^2 z}{\Sigma} \; , \quad V_0 = \mu \; , \quad P_0 = \mu a \nonumber \\
B_0 &=& \Sigma \; , \quad  S_0 = \frac{\left( \mu^2 + a^2 z^2 \right)^2 - z^2 \Delta a^2 \sin^2\theta }{\Sigma}
\end{eqnarray}
with $\Delta = \mu^2 - 2 \mu^2 z + a^2 z^2$ and $\Sigma = \mu^2 + a^2 z^2 \cos^2\theta$. 
Here $\mu$ is the mass of the reference metric Kerr spacetime and its angular momentum is $a \, \mu$. 
However, in the example of finding Kerr itself in vacuum gravity it would be `cheating' to take Kerr as the reference already! Thus we choose the simpler $\mathcal{F}_0$;
\begin{eqnarray}
\label{eq:ExampleRef}
T_0 &=& 1 - 2\, z \; , \quad W_0 = 2\,a\,z \; , \quad V_0 = \mu \; , \quad P_0 = \mu a \nonumber \\
B_0 &=& \mu^2 \; , \quad  S_0 = \mu^2
\end{eqnarray}
so that the reference metric is not Kerr, and we have not built in Kerr to our metric ansatz.
In either case of $\mathcal{F}_0$ above we compute the mass and angular momentum of the metric $g$ to be;
\begin{eqnarray}
M & = & \frac{\mu}{2}  \partial_z g_{vv} \bigg|_{z=0} \nonumber \\
J & = & - \frac{\mu}{2 \sin^2\theta}  \partial_z g_{v\phi} \bigg|_{z=0} 
\end{eqnarray}
and hence these are determined by the normal gradients of the functions $T$ and $W$ at the $z=0$ boundary. For a solution these quantities should be independent of $\theta$, and we indeed see this in all numerical solutions found. 
We also define the dimensionless `spin' of the black hole,
\begin{eqnarray}
j & = & \frac{J}{M^2}
\end{eqnarray}
which for Kerr equals $a / \mu$ and obeys $| j | \le 1$. 
An important point is that since the metric and reference metric agree asymptotically as $z \to 0$, this implies that the vector $\xi$ vanishes there. As discussed above, it is crucial to ensure that $\xi$ may vanish on boundaries of the domain where boundary conditions are imposed. 

In our ingoing method we construct a portion of the spacetime interior to the horizons.
Smoothness at these and elsewhere is imposed when solving by the Newton method provided our initial guess is smooth since, as discussed above, we expect each update of the metric will preserve smoothness. We generally take this initial guess simply to be the reference metric.
While the reference metric above has a particular mass and angular momentum, determined by $\mu$ and $a$, the solutions (which in this simple example should be Kerr) still admit a two parameter moduli space given by their mass $M$ and angular momentum $J$. The metric and reference metric need not have the same values for these. As discussed above, in order for the problem to be well posed we must constrain the physical data $M$ and $J$. There are many such ways to do this, but we have found a convenient and reasonably stable one is to simply fix the value of the metric functions $T$ and $W$ at the innermost point of the equatorial plane, $z = z_{max}$ and $\theta = \pi/2$. 
We call the values there $T_{data}$ and $W_{data}$ respectively.
Using the Newton method to solve the system we have found this much more stable than trying to  fix the point in moduli space by constraining the actual mass and angular momentum. Having found solutions for given $T_{data}$ and $W_{data}$ one can then tune these to obtain the desired mass and angular momentum by using a Newton method wrapped around the Newton method that solves the harmonic Einstein equation.

Finally one must choose a sufficiently large value of $z_{max}$ that the coordinates penetrate inside the horizon, but not too near the singularity that the numerical system becomes destabilized. Of course one does not know where this coordinate position of the horizon will be a priori, but in practise it is straightforward by trial and error to find values of $z_{max}$.

This method is pragmatic and works well, and accurately reproduces the Kerr family of metrics using the Newton method starting with an initial guess where the metric $g$ is equal to the reference metric $\bar{g}$.
We use sixth order finite differencing to compute derivatives of the metric functions in their rectangular domain. Interestingly we have found higher order pseudospectral methods render our scheme numerically unstable, presumably due to their non-local nature.
At the asymptotic boundary the metric functions $\mathcal{F}$ are fixed to zero. We choose to include the boundary points $\theta = 0, \pi/2$ in the domain (one could choose not to) and there we impose Neumann boundary conditions (one could also choose to impose the equations at these points instead) with the exception of the functions $T$, $W$ at the innermost equatorial point whose values we fix to $T_{data}$ and $W_{data}$. We then impose the harmonic Einstein equation at all interior points and also at the remaining boundary of the domain, $z = z_{max}$. Having found a solution we check that it is indeed consistent with having $\xi = 0$. In this simple vacuum gravity setting `solitons' with $\xi \ne 0$ cannot exist~\cite{Figueras:2016nmo}. While there is no proof for general matter that solitons cannot exist, later when we construct black holes in Einstein-aether we will only find solutions with $\xi = 0$ in the continuum.

We wish to prescribe the dimensionless spin $j = J/M^2$ of the black hole solution. However, due to scale invariance of vacuum GR we do not need to fix the overall mass as we will be interested in only dimensionless measures of the solution.
We arbitrarily fix the scale choosing $\mu = 1$ in the reference metric and ansatz. 
We take $a$ to be equal to the desired ratio $a = J/M$. As discussed above, the metric solution found will generally not have the same values of mass and angular momentum as the reference metric. Instead the values of $T_{data}$ and $W_{data}$ will determine the physical data of the solution. 
Due to the scale invariance we choose to fix $T_{data}$ to be zero. Then we find solutions varying $W_{data}$ using a Newton method until the one with the desired dimensionless spin $j$ is found. See Appendix \ref{app:details} for more details.

We provide details of convergence to the continuum for these Kerr solutions in Appendix~\ref{app:conv}, but in summary here using $40 \times 40$ points for the given reference metric allows a wide family of Kerr solutions to be found with a pointwise metric accuracy of $\sim 10^{-6}$.


\section{Rotating black holes in Einstein-aether theory}
\label{sec:results}

Having demonstrated an implementation of the method in the simple vacuum gravity setting, we now turn to the case of interest,  construction of rotating stationary black holes in the Einstein-aether theory. 
We are interested in the case where the spacetime is asymptotically flat and the aether asymptotes to be oriented in the future time direction, and further we restrict to solutions that have a reflection symmetry in their equatorial plane.
We use the same ansatz as above for the metric in equation~\eqref{eq:metricansatz} and take the functions $\mathcal{F}_0$ in equation~\eqref{eq:KerrRef} so that the reference metric is Kerr. 
Note that we then are implicitly assuming the Einstein-aether black holes, like Kerr, have the rigidity property that they possess both the  stationary asymptotically timelike Killing vector $\partial_v$ and also the commuting azimuthal Killing vector $\partial_\phi$. We also assume the solutions possess the same reflection symmetry that Kerr has in its equatorial plane.
In addition we require an ansatz for the aether and the Lagrange multiplier which we take as follows;
\begin{eqnarray}
\label{eq:aetheransatz}
u & = & H dv + K \sin^2\theta d\phi - \frac{X}{z^2} dz + \frac{Y \sin\theta \cos\theta}{z} d\theta \\
\lambda & = & z^3 L
\end{eqnarray}
We solve for the functions in the metric ansatz, together with those of the aether vector and $\lambda$, so $H, K, X, Y, L$ 
which again are functions of $z$ and $\theta$, making 15 functions altogether. 
The boundary conditions for the metric functions are as described above, and for these five new functions we impose asymptotically at $z = 0$ that,
\begin{eqnarray}
X \to \mu \, , \quad H \to -1 
\end{eqnarray}
with the remaining $K, Y, L$ vanishing there, corresponding to an aether vector which asymptotically is $u = \frac{\partial}{\partial v}$ as we require. 
The factors of $z$, $\sin\theta$ and $\cos\theta$ ensure that these five functions are smooth at $z = 0$ and also even in $\theta$ and $(\pi/2 - \theta)$ as for the functions in the metric.
We then proceed to find black holes, where we solve the harmonic Einstein equation with the Einstein-aether stress tensor, together with the constraint $u^2 = -1$ and the aether vector equation (part of which determines the Lagrange multiplier). We have implemented these equations using the original formulation given in Appendix~\ref{app:aetherTheory} rather than using the irreducible parameterization in the main text. These differ only in the Lagrange multiplier variable.

In the previous toy example of finding Kerr we know from uniqueness theorems that there are two moduli that specify a solution. These can be thought of as the mass and angular momentum of the black hole which, given a reference metric, translate into the values $T_{data}$ and $W_{data}$.
For Einstein-aether black holes there are no such uniqueness theorems known. In the case of spherically symmetric static black holes one can count the data specifying solutions and deduce that 
for asymptotically flat solutions with regular horizons
there is one modulus corresponding to the mass~\cite{Eling:2006ec}. In particular, once smoothness of horizons for the various degrees of freedom is imposed, there is no additional continuous data associated to the aether. In the axisymmetric rotating case here we assume that the continuous moduli are mass and angular momentum as in the vacuum GR case, once asymptotic flatness and smoothness of all horizons is imposed. This is compatible with our numerical results results, since
to obtain a numerical system that is solved by the Newton method we must fix two pieces of data as in the toy Kerr example. If we do not fix this no solution is found, as a continuum of solutions then exists and the Newton algorithm does not converge to any one of these. Furthermore, starting with different initial guesses we always find precisely the same solution, which indicates that there are no more continuous moduli than the two we expect. 
We note that while it is possible there is
a discrete set of black hole solutions for a given spin, in all cases we studied we only found one solution.
It would clearly be interesting to prove that stationary black holes have two continuous moduli. 

As for pure gravity the vacuum Einstein-aether theory is scale invariant. Thus again we do not need to fix the overall mass and are interested only in dimensionless measures of the geometry and aether.
As detailed above for pure gravity we again take $\mu = 1$, choose $T_{data}$ to be zero.
Making a reasonable initial guess for the aether functions $H, K, X, Y$ and $L$, we are then able to find rotating Einstein-aether black holes.
We may then use the Newton method to vary $W_{data}$ to move between solutions to find one with the desired spin $j = J/M^2$.

Solutions presented here were computed using a resolution of $40 \times 40$. 
In the Appendix~\ref{app:conv} we give detailed information about convergence tests for Einstein-aether black holes. 
The differencing is implemented using sixth order accurate stencils, and we find the method achieves this order of convergence in practice. We note that the order of convergence is sensitive to the relative location of the various horizons and the extent of our computational domain in the radial direction; if $z_{max}$ is too large compared to the coordinate location of the innermost horizon, we observe the convergence order can drop down to fourth order.
From the convergence tests we estimate $40 \times 40$ typically gives a pointwise accuracy in the metric functions of better than $10^{-5}$. 
For all solutions we checked that the vector used to define the harmonic Einstein equation, $\xi^\mu$ in equation~\eqref{eq:harmonic}, was numerically consistent with vanishing. Interestingly we found no `soliton' solutions where $\xi^\mu$ did not vanish in the continuum limit. 

We note that it is sometimes convenient in Einstein-aether theory to work with a redefined metric which can be chosen as one of the effective metrics governing the propagation of the various wave modes. This was used when constructing static spherically symmetric black holes in \cite{Eling:2006ec,Barausse:2011pu} to set the spin-0 effective metric to be the metric of the redefined theory, ie. to ensure $s_0^2 = 1$. In that context the only active degree of freedom is the spin-0 mode and hence this considerably simplifies construction of the black holes, as there is really only one horizon to be concerned with. However for rotating black holes which have only axisymmetry all the wave modes modes are active, and thus we have not used this freedom. It would allow one degree of freedom to have its horizon made to coincide with the metric horizon but there will remain ones that generally do not, and so no fundamental simplification would be made.

\subsection{Adding rotation to previously found static black holes}

Before turning to the phenomenologically allowed couplings we first verify our numerical code by reproducing results for static spherically symmetric black holes found previously in~\cite{Eling:2006ec,Barausse:2011pu} where $c_1$ is varied taking $c_3 = c_4 = 0$ and $s_0^2 = 1$, with the last condition fixing $c_2$. While in the rest of this paper we take vanishing $c_\sigma$ motivated by the strong LIGO constraints, in order to compare with these older results we take non-vanishing $c_\sigma$.
In terms of the irreducible couplings $c_{\omega,\sigma,\theta,a}$ this corresponds to varying $c_\omega$ while fixing $c_\omega = c_\sigma = c_a$, again with $s_0^2 = 1$ which determines $c_\theta$.

For that parameter range the static black holes tend to Schwarzschild as $c_\omega \to 0$, as then all the aether couplings become small. On the other hand taking $c_\omega \to 1^-$ gives strong deviations from GR.
It is interesting to then consider intermediate $c_\omega$, so that deviations from GR behaviour are quite small in the static case. Do rotating solutions deviate more from GR behaviour and hence from Kerr? 

\begin{figure}
\centerline{  \includegraphics[width=10cm]{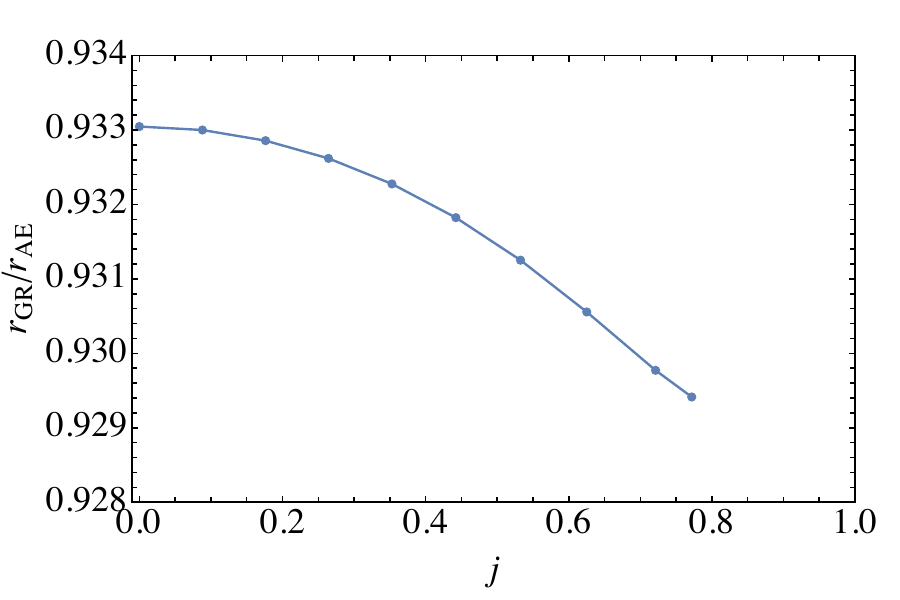}
   \includegraphics[width=10cm]{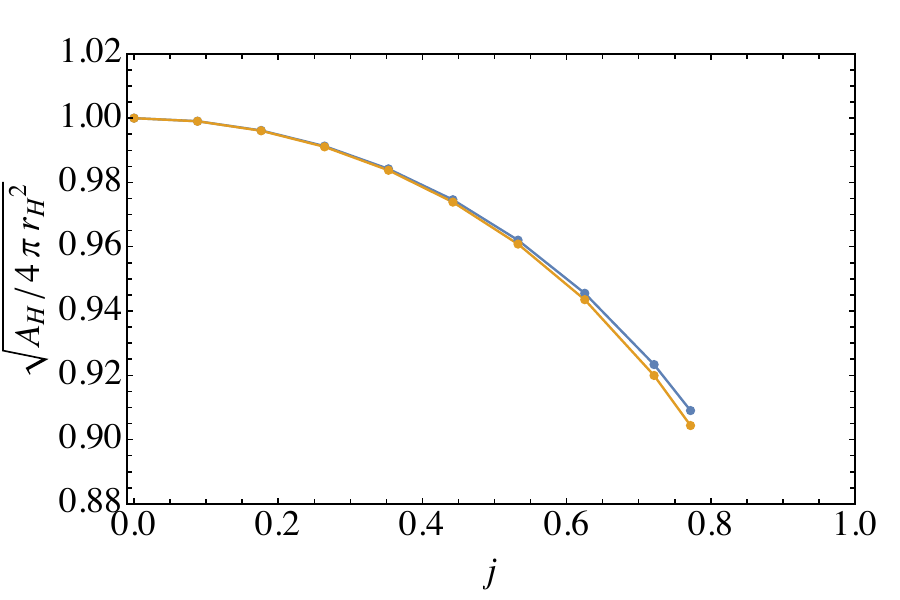}}
\centerline{    \includegraphics[width=10cm]{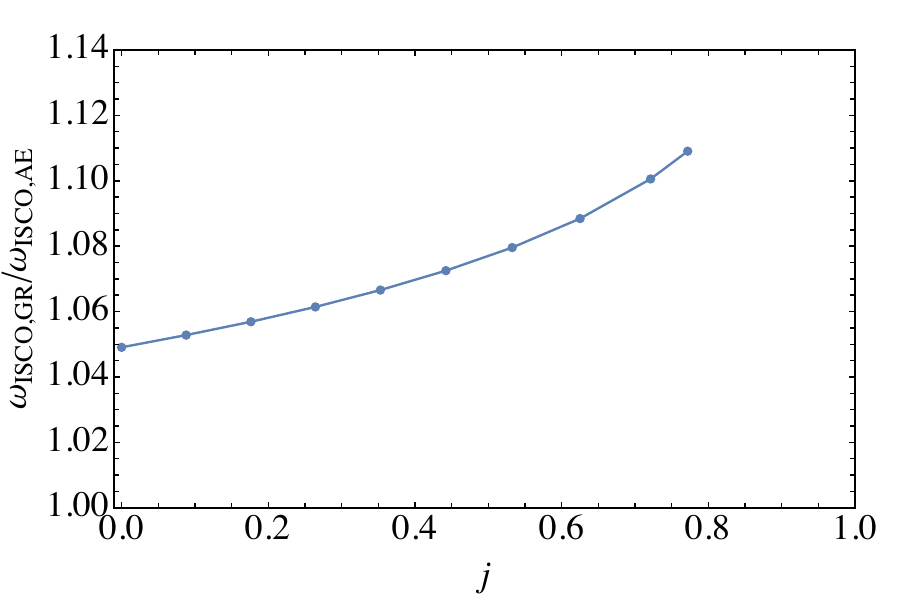}
 \includegraphics[width=10cm]{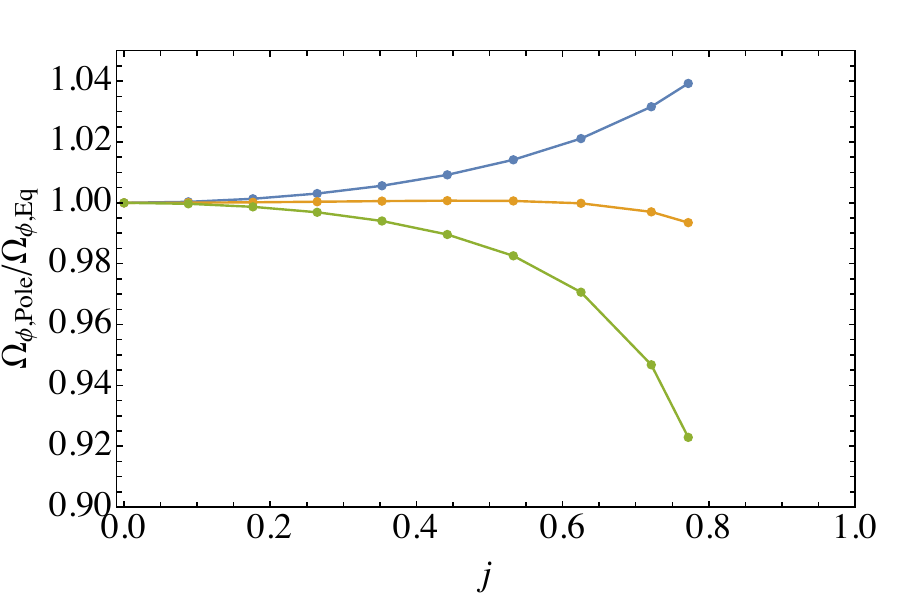} }
  \caption{\label{fig:comparison}
Comparison of properties of rotating Einstein-aether black holes with Kerr for the same spin and mass, for coupling parameters
  $c_\omega=c_\sigma=c_a=0.5$ and $c_\theta$ determined by $s_0^2=1$ (the same couplings 
  as a case previously considered 
  for static black holes 
   in~\cite{Barausse:2011pu}).
  The top left plot shows the ratio of the equatorial horizon radius for Kerr to that of the aether black holes, vs.\ spin $j= J/M^2$. The top right plot shows $\sqrt{A_H/4\pi r_H^2}$ with aether black holes in blue and Kerr in orange. 
  The lower left plot shows the ratio of the ISCO frequency for Kerr to the aether solutions. The lower right plot shows the ratios of the angular velocity $\Omega_\phi$ at the pole to that at the equator for the spin-0 (blue), 1 (orange) and 2 (green) wave mode horizons. These deviate from one since, unlike for the Kerr metric horizon, these are not Killing horizons. 
 }
\end{figure}

We show data here for the intermediate value $c_\omega = c_1 = 0.5$. We find qualitatively similar results for other intermediate values where we are able to find solutions. Varying the spin $j$ up to $\sim 0.8$ it is straightforward to find rotating black hole solutions.
In figure~\ref{fig:comparison} we show various quantities computed for these Einstein-aether black holes and compared to Kerr for the same mass and spin. We show the ratio of the equatorial radius of the horizon (defined as the circumference divided by $2\pi$) of Kerr to that of the metric horizon in the aether theory, $r_{GR}/r_{AE}$. For zero spin this value is given in~\cite{Barausse:2011pu} as $r_{GR}/r_{AE} = 0.93304$. We reproduce this value for zero spin, and see that at least out to spins of $\simeq 0.8$ this ratio is remarkably independent of the spin, indicating a larger equatorial radius for the aether black holes than for Kerr of the same mass and spin. 
We also plot the quantity  $\sqrt{A_H/4\pi r_H^2}$, with $A_H$ the horizon area and $r_H$ the equatorial radius, which measures the deviation from sphericity, both for Kerr and the aether black holes. Again we see little variation between these as spin is added, the aether black holes being marginally more spherical than Kerr. 
Interestingly, taken together these results suggest the aether carries a net negative energy for these solutions, allowing a larger black hole for the same mass and spin than for Kerr.
Finally we show the ratio of the ISCO frequency of Kerr to that of the aether black hole with the same mass and spin. Here we see that a greater deviation from GR develops as we increase the spin of the solutions, although it is not a dramatic change in behaviour, at least out to spin $\simeq 0.8$.
Since the ISCO is closer to the horizon for higher spin, it is natural that the deviation in the frequency is greater for higher spin, since it's more sensitive to the change of horizon radius. Note also that the larger horizon for the aether black holes relative to Kerr naturally leads to a lower ISCO frequency.

An interesting possibility is that closed timelike curves may be present in the effective metrics for the various degrees of freedom behind their horizons. A simple class of such curves are those generated by the $\phi$ circle, such as occur inside the inner horizon of Kerr. We have checked whether such curves occur by examining the component $(g_{\rm eff})_{\phi\phi}$ which if negative would yield such a closed timelike curve, and found it to be positive
within the coordinate domain of all obtained solutions. Of course this does not preclude closed timelike curves of this form further in the interior of the spacetimes than we have constructed, or the existence of more general closed timelike curves than those purely generated by $\partial_\phi$.

\subsection{Horizons}
\label{sec:nonKillinghorizons}

A black hole horizon is usually defined as the boundary of the causal past of future null infinity, where the causal structure is determined by the spacetime metric. 
In Einstein-aether theory, each of the wave modes is associated with 
an effective metric $g^{\rm eff}_{\mu\nu}$ \eqref{eq:effmetric}, which defines
a corresponding notion of event horizon.\footnote{For all the 
cases studied here either 
the spin-0 or the spin-2 horizon coincides with the metric horizon.}
These horizons are all 
hypersurfaces that are null with respect to the appropriate metric. 
For stationary axisymmetric solutions, as considered here, the Killing 
vectors $\partial_v$ and $\partial_\phi$ are tangent to any horizon, 
so the horizon is the hypersurface which, at least locally, can be defined by $z = z_H(\theta)$.
Then defining the normal 1-form, 
\begin{equation}\label{n}
n \equiv d(z - z_H(\theta)),
\end{equation} 
the condition that the horizon is a null hypersurface is the condition that $n$ is null with respect to $g_{\rm eff}$, i.e.\ $(g^{-1}_{\rm eff})^{\mu\nu} n_\mu n_\nu = 0$.\footnote{We emphasize that $g^{-1}_{\rm eff}$ is the inverse of $g_{\rm eff}$, rather than that metric with indices raised with the usual spacetime metric $g$. Explicitly this is $(g^{-1}_{\rm eff})^{\mu\nu} = g^{\mu\nu} + \left( 1 - \frac{1}{s_{(a)}^2} \right) u^\mu u^\nu$.}

Stationary axisymmetric black hole horizons in general relativity are typically Killing horizons,
that is, there is a Killing vector tangent to their null generators.
This is automatically the case for static, spherically symmetric black holes, but for
rotating black holes it is a nontrivial property. For the Kerr horizon, for example, the
null generators are tangent to the Killing field $\partial_v + \Omega_H\partial_\phi$, where
$\Omega_H$, the angular velocity of the horizon, is a constant. 
A theorem of Hawking~\cite{Hawking:1971vc} proved 
that stationary black hole horizons
in vacuum or electrovac analytic solutions to Einstein's equation
are Killing horizons. That theorem is not relevant to us here, however,
since the Einstein-aether field equations are different, and since we have no reason to assume analyticity.
However, a different theorem does apply: without invoking field equations, Carter~\cite{Carter} showed that an event 
horizon must be a Killing horizon if the spacetime is 
stationary and axisymmetric, with two commuting
Killing fields whose integral
2-surfaces are themselves orthogonal to 2-surfaces. This integrability condition is called the $t$-$\phi$ orthogonality property. 
We find that in our 
Einstein-aether theory solutions
the $t$-$\phi$ orthogonality property
fails to hold, and 
the horizons are in fact {\it not} Killing
horizons! 
On each wave mode horizon the surface spanned by 
the two Killing vectors is spacelike with respect to the wavemode metric except at the poles and the equator,
and hence
does not include the 
null direction.
The null generators are therefore not parallel to a linear combination of
the Killing vectors; rather, they are parallel to
\begin{equation}\label{angvel}
    \chi =\partial_v + \Omega_{\phi} \partial_\phi + \Omega_{\theta} \partial_\theta + \chi^z \partial_z,
\end{equation}
with nonzero $\Omega_{\theta}$ and $\chi^z$.

We expect that in any solution found by our numerical method all the wave mode horizons are captured within the computational domain, 
since without imposing smoothness at the horizons the system is not expected to be well posed with a unique solution (or possibly a finite set of solutions). Indeed this is borne out in our computations. We generally find solutions only when the coordinate domain extends deep enough to capture all the horizons. If the coordinate domain ends too far out, so that it does not contain all the horizons, the Newton method does not converge.\footnote{Interestingly in certain cases we have found solutions with the coordinate domain just missing the innermost wave mode horizon, but when resolution is increased these do not have the correct sixth order convergence, and so presumably cannot be trusted.}

\begin{figure}
\centerline{ 
  \includegraphics[width=8cm]{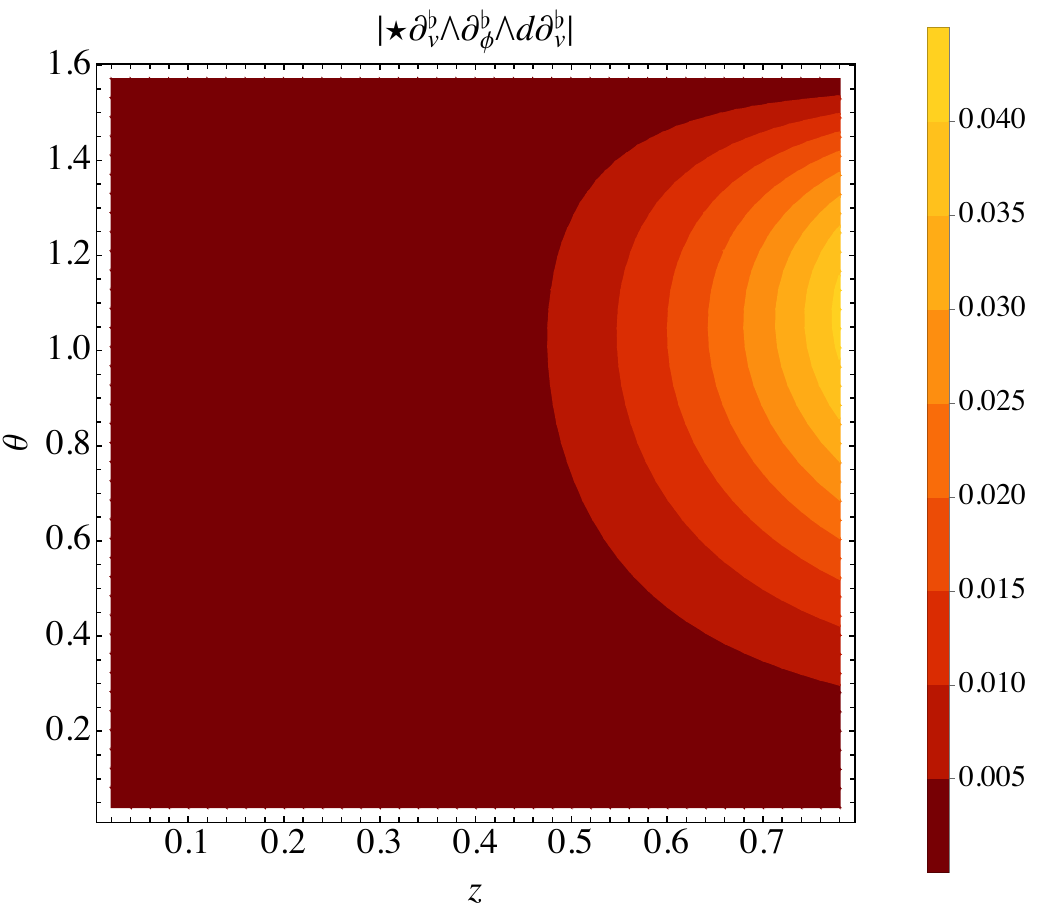}
  \hspace{0.1cm}
  \includegraphics[width=8cm]{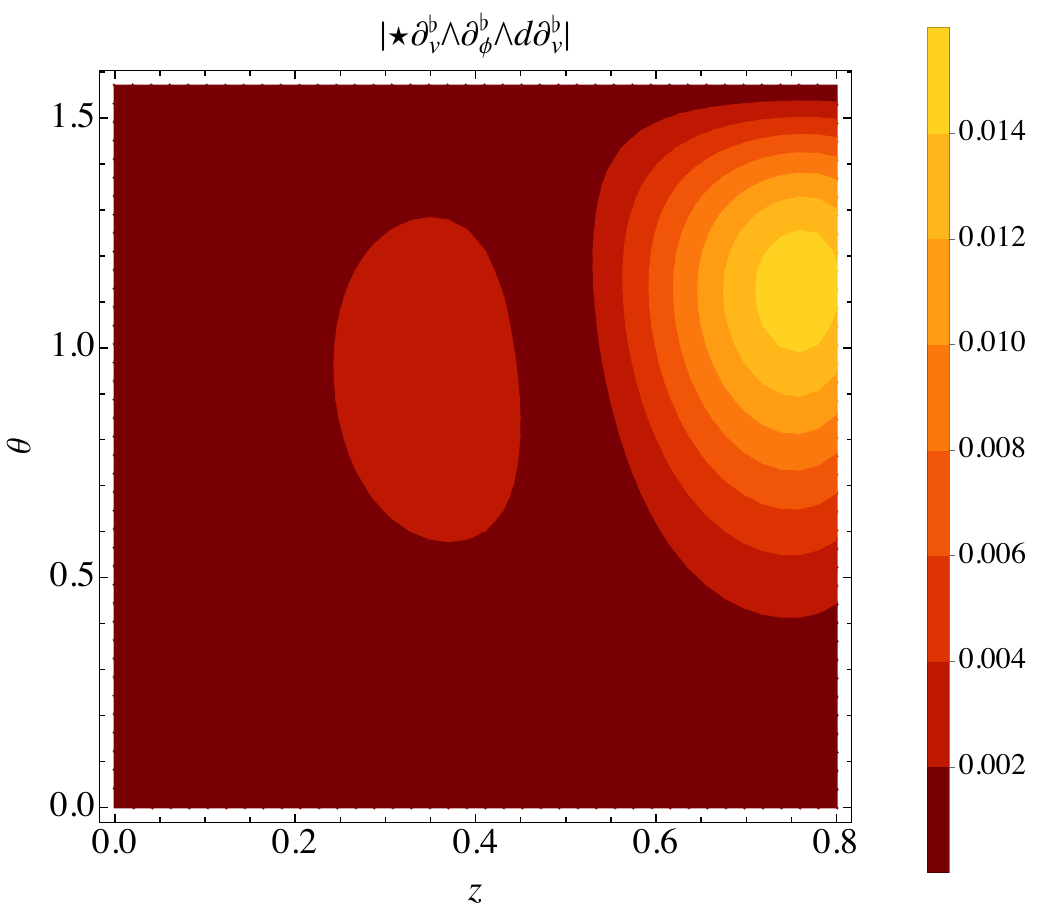}
  \hspace{0.1cm}
  }
  \caption{\label{fig:integ} 
  The lefthand plot shows $|\star \partial_v^\flat\wedge \partial_\phi^\flat\wedge d\partial_v^\flat|$
  for the metric (the same as the spin-0 wave mode metric) for the lefthand solution shown in  figure~\ref{fig:horizons}. The righthand plot shows the same quantity for the spin-1 wave mode metric for the righthand solution shown in  figure~\ref{fig:horizons}.
  These do not vanish, indicating these metrics do not have the $v$-$\phi$ orthogonality property.
  }
\end{figure}

\begin{figure}
\centerline{ 
  \includegraphics[width=10cm]{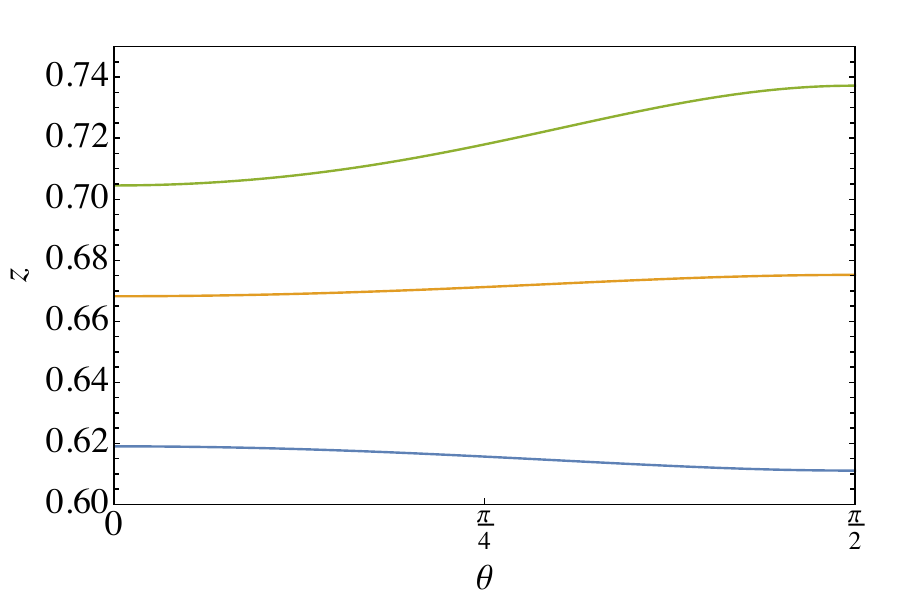}
  \hspace{0.1cm}
  \includegraphics[width=10cm]{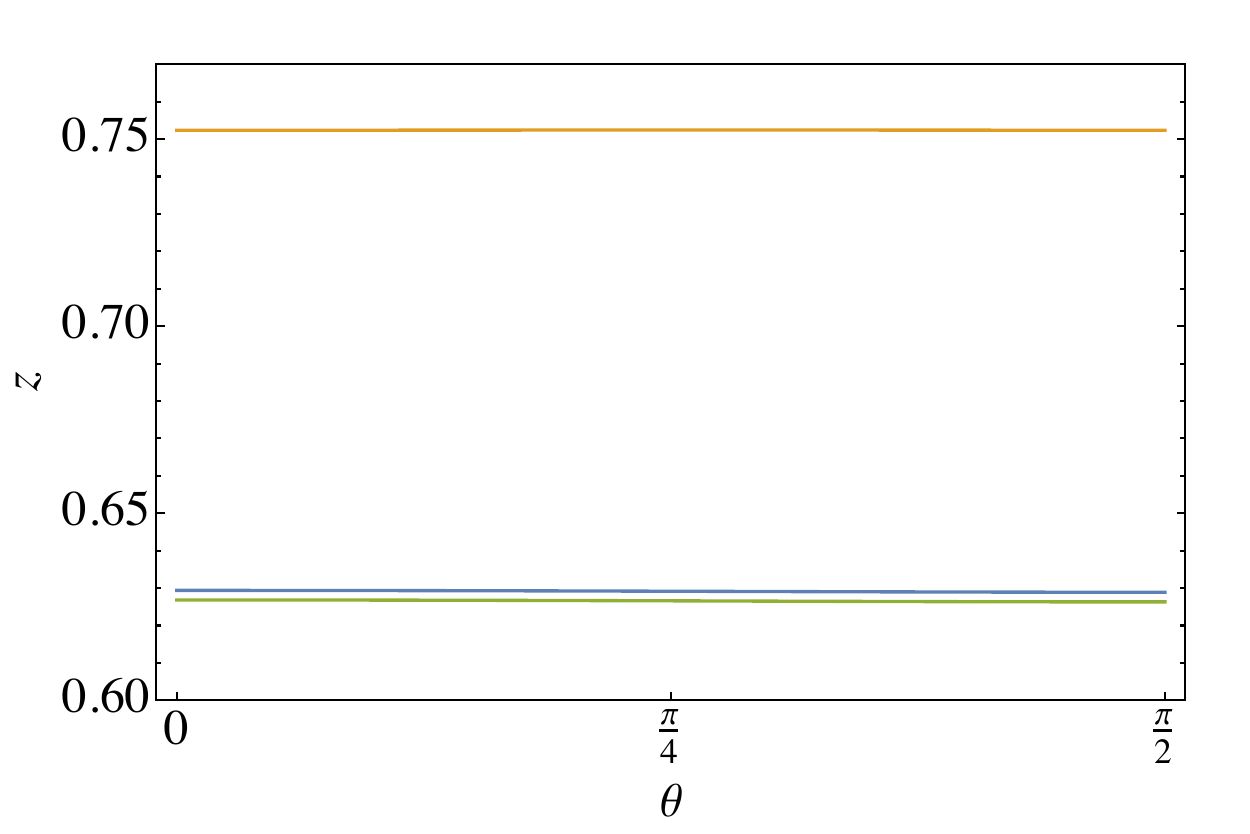}
  }
  \caption{\label{fig:horizons} 
  Location $z_H(\theta)$ of the spin-0 (blue), 1 (orange) and 2 (green) horizons in the coordinate domain. \textit{Left}: the $j = 0.77$ solution for $c_\omega = c_\sigma = c_a = 0.5$, with $s_0^2 = 1$. The coordinate chart extended to $z_{max} = 0.78$ in this case. \textit{Right}: the $j=0.80$ solution for $c_\theta=0.043$, $c_\sigma=0$, $c_\omega=0.264$ and $c_a=0.014$ in the Family IA (see Section \ref{sec:phenobh}). In this case, the speeds of the various modes are $s_0^2=1.038$, $s_1^2=9.753$ and $s_2^2=1$ respectively, and the coordinate chart extends to $z_{max}=0.8$.
  }
\end{figure}

The $t$-$\phi$ ---here actually $v$-$\phi$--- orthogonality property
discussed above corresponds (according to the Frobenius theorem) to the vanishing of the two 4-forms,
$\partial_v^\flat\wedge \partial_\phi^\flat\wedge d\partial_v^\flat$
and $\partial_v^\flat\wedge \partial_\phi^\flat\wedge d\partial_\phi^\flat$
(where $(\partial_v^\flat)_a:=g^{\rm eff}_{ab}\partial_v^b$).
In figure~\ref{fig:integ} we plot the norm of the Hodge dual of the first of these (which is a scalar), for the
metric horizon in the lefthand of the two solutions shown in figure~\ref{fig:horizons}. And for the righthand solution in that figure we plot the same quantity for the spin-1 horizon, where we note the quantity is very small for the metric horizon since that solution is close to Kerr.
As seen in the plots, this integrability condition
fails to hold, so there is no reason the horizon must
be a Killing horizon. 
We see similar non-vanishing behaviour for the second integrability condition.
In Appendix~\ref{app:conv} both these conditions
are checked for the Kerr solution, where it is shown they hold to very good numerical accuracy, despite the use of a non-adapted coordinate system.

The function $z_H(\theta)$ is determined by 
the condition that the normal 1-form \eqref{n} be null, 
$(g^{-1}_{\rm eff})^{\mu\nu} n_\mu n_\nu = 0$,
which defines a non-linear first order ordinary differential equation for $z_H(\theta)$. It is a quadratic in $z_H'(\theta)$ with two solutions,
\begin{equation}
    z'_H(\theta) = \frac{ (g^{-1}_{\rm eff})^{z\theta} \pm \sqrt{ \left( (g^{-1}_{\rm eff})^{z\theta} \right)^2 - (g^{-1}_{\rm eff})^{zz} (g^{-1}_{\rm eff})^{\theta\theta}}  }{ (g^{-1}_{\rm eff})^{\theta\theta}  } \; .
\end{equation}
At $\theta=0$ and $\theta=\pi/2$ the normal 1-form
reduces to $n = dz$, so
this equation reduces to the condition $(g^{-1}_{\rm eff})^{zz} = 0$, 
whose solution serves as the initial data for integration
of $z_H(\theta)$ from pole to equator or vice versa. 
In principle multiple solutions might exist, 
in which case the outermost, i.e. the one with the smallest $z$, would correspond to the event horizon. 
In practice we find only one root within our coordinate domains.
Away from the poles and equator 
the plane spanned by $\partial_v$ and $\partial_\phi$  is spacelike with respect to $(g_{\rm eff})_{\mu\nu}$ so
has two orthogonal null directions
which correspond to the two roots in the solution for  $z'_H(\theta)$. 
Only one of these null directions limits to 
the unique null normal at the poles and equator.\footnote{In practice we identify the appropriate root as the one that ensures the discriminant in the quadratic above becomes positive as one integrates away from the pole or equator. We find the other root leads to a negative discriminant, and hence to 
an unphysical, complex-valued solution.
Depending on the horizon, the physical solution may be the one with the positive square root or that
with the negative square root. 
For reasons we do not understand, numerical stability of the integration appears to require us to integrate the 
former from the equator to the pole and the latter from the pole to the equator.
The numerical solutions are then simple to find using Mathematica's {\tt NDSolve}, and they interpolate precisely from the location of the horizon at the starting point of the integration to that at the end point, as determined by $(g^{-1}_{\rm eff})^{zz} = 0$. Attempting to integrate in the opposite sense gives poor numerical behaviour, with strong sensitivity to the initial conditions.
}

In figure~\ref{fig:horizons} we plot the various horizons within the coordinate domain for an example solution, the spin $j = 0.77$ solution considered in the previous section. For that solution the chart was taken to extend to $z_{max} = 0.78$. The spin-0, 1 and 2 horizons are distinct, and the curves defining them are $\theta$ dependent.
Note that for those parameters the spin-0 horizon coincides with the metric horizon. 
In the figure we also present a similar plot for a typical solution taken from a family we construct in the next section~\ref{sec:phenobh}, where the ordering of the wave mode horizons is different, and now the metric horizon coincides with the spin-2 horizon as $c_\sigma$ is taken to vanish.
As discussed in detail later, the spacetime geometry for this solution is close to the Kerr metric with the same spin, and the spin-0 and spin-2 horizons are close to each other and lie close to the corresponding Kerr value $0.625$. 
Interestingly for our choice of reference metric the coordinate positions, $z_H(\theta)$, of the various wave mode horizons appear to have a surprisingly weak dependence on the angle $\theta$ for all the Einstein-aether black holes we have found in this work.

\begin{figure}
\centerline{ 
  \includegraphics[width=10cm]{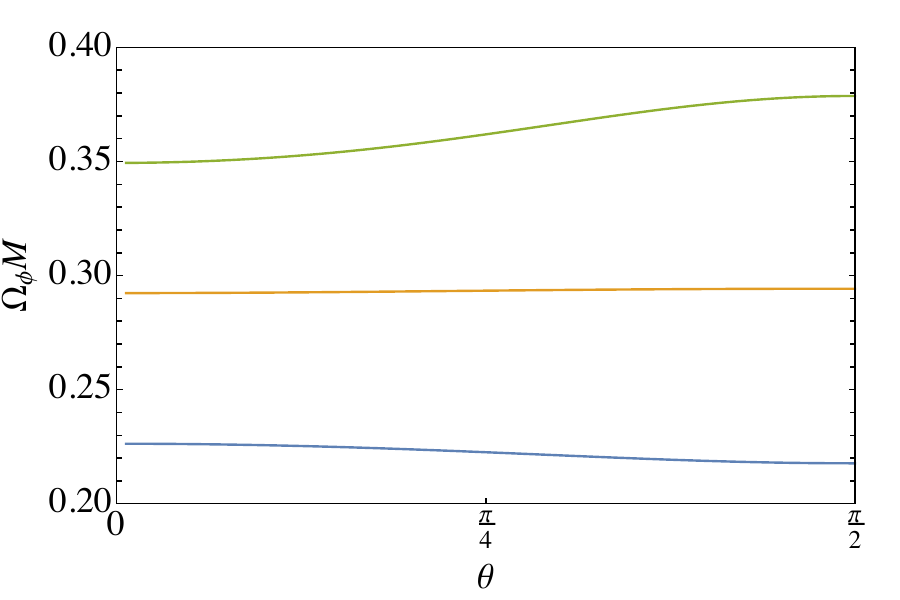}
  \hspace{0.1cm}
  \includegraphics[width=10cm]{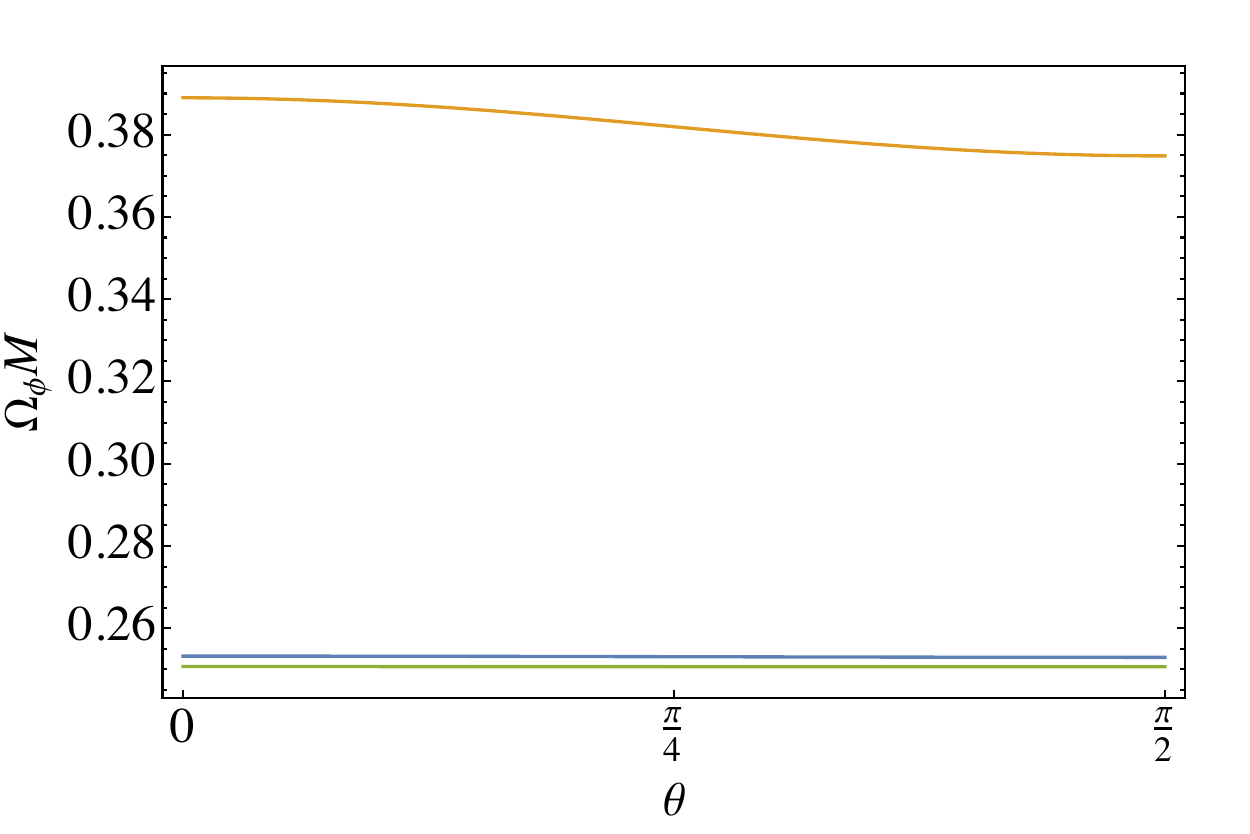}
  }
  \centerline{ 
  \includegraphics[width=10cm]{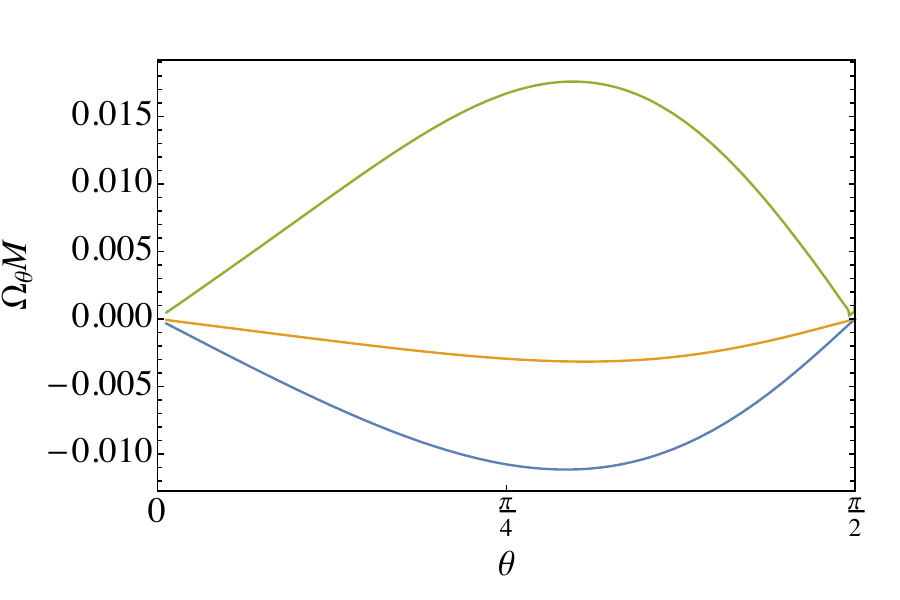}
  \hspace{0.1cm}
  \includegraphics[width=10cm]{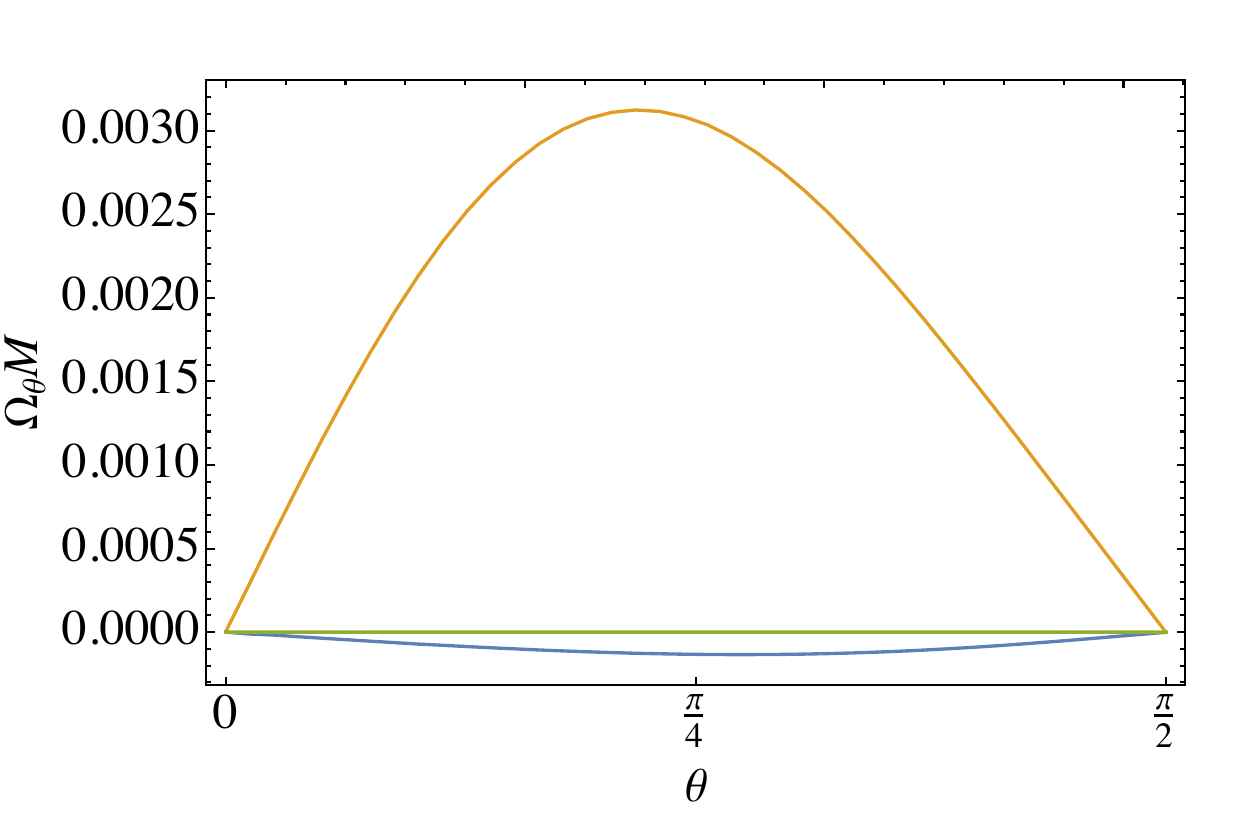}
  }
  \caption{\label{fig:omega} 
  The angular velocities $\Omega_{\phi}(\theta)$ and $\Omega_{\theta}(\theta)$, made dimensionless with the black hole mass $M$, for the wave mode horizons shown in  figure~\ref{fig:horizons}. The non-vanishing $\Omega_{\theta}(\theta)$, and the explicit dependence of $\Omega_{\phi}$ on $\theta$, show that these are not Killing horizons. The corresponding values for Kerr with the same 
  mass and angular momentum are 
  $\Omega_H M =\{0.235, 0.250\}$ for the black holes on the \{left,right\}, respectively.
  }
\end{figure}

In figure~\ref{fig:omega} we plot the angular velocities 
$\Omega_\phi$ and $\Omega_\theta$, corresponding to the horizons shown in figure~\ref{fig:horizons}.
The presence of a nonzero $\Omega_\theta$,
and the explicit dependence of $\Omega_\phi$ on $\theta$, indicate
 that the horizons are not 
Killing horizons. It goes linearly to zero at the pole and the equator, and in each case has one sign everywhere else. Thus on any given horizon the generators spiral either from 
the pole to the equator, or vice versa, over an infinite range of Killing time, with exponential behavior of $\theta$ with $v$ at the two ends.

Since the rotating black hole horizons are not Killing, the velocity $\Omega_\phi$ is not constant on them.
For small spins we expect the variation in $\Omega_\phi$ across the horizon to be quadratic in the spin, rather than linear. This is because the degree of variation should have an expansion in powers of the spin and yet be independent of the sense of the black hole 
rotation, hence independent of the sign of the spin. We confirm this by computing the spin dependence of
the ratio of the value of $\Omega_\phi$ at the pole 
to that at the equator, for the various wave mode horizons. This quantity is shown in the previous figure~\ref{fig:comparison} for the same solutions that are discussed there. It deviates from 1 with precisely $j^2$ dependence for small spins. We see a similar quadratic dependence on spin for the maximum absolute value of the angular velocity $\Omega_\theta$, as the above argument also predicts.

While black holes with non-Killing horizons have been numerically constructed in asymptotically AdS spacetimes~\cite{Figueras:2012rb,Fischetti:2012vt,Sonner:2017jcf}, we believe our solutions are the first instance of black holes which are asymptotically flat and lack $t$-$\phi$ orthogonality, and further have non-Killing horizons of the effective metrics for the physical degrees of freedom,
including in this case the usual spacetime metric horizon.
This was not previously evident 
for Einstein-aether black holes since 
a static, spherically symmetric horizon is always
a Killing horizon. Presumably it
is also not evident in the slow rotation approximation,
working to first order in the spin as in~\cite{Barausse:2015frm}, since the non-constancy of the angular velocity $\Omega_\phi$ and non-vanishing $\Omega_\theta$ appear only at second order.

\subsection{Black holes in phenomenological regimes I and II}
\label{sec:phenobh}

Following our earlier discussion showing that in the phenomenological parameter regions I and II  cf.\ Fig.~\ref{fig:regions} a solution may take the approximate form of a GR solution for the metric, with a twist free aether `painted' on top we might expect to find that the metric is close to Kerr.
Indeed we will show numerically that this does occur, by examining three families of aether parameters that tend towards region I when taking $|c_{a}|$ small, and three that tend towards region II. 
We will call these the region I and region II families respectively.
We write
\begin{eqnarray}
c_{a} = \epsilon
\end{eqnarray}
and then explore the black hole behaviours in the $c_a = \epsilon \to 0$ limit, to understand whether the behaviour matches the expected behaviour in equation~\eqref{eq:GRbehaviour}. In particular we will demonstrate that Kerr with a twist free aether `painted' on emerges in the $\epsilon \to 0$ limit, and also we will study the leading $O(\epsilon)$ approach to this limit. 
For both families I and II we make the choice
\begin{eqnarray}
c_\omega = \lambda + \epsilon \, ,
\end{eqnarray}
so that as $\epsilon \to 0$ we have $c_\omega \to \lambda$.
For the region I families we take
\begin{itemize}
    \item Family IA: $\lambda = 0.25$ and $\frac{c_\theta}{3} = \epsilon +5 \, \epsilon^2$
    \item Family IB: $\lambda = 1.25$ and $\frac{c_\theta}{3} = \epsilon +5 \, \epsilon^2$
    \item Family IC: $\lambda = 0.5$ and $\frac{c_\theta}{3} = \epsilon +3 \, \epsilon^2$
\end{itemize}
so that for $| \epsilon | = O(10^{-5})$ these parameters are within the viable region I. For region II families  we take
\begin{itemize}
    \item Family IIA: $\lambda = 1$ and $\frac{c_\theta}{3} = 0.7 - \epsilon$   
    \item Family IIB: $\lambda = 0.2$ and $\frac{c_\theta}{3} = 0.475  - \epsilon$  
    \item Family IIC: $\lambda = 0.5$ and $\frac{c_\theta}{3} = 1 + 0.5\epsilon$  
\end{itemize}
so that for $| \epsilon | = O(10^{-7})$ these lie within the viable region II. The families above are representative of a more general behaviour.
We have computed black holes for other parameter choices that limit to region I and II for small $\epsilon$, and these give the same behaviours in the limit $\epsilon \to 0$.

For these parameter families we have constructed numerical black hole solution with various spins $j$ starting with $\epsilon = O(1)$ and reducing it to see the limiting behaviour as $\epsilon \to 0$. The largest spin we were able to comfortably find for a good range of $\epsilon$ was $j = 0.8$. With more effort it would presumably be possible to push to greater spins. The results we present in what follows are for this $j = 0.8$ case, but we emphasize that we see exactly the same behaviour for the smaller spins as well.

As we construct the solutions we must make sure $z_{max}$ is sufficiently large to capture all the horizons. In the appendix~\ref{app:details} we discuss details of the construction of these solutions, and in particular the values of $z_{max}$ chosen for them. 
An important point is that, for both the region I and II families, as $\epsilon \to 0$ the spin-1 speed diverges. Thus the spin-1 horizon becomes a universal horizon in this limit. We find that constructing such black holes becomes numerically harder for small $\epsilon$, precisely as to reach a given $\epsilon$ one must ensure the extent of the coordinate chart, $z_{max}$, is large enough to still capture this spin-1 horizon. Typically we find it difficult to extend the families to $\epsilon$ smaller than $\sim O(10^{-2})$. However, as we shall see, this is certainly sufficient to see the limiting $\epsilon \to 0$ behaviour emerging. For $O(1)$ values of $\epsilon$, and hence away from the phenomenological regime, it also becomes hard to find solutions. In both limits, i.e., small and large $\epsilon$, the difficulties seem to arise from the fact that the speeds of the various modes differ significantly, and consequently the coordinate locations of the corresponding horizons become widely separated in our grid. Then the computational domain necessarily covers a large part of the interior of some of the horizons, and we observe it becomes harder to obtain convergence to a solution.

\begin{figure}
\centerline{  
  \includegraphics[scale=0.8]{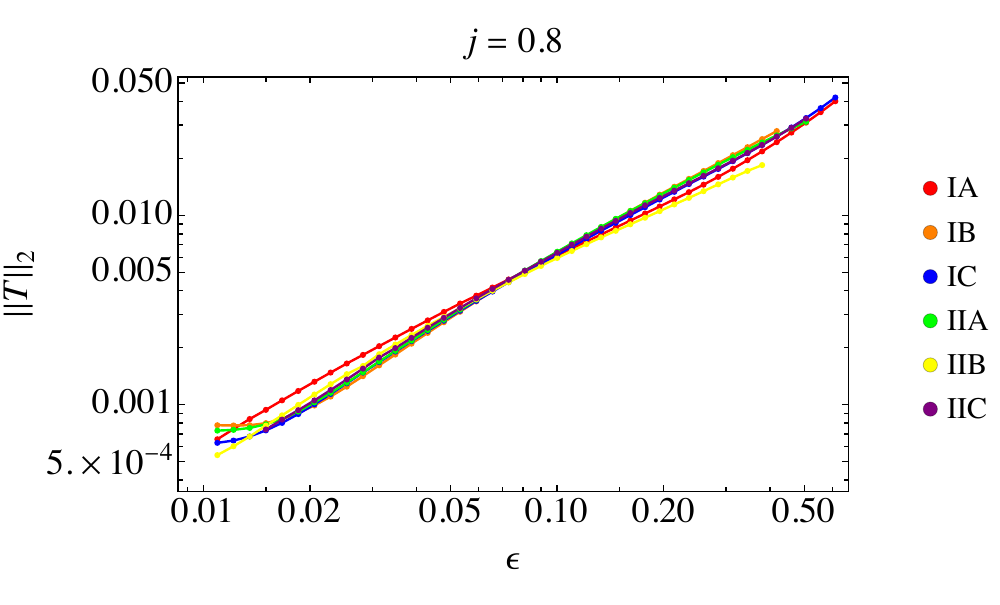}
  }
  \caption{\label{fig:MetricLimit}
  $L^2-$norm of the metric function $T$ taken over the coordinate domain up to $z_{cut}=0.6$ plotted against $\epsilon$. This domain includes the spin 2 horizon for all families considered and some of the other horizons, depending on the parameters. This norm tends to zero as $\epsilon\to 0$ for all the families of black holes, consistent with the spacetime approaching Kerr.
  }
\end{figure}

We wish to show the metric~\eqref{eq:metricansatz} tends to Kerr in the limit $\epsilon \to 0$.
Due to our choice of the reference metric being Kerr with $a = J/M$, and since we further fix $T_{data} = 0$, then conveniently if the metric indeed tends to Kerr it does so by becoming equal to the reference metric so we may see this simply by showing that the various metric functions $\mathcal{F}$ all tend to zero.\footnote{Had we taken $T_{data} \ne 0$ the metric would have tended to Kerr but not in the same coordinate presentation as the reference metric, and so the functions in the metric ansatz would be non-trivial.}
The $L^2-$norm of the metric function $T$ over the coordinate domain is displayed in figure~\ref{fig:MetricLimit} for black holes with spin $j = J/M^2 = 0.8$, where
\begin{eqnarray}
|| T ||_2= \left[ \int_0^{z_{cut}} dz \int_0^{\pi/2} d\phi\, | T |^2\right]^{\frac{1}{2}}\,.
\end{eqnarray}
To compute this norm, we first interpolate the function
$T$ 
to sixth order (in accordance with the order of our differencing scheme) and then calculate the integral using {\tt Mathematica}'s {\tt NIntegrate} function.
Figure~\ref{fig:MetricLimit} clearly shows that $||T||_2\to 0$ as $\epsilon \to 0$ for all the families.
Furthermore the gradient of the linear behaviour in this log-log plot is precisely consistent with $|| T ||_2 \propto \epsilon$ for small $\epsilon$, agreeing with the `painting' picture encapsulated in equation~\eqref{eq:GRbehaviour}.
While we must take different values of $z_{max}$ for the various solutions in a family as $\epsilon$ is varied, we choose $z_{cut} \le z_{max}$ to be the largest value common to all solutions found so that the norms of these are taken over the same coordinate range and hence can be compared. For the data in the figure we have taken $z_{cut} = 0.6$ which is the smallest value of $z_{max}$ we used in constructing that family of solutions.

\begin{figure}
\centerline{  
\includegraphics[scale=0.8]{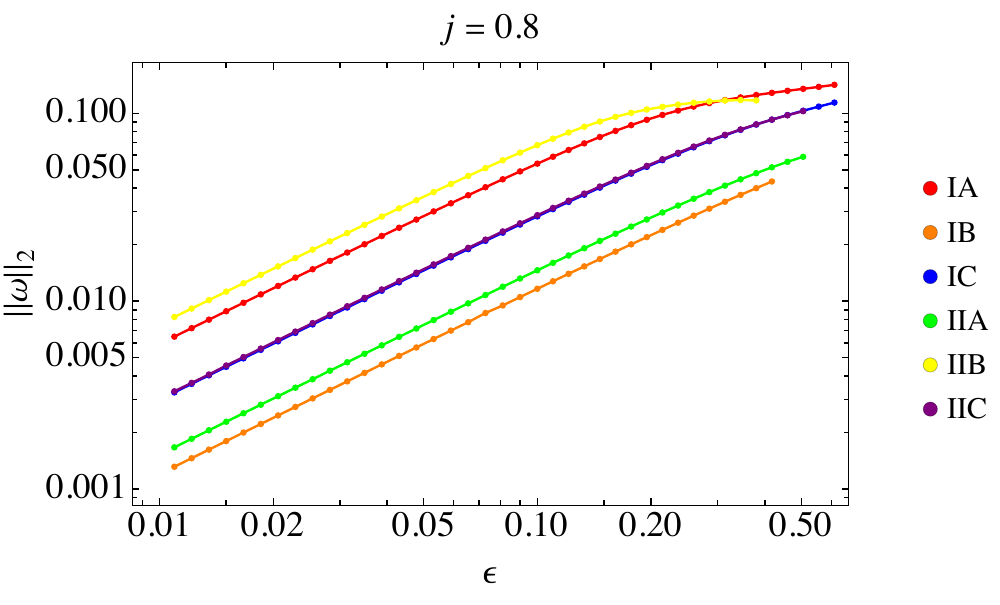} 
}
  \caption{\label{fig:aetherTwist}
  Figure showing the aether twist vanishes in the limit $\epsilon \to 0$.
  We plot the $L^2-$norm of $\omega_{\mu\nu}$, over the coordinate domain extending up to $z_{cut}=0.6$ for different families of solution, here all with $j = 0.8$, against $c_a = \epsilon$. We see the twist tends to zero as $\epsilon \to 0$ for each family. 
  We see analogous behaviour for other spin values of $j$ tested.
  } 
\end{figure}

We see the same behaviour for all the other metric functions  for each of the families of solutions with various spin values of $j$ for which we have constructed solutions. Thus we indeed confirm that Kerr emerges in the limit $\epsilon \to 0$ of the families of black holes we study.
We then expect the aether to be `painted' on to the Kerr spacetime in a twist free configuration as discussed in sections~\ref{sec:paintingI} and~\ref{sec:paintingII}. 
We confirm this in figure~\ref{fig:aetherTwist} by plotting the $L^2-$norm of the twist field,
\begin{eqnarray}
|| {\omega} ||_2 = \left[ \int_0^{z_{cut}} dz \int_0^{\pi/2} d\phi\, | \omega_{\mu\nu} \omega^{\mu\nu} |\right]^\frac{1}{2}\, ,
\end{eqnarray}
for the different families for a particular spin $j=0.8$. This figure  shows the twist vanishing linearly with $\epsilon$ as we take $\epsilon \to 0$, again compatible with the `painting' picture in equation~\eqref{eq:GRbehaviour}, and shows that $u^{(1)}_\mu$, which gives the $O(\epsilon)$ correction to the leading twist free aether  $\bar{u}_\mu$,  has non-vanishing twist. The same behaviour is seen for all the spins we have constructed. Note that
in this figure we see the expected deviation from the small $\epsilon$ linear behaviour, corresponding to the leading term in the expansion~\eqref{eq:GRbehaviour}, as $\epsilon$ is increased towards $O(1)$ values. This is seen also for the other quantities we shortly discuss.

\begin{figure}
\centerline{  
  \includegraphics[scale=0.8]{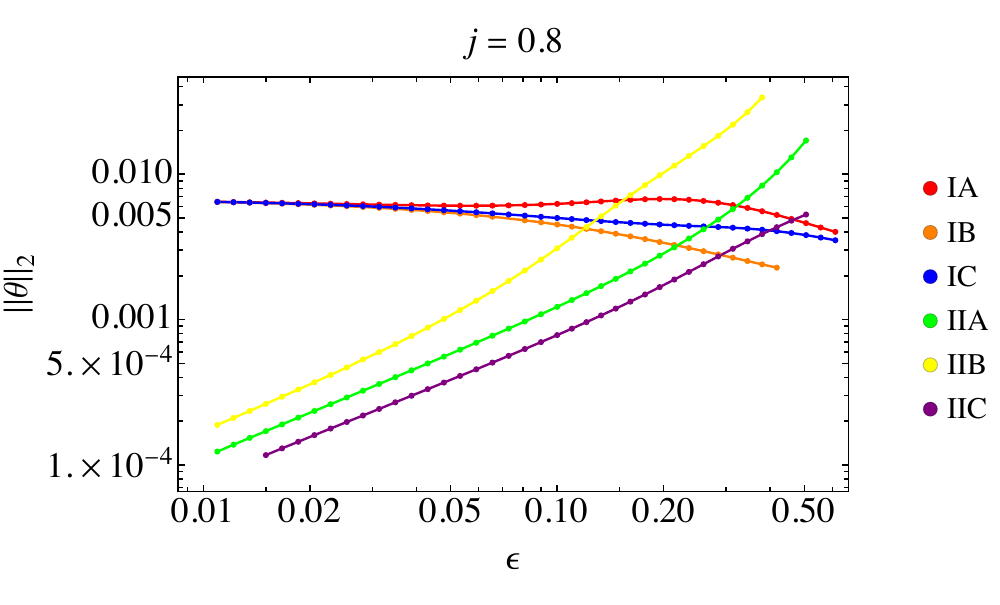}
  }
  \caption{\label{fig:aetherExpansion}
  The $L^2-$norm of the aether expansion over the coordinate range extending up to $z_\textrm{cut}=0.6$ with $j=0.8$, vs.\  $\epsilon$. For the region II families this norm vanishes linearly with $\epsilon$. For region I families the aether has a non-zero expansion  as $\epsilon\to 0$, which depends on the spin of the black hole  but is independent of the  coupling $c_\omega$.} 
\end{figure}

From our earlier discussion in  sections~\ref{sec:paintingI} and~\ref{sec:paintingII}, as $\epsilon \to 0$  we expect the aether to obey equation~\eqref{eq:aetherRegion1} in region I and~\eqref{eq:aetherRegion2} in region II. These are different relations, the latter being the expansion free condition, the former allowing non-trivial expansion. Indeed in figure~\ref{fig:aetherExpansion} we see the $L^2-$norm of the expansion $\theta$ vanishing for the region II families linearly in $\epsilon$ as $\epsilon \to 0$, whereas for the region I families it does not vanish.

\begin{figure}
\centerline{
  \includegraphics[scale=0.8]{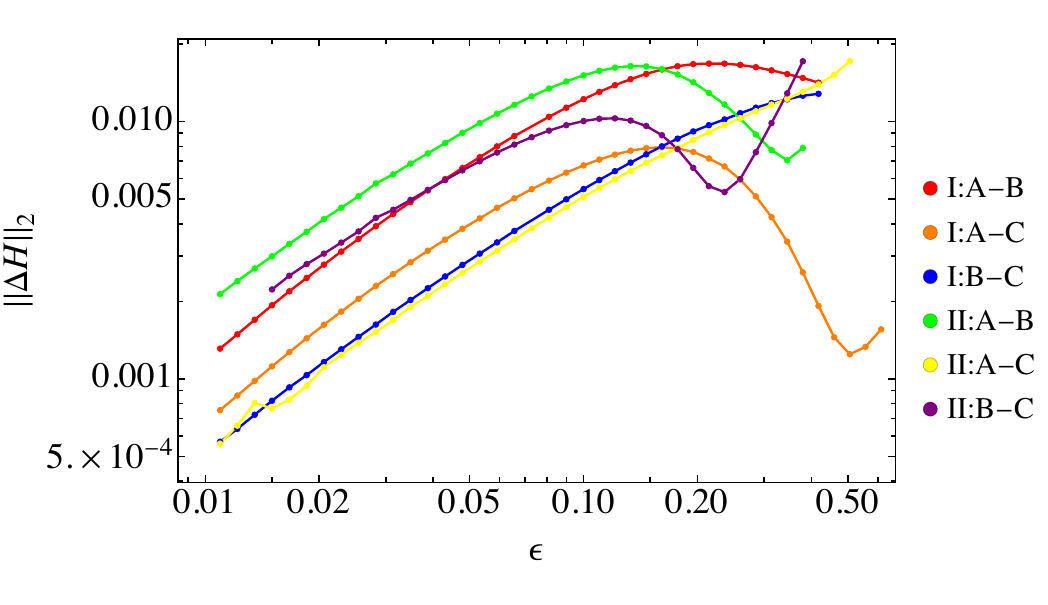}
  }
  \caption{\label{fig:aetherLimit}
 The $L^2$-norm difference of the function  $H$ in the aether vector between the different region I families as a function of  $\epsilon$ for black holes with spin $j=0.8$, and the same for the region II families. 
     The aether tends to a universal form as $\epsilon \to 0$ for both families.
     }
\end{figure}

We now proceed to confirm another important aspect of our `painted' on aether picture. We expect the aether~\eqref{eq:aetheransatz} in the limit $\epsilon \to 0$, when $u_\mu \to \bar{u}_\mu$, to be universal in the sense that it depends only on the GR spacetime that it is `painted' onto, $\bar{g}_{\mu\nu}$, and whether one is in region I or II, but does not depend on the $O(1)$ couplings that define the theory in that region in the $\epsilon \to 0$ limit. Recall this is because the aether potential is determined by the equation~\eqref{eq:aetherRegion1} for region I in the limit $\epsilon \to 0$, or equation~\eqref{eq:aetherRegion2} for region II in that limit, and these are explicitly independent of all the couplings. 
Furthermore for these stationary black holes we can also determine the answer to our earlier question of whether there is a moduli space of solutions for the limiting aether $\bar{u}_\mu$ or not.
Our numerical solutions demonstrate that as $\epsilon \to 0$ there exists a unique and universal aether for both region I and II families given our asymptotic boundary conditions.
For a given mass and spin, the region I families give the same aether vector field as $\epsilon \to 0$. Likewise one obtains the same aether for the region II families in this limit, although this is a different aether from that for the region I families.
This is suggested in figure~\ref{fig:aetherExpansion} in the case of region I families where the non-zero value of the norm of the expansion is the same in the limit $\epsilon \to 0$ for the different families shown. 
In figure~\ref{fig:aetherLimit} we confirm this in detail by giving the $L^2-$norm 
(with the same $z_{cut}$ as previously)
of the difference of the aether vector function $H$ between different region I families. We see these norms vanish as $\epsilon \to 0$. In that figure we plot the same for different region II families, and again see that the function $H$ tends to the same form as $\epsilon \to 0$. The same holds true for the other aether functions $K$, $X$ and $Y$. 
The expansion free condition associated to region II given in equation~\eqref{eq:aetherRegion2}, and likewise equation~\eqref{eq:aetherRegion1} for region I,  locally govern the leading twist free aether behaviour in a manner independent of the large aether couplings. However there could have been global data which would be reflected in the different families giving different limiting forms of the aether. Since we  see the same limiting aether for all the families within region I, and correspondingly for region II, in fact it seems that regularity of the various aether horizons is sufficient to uniquely fix the solution. Thus, the aether behaviour in the limit $\epsilon \to 0$ is universal, ie. different for region I or II but not depending on the large aether couplings.

\begin{figure}
\centerline{
  \includegraphics[scale=0.8]{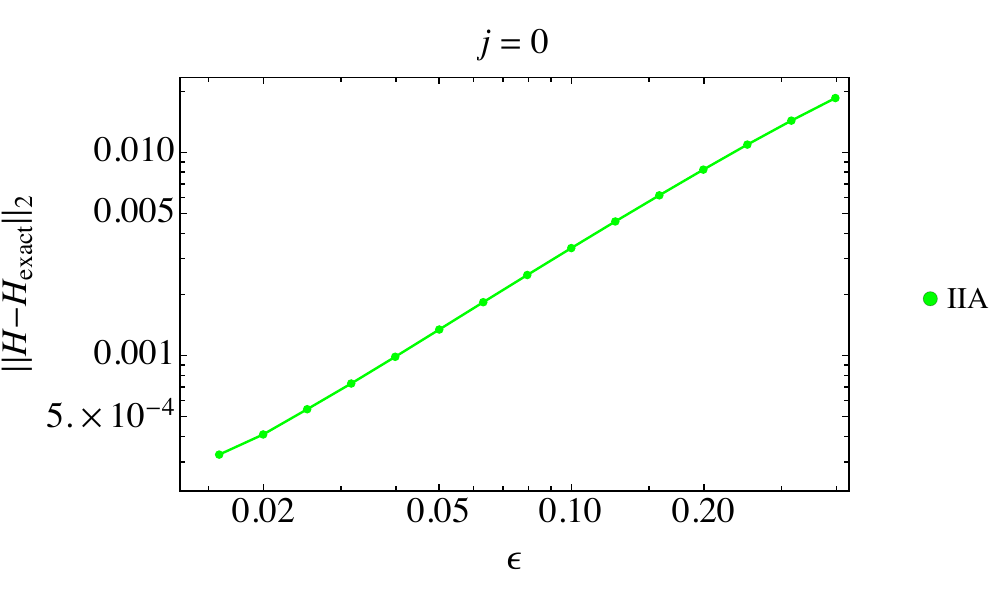}
  }
  \caption{\label{fig:checkExact}
  The $L^2$-norm of the difference of the aether function $H$ from its $\epsilon = 0$ analytic form $H_{\rm exact}$  (determined from equation~\eqref{eq:exact2})
  for static spherically symmetric family IIA solutions with no spin. The correct analytic form is indeed reproduced in this spinless case. 
  }
\end{figure}

Finally, we compare with some analytical results available in the literature. Recall that for the region II families in the static spherically symmetric case we expect that as $\epsilon \to 0$ the solution tends to the exact analytic solution with Schwarzschild metric as in equation~\eqref{eq:exact1} and aether as in equation~\eqref{eq:exact2}.
If we compute these static family II solutions, and rescale them to have mass $M = 1$, so that $r_0 = 2$, then in the limit $\epsilon \to 0$ the aether function $H$ should take the form,
\begin{eqnarray}
H_{\rm exact} = - \sqrt{1 - 2 z + r^4_{\text{\ae}} z^4 }
\end{eqnarray}
in our coordinate system for the resulting  Schwarzschild metric, which is
related to that in  
equation~\eqref{eq:exact1} by $z = 1/r$. As discussed earlier, the parameter $r_{\text{\ae}}$ is determined as $r_{\text{\ae}} = \frac{3^{3/4}}{2}$ (for $r_0 = 2$) to ensure existence of a regular spin-0 and spin-1 horizon (or equivalently ensure existence of a universal horizon). 
We indeed see precisely this behaviour emerge as $\epsilon \to 0$. For family IIA we plot in Fig.~\ref{fig:checkExact} the $L^2$-norm of the difference $H - H_{\rm exact}$ for solutions as we vary $\epsilon$, where we scale the $z$ coordinate for each solution so that it has mass $M = 1$, and we use a cut-off $z_{cut} = 0.6$ for the norm in this rescaled coordinate (note that as $\epsilon \to 0$ the metric horizon is at $z = 0.5$ so the norm is computed including some interior to this horizon).

The comparison with analytic solutions can be extended a bit
further, to first order in the spin of rotating solutions. In~\cite{Barausse:2015frm} the linear in spin correction to a static solution was considered analytically in the case $c_a = 0$. For $c_\sigma = 0$ this is precisely the slowly spinning correction to our limiting family II static solution. In this case the metric found in~\cite{Barausse:2015frm} is the Kerr metric, to linear order in the spin. As discussed above, we indeed find Kerr in the $\epsilon \to 0$ limit for the family II solutions for all spins
(as expected due to the vanishing of the aether stress tensor in this limit for an aether orthogonal to a maximal foliation), so our numerical solutions for the metric match the approximate analytic ones for slow rotation.  As for the aether covector $u_a$, Ref.~\cite{Barausse:2015frm} found that it is not perturbed at linear order in the spin. This is also consistent with our numerical limiting family II solution. As discussed in the next section (see figure~\ref{fig:KLimit}), the $u_\phi$ aether component vanishes as $\epsilon \to 0$ for any spin. 
This is a consequence of the aether being orthogonal to an axisymmetric
maximal foliation, hence in particular orthogonal to $\partial_\phi$.
The remaining components, $u_v$, $u_z$ and $u_\theta$,  should be even functions of spin (as they should be invariant under flipping the spin $j \to -j$), and hence have no linear deviation from their static form, which is indeed what we see numerically.

\subsection{Universal aether for regions I and II}
\label{sec:aetherregionsIandII}

As discussed above we have seen the limiting behaviour we expect for small $\epsilon$, namely, an aether `painted on' to  a near-Kerr spacetime, independent of the coupling 
constants for regions I and II. In this subsection we shall characterize the configuration of these universal aether fields.

For reasonable spins, such as $j = 0.8$ as presented above, it becomes difficult numerically to find the  black hole solutions when $\epsilon \lesssim 0.01$, as one must solve the system far inside the metric horizon as the spin-1 wavespeed diverges (as does the spin-0 wavespeed for region II). We believe our failure to find solutions is a technical problem due to stability of the Newton solver, rather than a reflection that such solutions do not exist. 
To find solutions truly in region I we require $\epsilon \sim 10^{-5}$ or less, and for region II we need $\epsilon \sim 10^{-7}$, and this is not possible with our numerical code. However already for $\epsilon \sim 10^{-2}$ we clearly see the solutions take the form of the expansion in $\epsilon$  in equation~\eqref{eq:GRbehaviour}, and we can accurately extrapolate the leading behaviours $\bar{g}_{\mu\nu}$ and $\bar{u}_\mu$. We have seen the leading metric $\bar{g}_{\mu\nu}$ is Kerr. By linear extrapolation of our solutions we can deduce the leading aether $\bar{u}_\mu$ for both the region I and region II families as a function of the Kerr parameters. We obtain the same result, up to expected numerical accuracy, for all the region I families, confirming the universal behaviour. Likewise for the region II families such extrapolation yields the same aether $\bar{u}_\mu$. 

\begin{figure}
\centerline{  
  \includegraphics[scale=0.8]{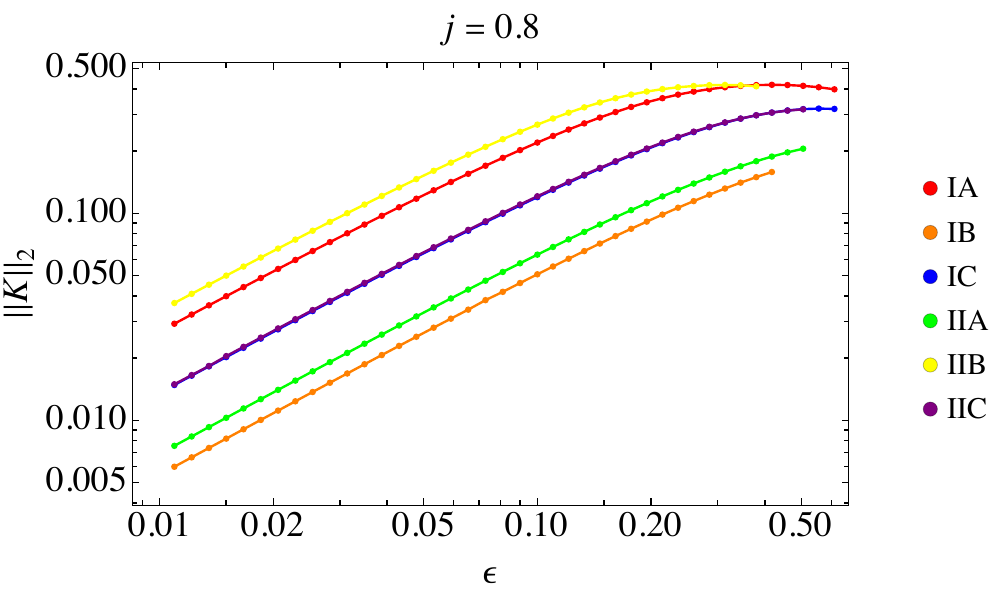}
  }
  \caption{\label{fig:KLimit}
  $L^2-$norm of the aether function $K$ taken over the coordinate domain up to $z_{cut}=0.6$. This domain includes the spin 2 horizon for all families considered and some of the other horizons, depending on the parameters. This norm tends to zero as $\epsilon\to 0$ for all the families of black holes, indicating that the $\phi$-component of the aether co-vector vanishes in this limit. 
  }
\end{figure}

\begin{figure}
\centerline{  
  \includegraphics[scale=0.8]{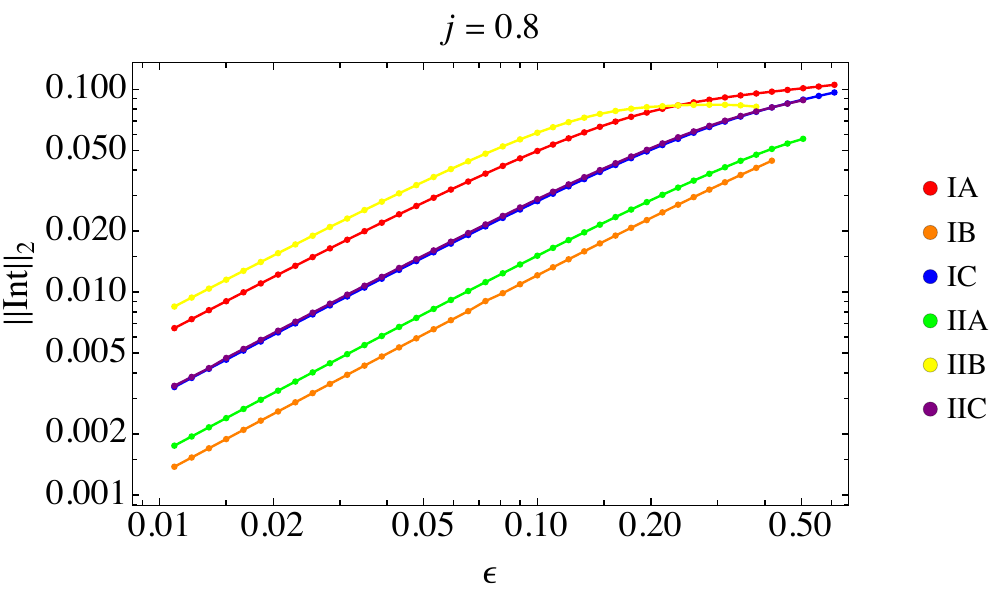}
  }
  \caption{\label{fig:IntLimit}
  $L^2-$norm of the integrability condition $\textrm{Int}\equiv z^2H\left[\partial_\theta \left( \frac{X}{z^2 H} \right) + \partial_z\left( \frac{Y \sin\theta \cos\theta}{z H} \right)\right] = 0$,  taken over the coordinate domain up to $z_{cut}=0.6$. This norm tends to zero as $\epsilon\to 0$ for all the families of black holes. Note that we have multiplied the integrability condition by an extra factor of $z^2\,H$ to improve its behaviour near infinity $(z=0)$.
  }
\end{figure}

This twist free leading aether takes the form $\bar{u} = k \partial f$. Recalling our aether ansatz in equation~\eqref{eq:aetheransatz} 
and the asymptotic boundary conditions
the potential must take the form
\begin{eqnarray}
f = v + \Phi(z,\theta)
\end{eqnarray}
and so we should have,
\begin{eqnarray}
k = H \; , \quad \partial_z \Phi = - \frac{X}{z^2 H} \; , \quad \partial_\theta \Phi = \frac{Y \sin\theta \cos\theta}{z H}
\end{eqnarray}
in the $\epsilon \to 0$ limit. 
We note that the requirement that the aether asymptotically tends to $\partial_v$ implies there is no linear term in $\phi$, so that the aether congruence consists of zero angular momentum worldlines. 
By linearly extrapolating the functions $H, X$ and $Y$ for our small $\epsilon$ solutions we can accurately deduce this limiting form for $\epsilon \to 0$.
Comparison of the aether ansatz with the form $\bar{u} = k \partial f$ also implies that in the limit $\epsilon \to 0$ the aether function $K$ should vanish, and we indeed see this in the numerical solutions. In figure~\ref{fig:KLimit} we display the $L^2$-norm of the function $K$ in the aether co-vector, confirming that indeed $K$ vanishes in the $\epsilon\to 0$ limit. 

\begin{figure}
\centerline{  
  \includegraphics[width=8cm]{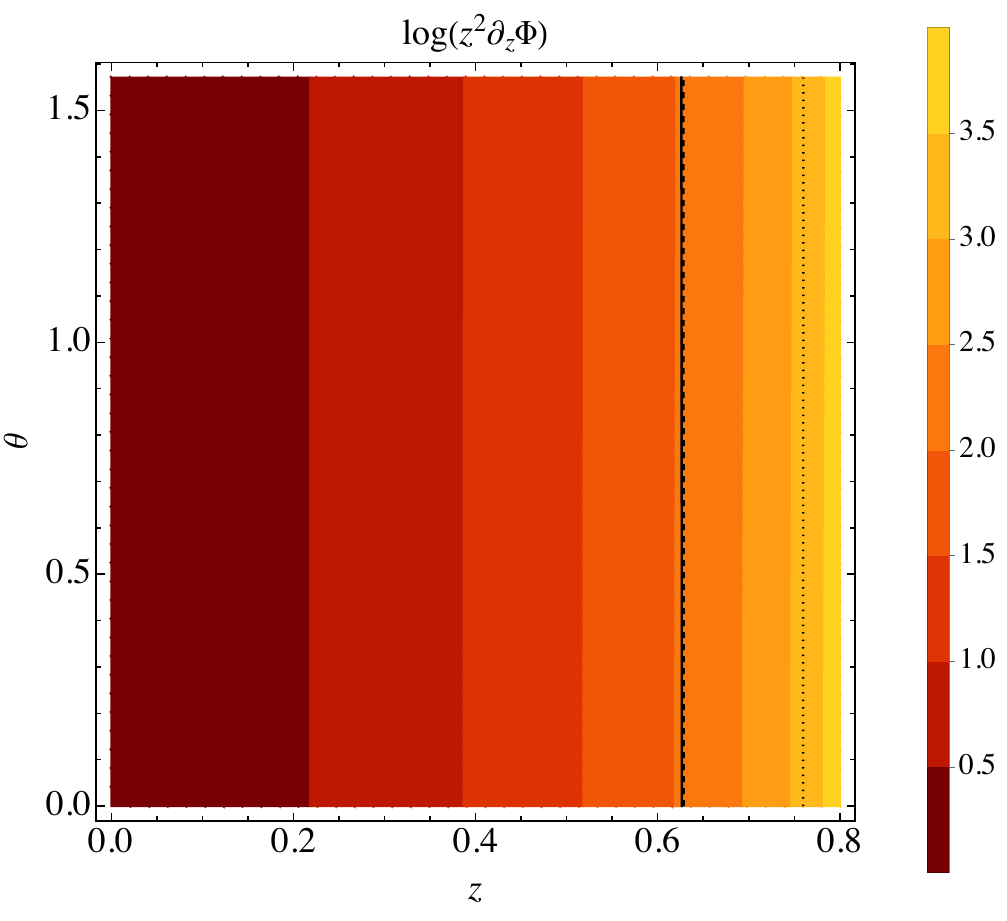}
  \includegraphics[width=8cm]{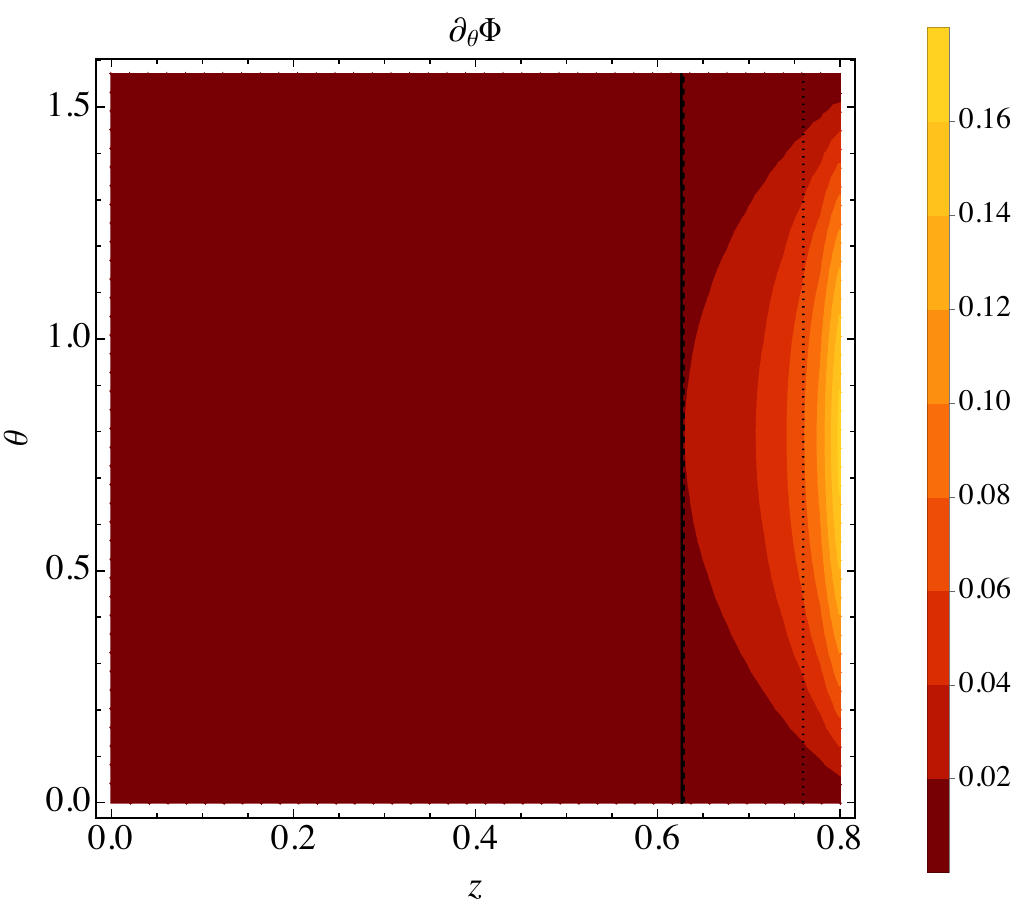}
  }
  \centerline{  
  \includegraphics[width=8cm]{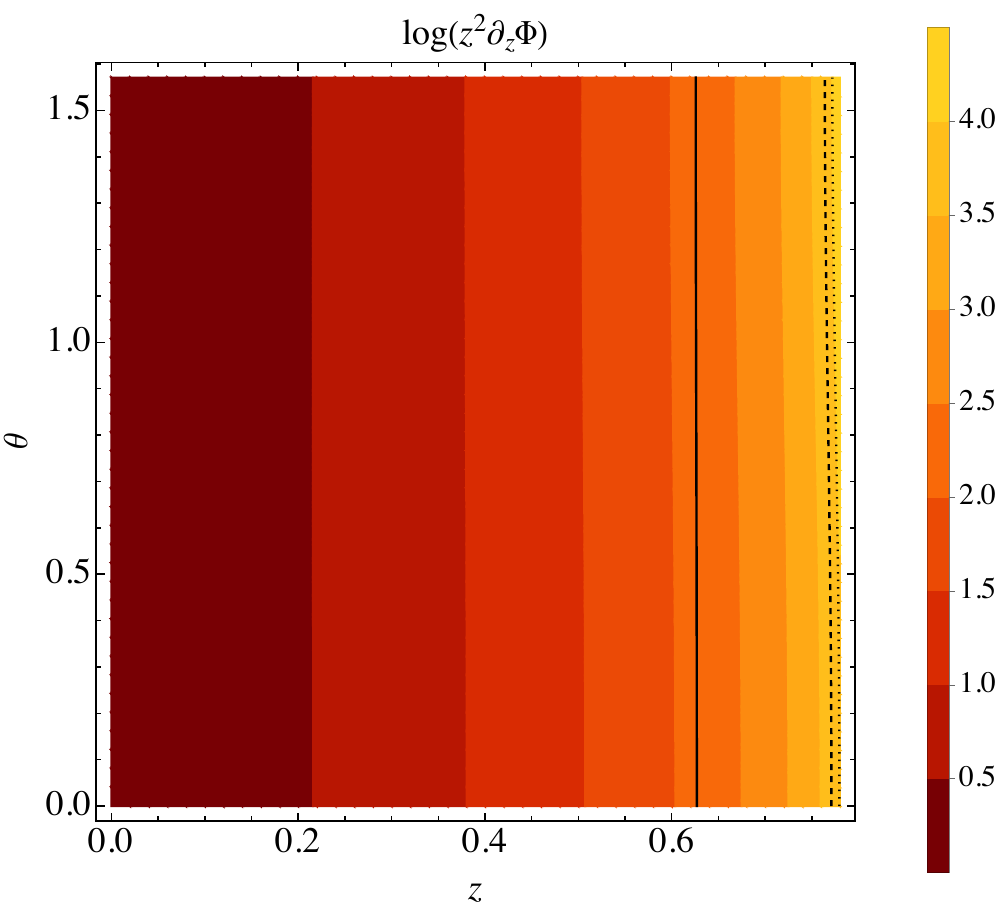}
  \includegraphics[width=8cm]{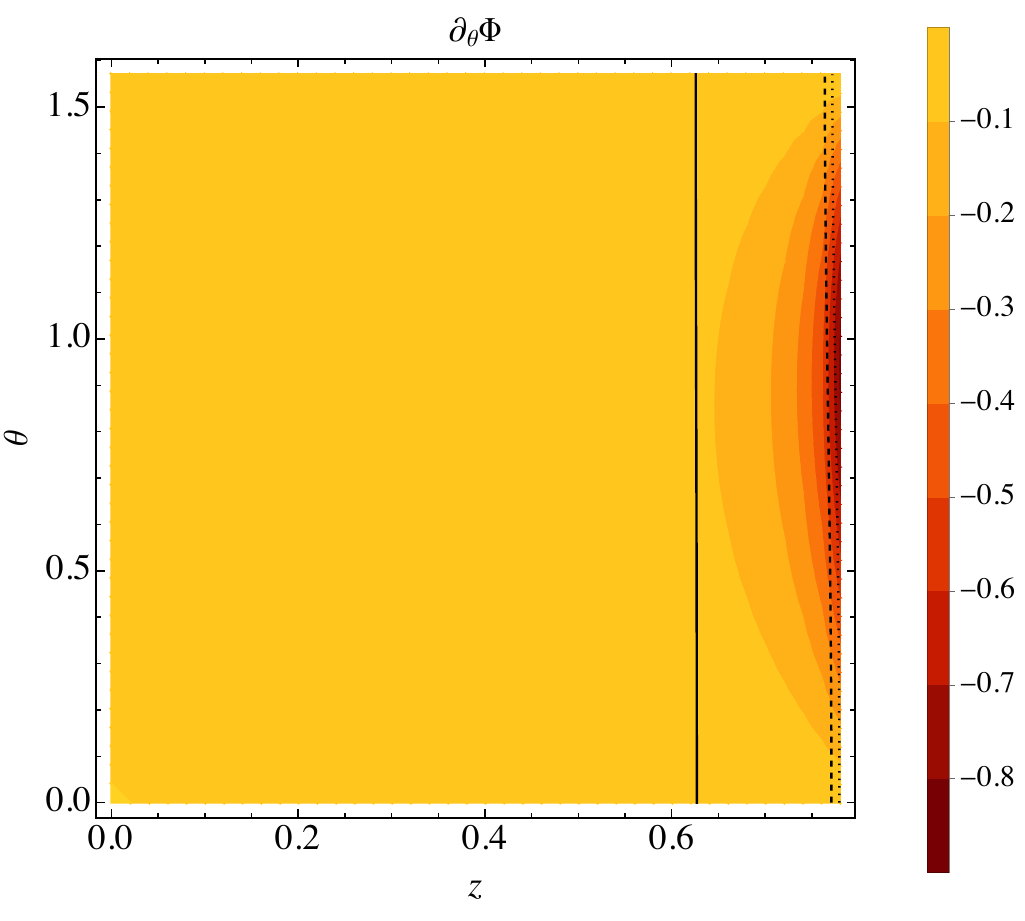}
  }
  \caption{\label{fig:aethervector}
  Figure showing the spatial $z$ and $\theta$ derivatives of the potential for the universal twist free aether $\bar{u} = k \partial f$ for spin $j=0.8$ black holes for region I (top frames) and II (bottom frames). The potential is computed by using the functions $H$, $X$ and $Y$ from our small $\epsilon$ solutions to extrapolate $\epsilon \to 0$. The dashed, dotted and solid black lines show the spin 0, 1 and 2 horizons respectively of the last black hole solution with $j=0.8$ that we have constructed, corresponding to $\epsilon=0.01$.
  }
\end{figure}

We have also checked from the numerical solutions that the integrability condition, $\partial_\theta \left( \frac{X}{z^2 H} \right) + \partial_z\left( \frac{Y \sin\theta \cos\theta}{z H} \right) = 0$, holds in the limit $\epsilon \to 0$, see figure~\ref{fig:IntLimit}.
Using our small $\epsilon$ solutions to extrapolate the potential, in  figure~\ref{fig:aethervector} we show the $z$ and $\theta$ derivatives of the potential, $\partial_z \Phi$ and $\partial_\theta \Phi$, over the coordinate domain for solutions with $j = 0.8$ for region I and for  region II. 
 We see that for both region I and region II the radial derivatives, $\partial_z \Phi$, are far greater than the angular ones, $\partial_\theta \Phi$, in magnitude.
 One can see by eye that, as expected, the radial dependence is different in detail for region I and region II. We note that $\bar{g}_{\mu\nu}$ is the same Kerr metric in the same coordinates for both region I and II solutions, and hence this difference is physical and not related to a difference of coordinates. There is a small $\theta$ dependence in the aether as we see from the plots of $\partial_\theta \Phi$, and it is again quite different for the two regions.

\begin{figure}
\centerline{  
  \includegraphics[width=8cm]{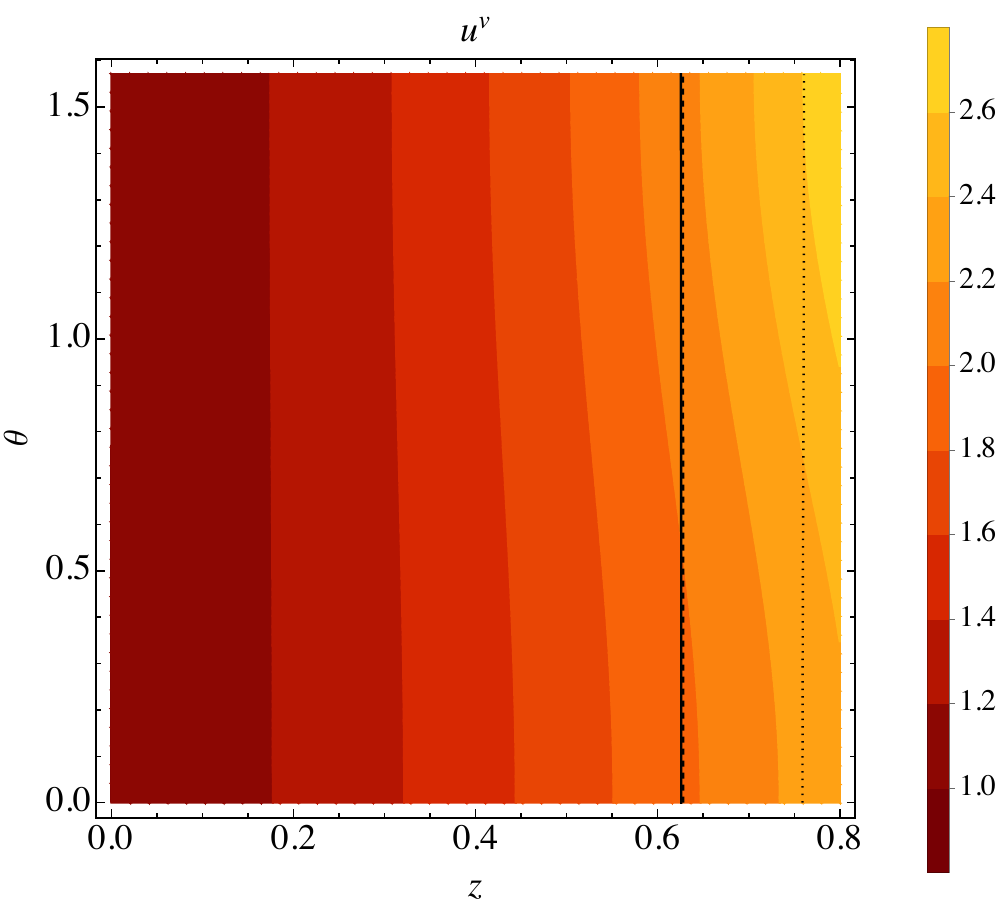}
  \includegraphics[width=8cm]{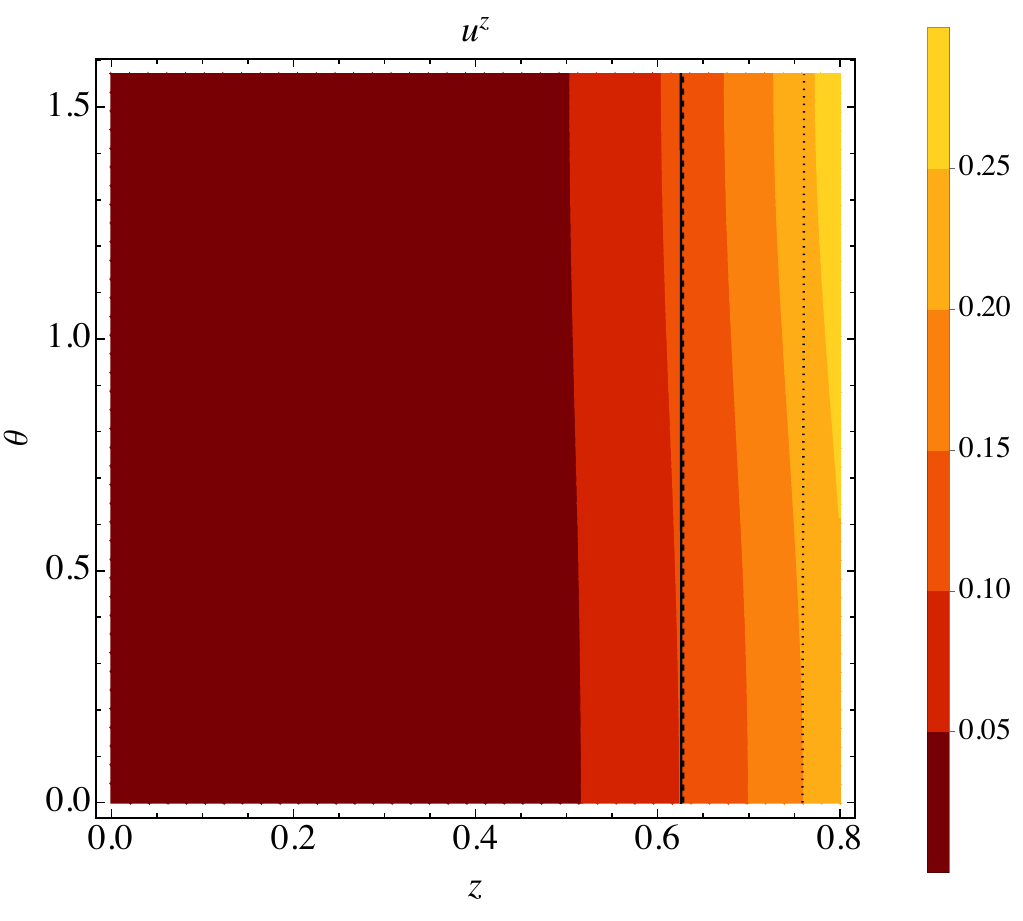}
  }
  \centerline{  
  \includegraphics[width=8cm]{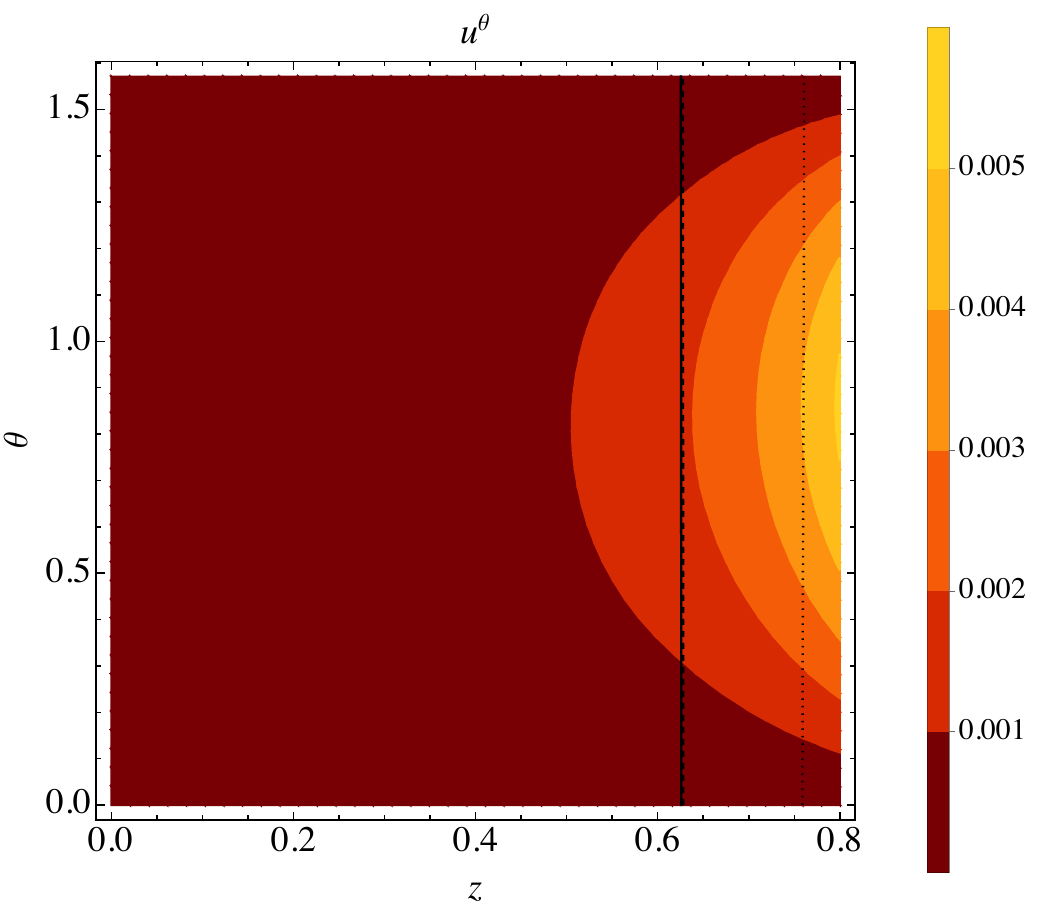}
  \includegraphics[width=8cm]{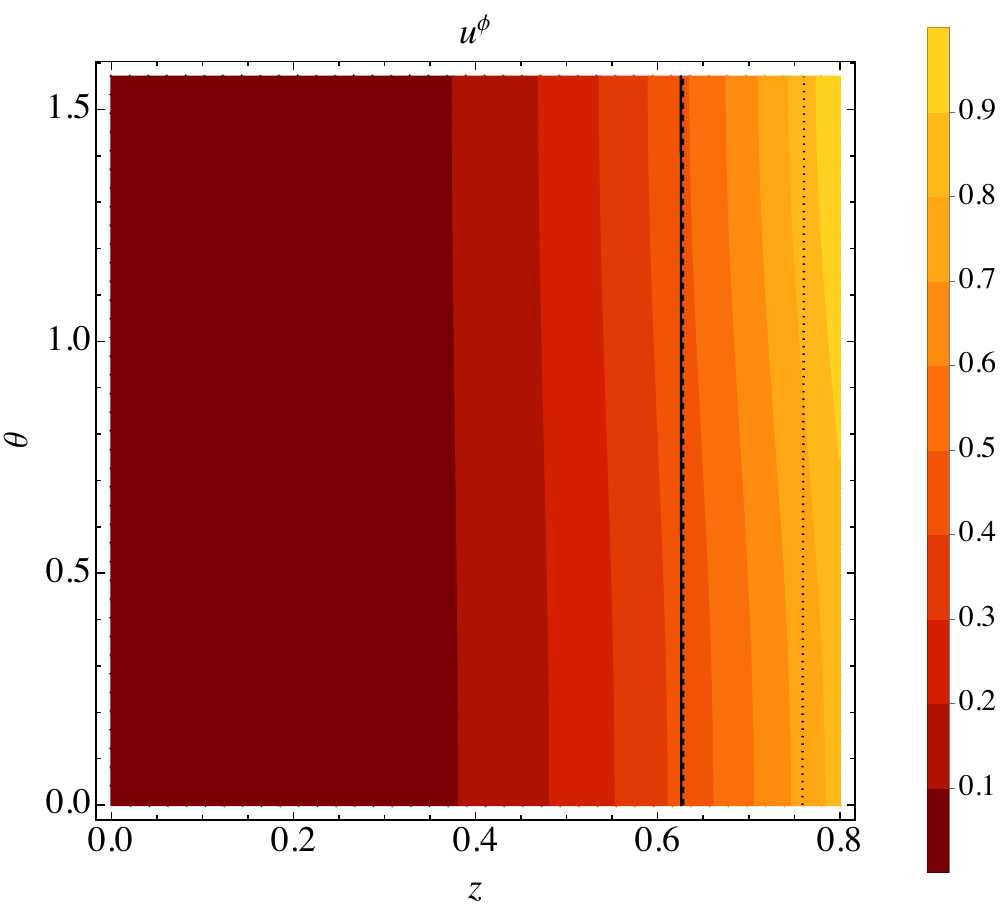}
  }
  \caption{\label{fig:aethervector2}
  The aether components $u^v$, $u^z$, $u^\theta$ and $u^\phi$ for the region I family for spin $j=0.8$ extrapolated to $\epsilon \to 0$ using our small $\epsilon$ solutions. These `upstairs' components control integral curves of the aether. 
   The positivity of $u^v$ and $u^z$ shows that the aether worldlines fall into the black hole as $v$ increases along them.
  Due to the relative smallness of $u^\theta$, integral curves are approximately radial with little tilt in the $\theta$ direction when projected into the $(z,\theta)$ plane -- if plotted on these figures they would be indistinguishable from horizontal lines. As for the previous figure, the dashed, dotted and solid black lines depict the horizons of our smallest $\epsilon$ solution. The region II family yields qualitatively similar behaviour for these aether components, although they differ in detail.
  }
\end{figure}

It is not easy to see from the aether potential how the aether vector behaves in the $\epsilon \to 0$ limit. Aether worldlines are determined by the `upstairs' components of the aether, $u^\mu$, and while the metric approaches Kerr for $\epsilon \to 0$, 
the off-diagonal components of the metric 
entail a complicated relationship of
these `upstairs' aether components 
with the derivatives of the aether potential. 
Thus in figure~\ref{fig:aethervector2} we plot these upstairs aether vector components for the region I family black holes with spin $j=0.8$ extrapolated to $\epsilon \to 0$ using our small $\epsilon$ solutions. 
We note that in this $\epsilon \to 0$ Kerr limit our coordinate $z = 1/r$ where $r$ is the corresponding Boyer-Lindquist radial coordinate, and $\theta$ is the same as the Boyer-Lindquist  coordinate.
We see that $u^v$ and $u^z$ are everywhere positive, so $dz/dv>0$ along the aether worldlines, and hence they `fall into' the black hole spin-0 and spin-2 horizons.
It is very likely they also fall into the spin-1 horizon too, although our smallest $\epsilon$ solution, with $\epsilon = 0.01$, is constructed with a coordinate domain that does not extend as far as is required to cover the extrapolated position of the spin-1 horizon in the $\epsilon \to 0$ limit, and hence we cannot say for sure.  We have computed the aether worldlines and do not display them on these diagrams as their motion in the $(z,\theta)$ plane appears indistinguishable by eye from a horizontal line, due to the relative smallness of $u^\theta$. 
We find very similar behaviour for the region II family in the $\epsilon \to 0$ limit, although the aether components are quantitatively different.

\begin{figure}
\centerline{  
  \includegraphics[width=8cm]{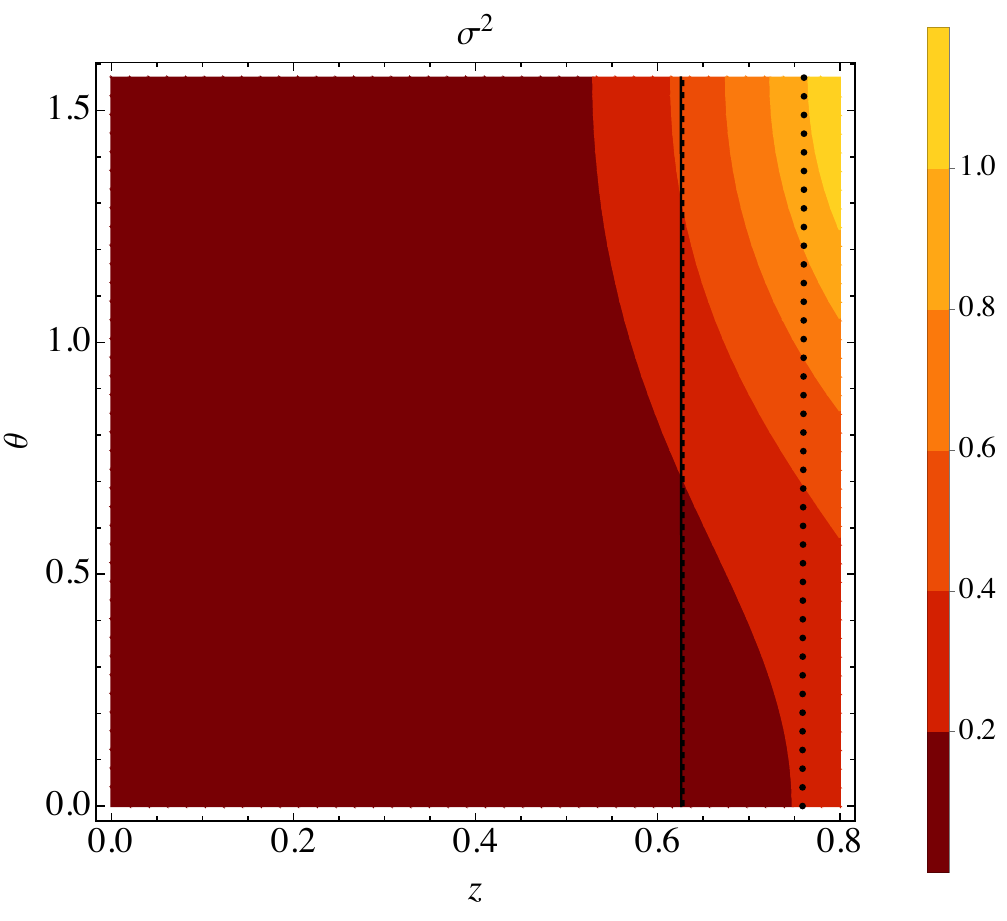}
  \includegraphics[width=8cm]{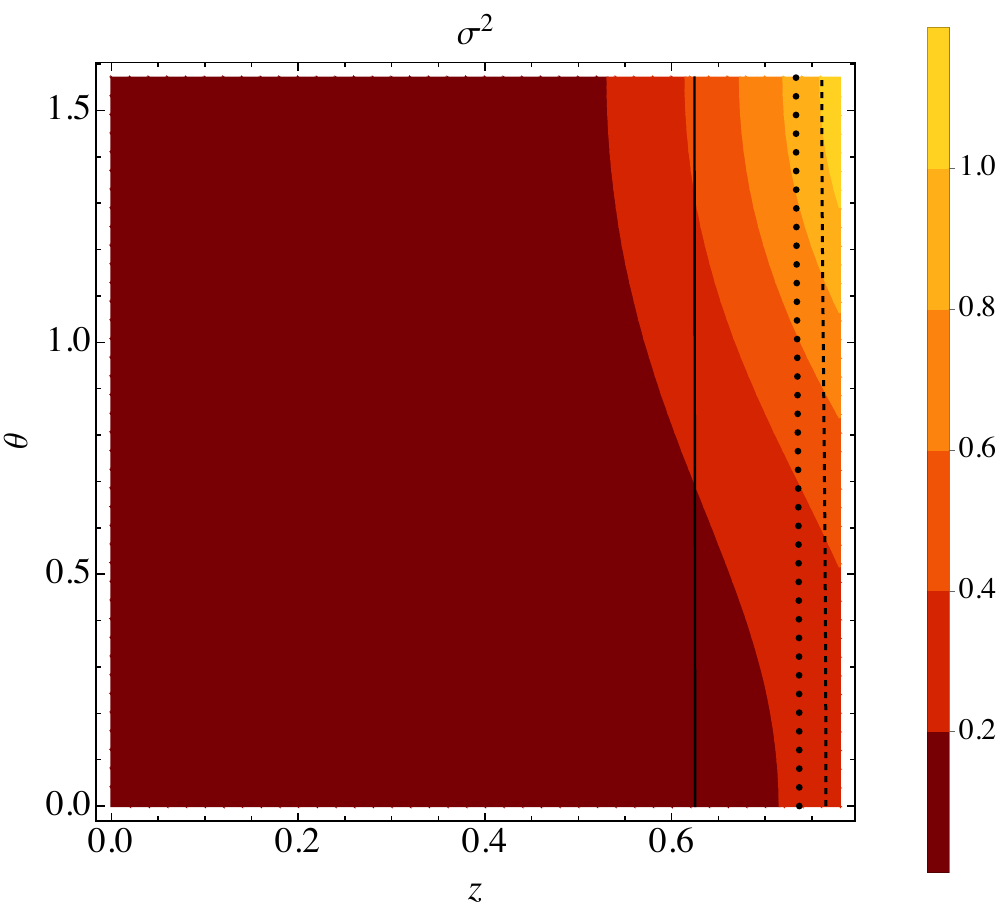}
  }
  \centerline{  
  \includegraphics[width=8cm]{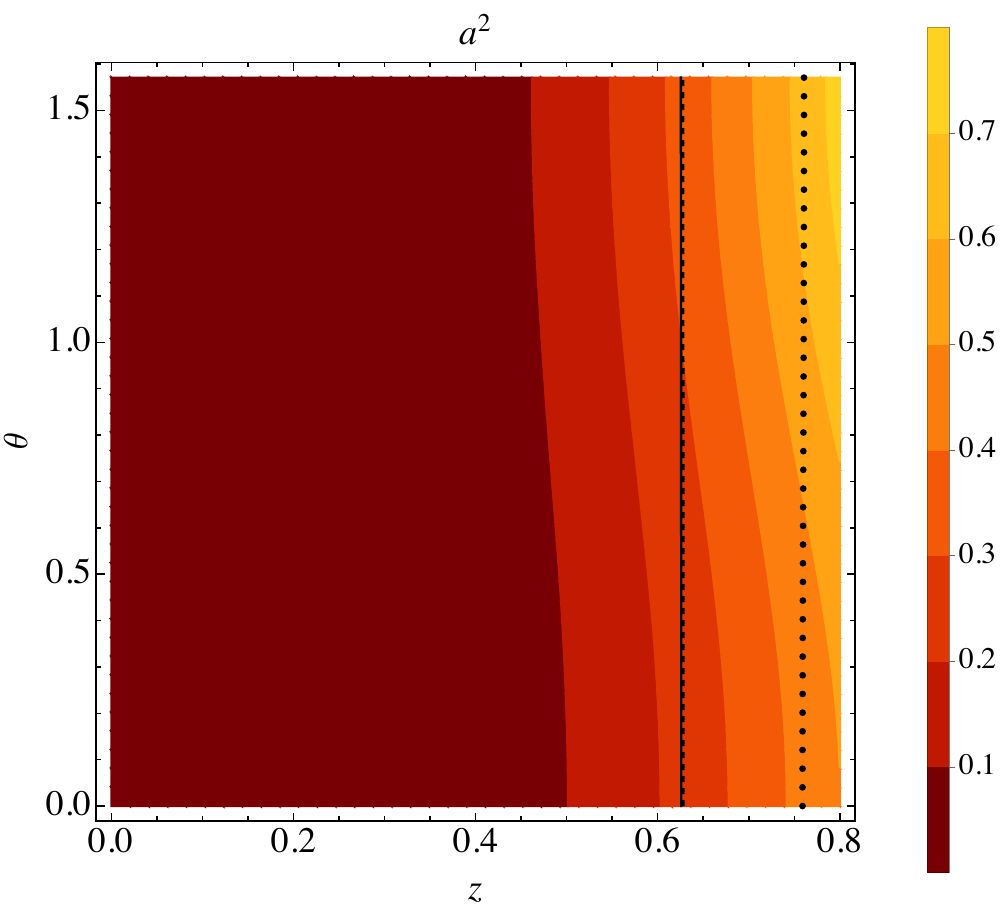}
  \includegraphics[width=8cm]{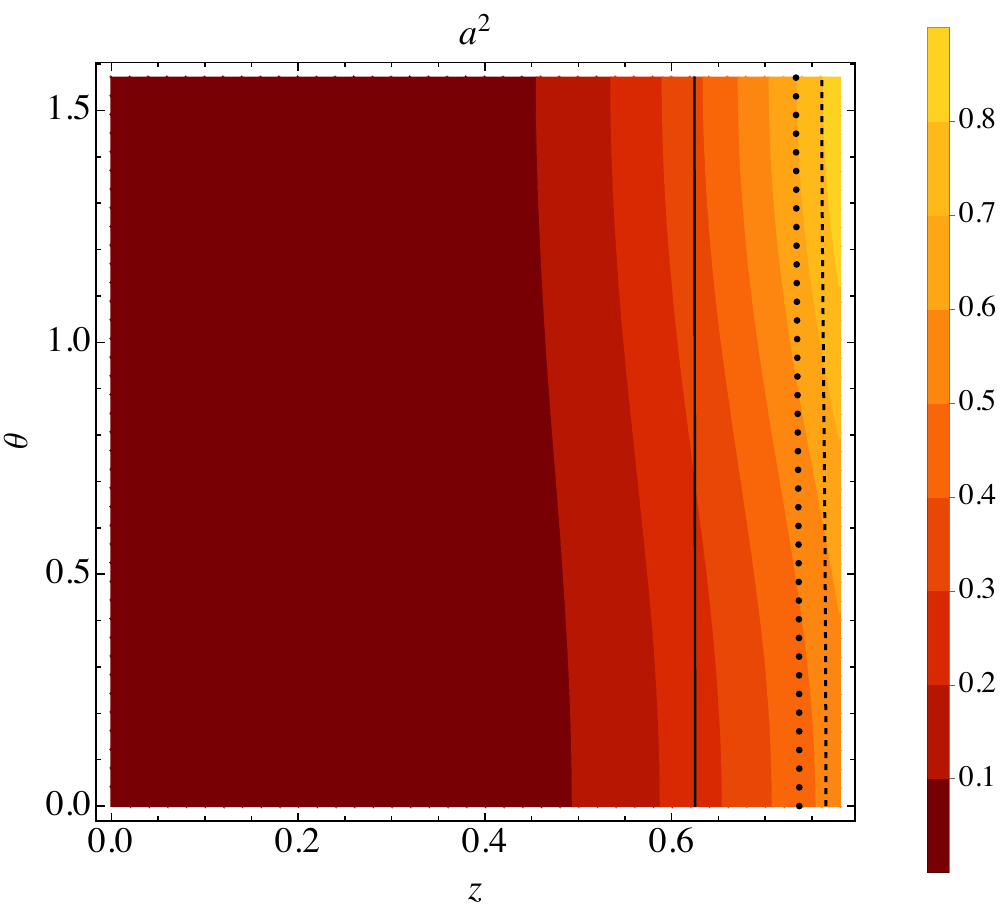}
  }
  \caption{\label{fig:lim_shear_and_a2}
  Figure showing the shear squared, $\sigma_{\mu\nu}\sigma^{\mu\nu}$, (top) and acceleration squared, $a_\mu\,a^\mu$ (bottom) for the families IA (left) and IIB (right) for black holes with $j=0.8$ in the limit $\epsilon \to 0$. These are calculated using extrapolation to $\epsilon=0$ from our small $\epsilon$ solutions. The dashed, dotted and solid black lines show the spin 0, 1 and 2 horizons respectively of the last black hole solution with $j=0.8$ that we have constructed, corresponding to $\epsilon=0.01$.
  }
\end{figure}

In figure~\ref{fig:lim_shear_and_a2} we show the limiting shear tensor (squared), $\sigma_{\mu\nu}\sigma^{\mu\nu}$, and the limiting acceleration (squared), $a_\mu\,a^\mu$, for Families IA and IIB for black holes with $j=0.8$. We have obtained the limiting shear and acceleration by pointwise quadratic extrapolation of the same objects calculated for finite values of $\epsilon$. As the figure shows, both $\sigma_{\mu\nu}\sigma^{\mu\nu}$ and $a_\mu\,a^\mu$ are smooth and look quite similar in regions I and II. In particular, these objects are $O(1)$ and develop non-trivial $\theta$-dependence near the horizons, while they decay quite fast near infinity.  

\begin{figure}
\centerline{  
  \includegraphics[scale=0.7]{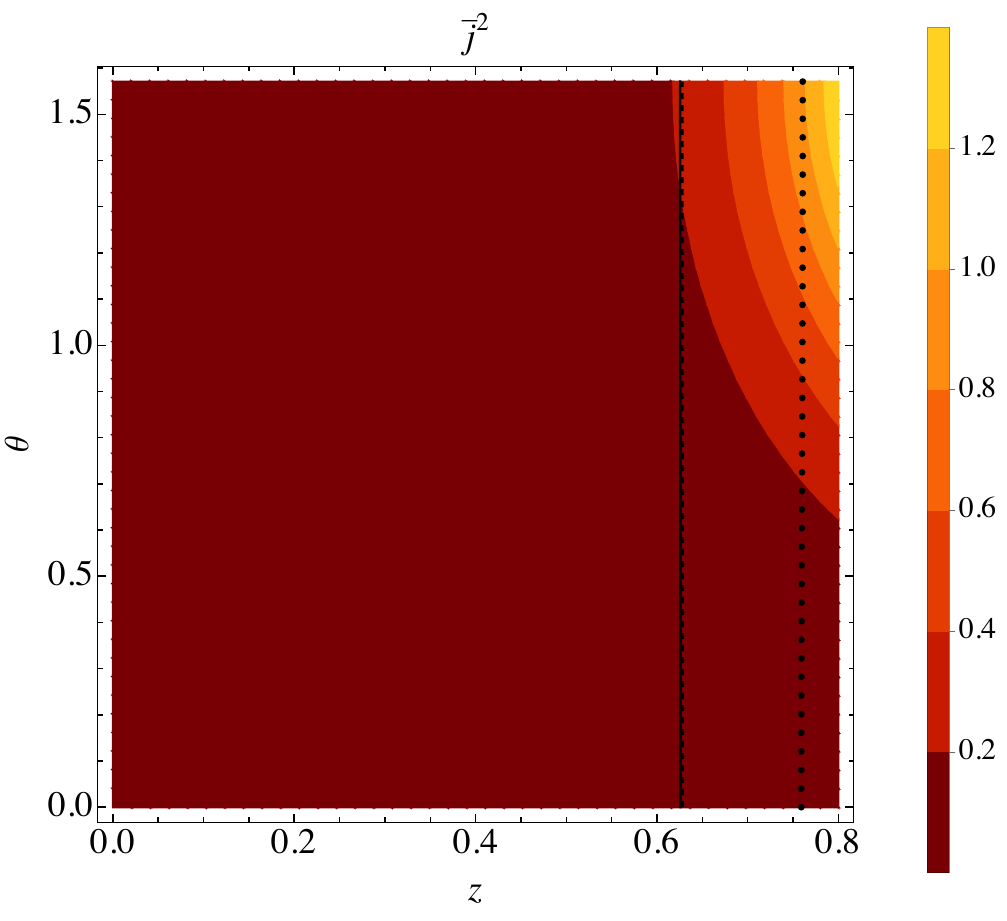}
  }
  \caption{\label{fig:barj2}
  Figure showing $\bar j^2$ for black holes with $j=0.8$ in the parameter region I extrapolated from our small $\epsilon$ solutions. This figure shows that $\bar j^\mu$ is non-vanishing. The dashed, dotted and solid black lines are as for the previous figures. 
  }
\end{figure}

Recall that for region I the limiting aether obeys the equation~\eqref{eq:aetherRegion1}. As discussed earlier this may be simply satisfied if $\bar{j}^\mu = 0$. Here we show that the limiting region I aether has a non-vanishing $\bar{j}^\mu$, confirming that the fourth order equation~\eqref{eq:aetherRegion1} for the twist potential is the true equation of motion, rather than the simpler condition $\bar{j}^\mu = 0$. 
From the earlier discussion we may compute $\bar{j}^\mu$ two different ways. Firstly we may compute it using equation~\eqref{eq:deltaomega} as;
\begin{eqnarray}
\bar{j}^\mu = \lim_{\epsilon \to 0}  \frac{ c_\omega \left( {\nabla}_\nu \omega^{\nu\mu} +  {a}_\nu \omega^{\nu\mu}   \right) }{\epsilon}
\end{eqnarray}
and secondly we may compute it using equation~\eqref{eq:fEqOneParam} as,
\begin{eqnarray}
 \bar{j}^\mu  = \lim_{\epsilon \to 0} \left(
 -
\left( {\nabla}^\mu {\theta} +  {u}^\mu {u} \cdot {\nabla} {\theta} \right) +  \left( {u} \cdot {\nabla} {a}^\mu - {u}^\mu {a}^2 + \frac{2}{3} {\theta} {a}^\mu  -  {\sigma}^{\mu\nu} {a}_\nu  \right)   \right)
\end{eqnarray}
We compute $\bar{j}$ both ways, using multiple small $\epsilon$ solutions to extrapolate to compute the   $\epsilon \to 0$ limits. Both methods agree, providing a further check on our numerical code. Since the form of the aether in this limit, $\bar{u}$, appears to be universal and hence the same for the different families I$A$, $B$ and $C$ we see the same function $\bar{j}$ for each. In figure~\ref{fig:barj2} we show $\bar{j}^2$ for these families and see clearly that it is non-vanishing and $O(1)$.

\subsection{Universal horizons}

A hypersurface everywhere
orthogonal to the aether,
and lying 
inside the fastest wave mode horizon
for a black hole,
is called a universal horizon, because 
it is a causal barrier for influences of {\it any} speed with respect to the aether rest frame. 
Universal horizons would be relevant if higher spatial derivatives are present in the field equations, producing unbounded propagation speeds. Such higher derivative terms are present  for example in Ho\v{r}ava-Lifshitz gravity, and in that theory the aether is
always hypersurface orthogonal, so black holes
in that theory presumably always possess universal horizons. 
The presence of universal horizons
in static spherically symmetric black holes
in Einstein-aether theory is also
generic, since in spherical symmetry the
aether is always hypersurface orthogonal,
and indeed they were found 
in numerical solutions of this sort~\cite{Barausse:2011pu}. 
For rotating black holes, however, 
their existence is not guaranteed.
In fact, already in~\cite{Barausse:2015frm} it was shown perturbatively that slowly rotating Einstein-aether black holes have no universal horizons. 
Given our numerical solutions for rotating black holes, what can be said at the nonperturbative level?

If a universal horizon does exist then, since the aether is orthogonal to that hypersurface, its twist must vanish there. 
To rule out the presence of a universal horizon, it thus suffices to show that the twist is nowhere vanishing. This is the method that was used in the perturbative analysis of~\cite{Barausse:2015frm}. For the various rotating black holes we have constructed here, this criterion allows
us to conclude that there is no universal horizon within our coordinate domains, which do cover at least the spin-0,1, and 2 horizons.
If a universal horizon were to be present it would of course lie inside these horizons, however, and 
we cannot presently rule out the possibility that one might be lurking deeper in the interior. Nevertheless, we do have some evidence 
against that possibility.

\begin{figure}
\centerline{  
  \includegraphics[scale=0.5]{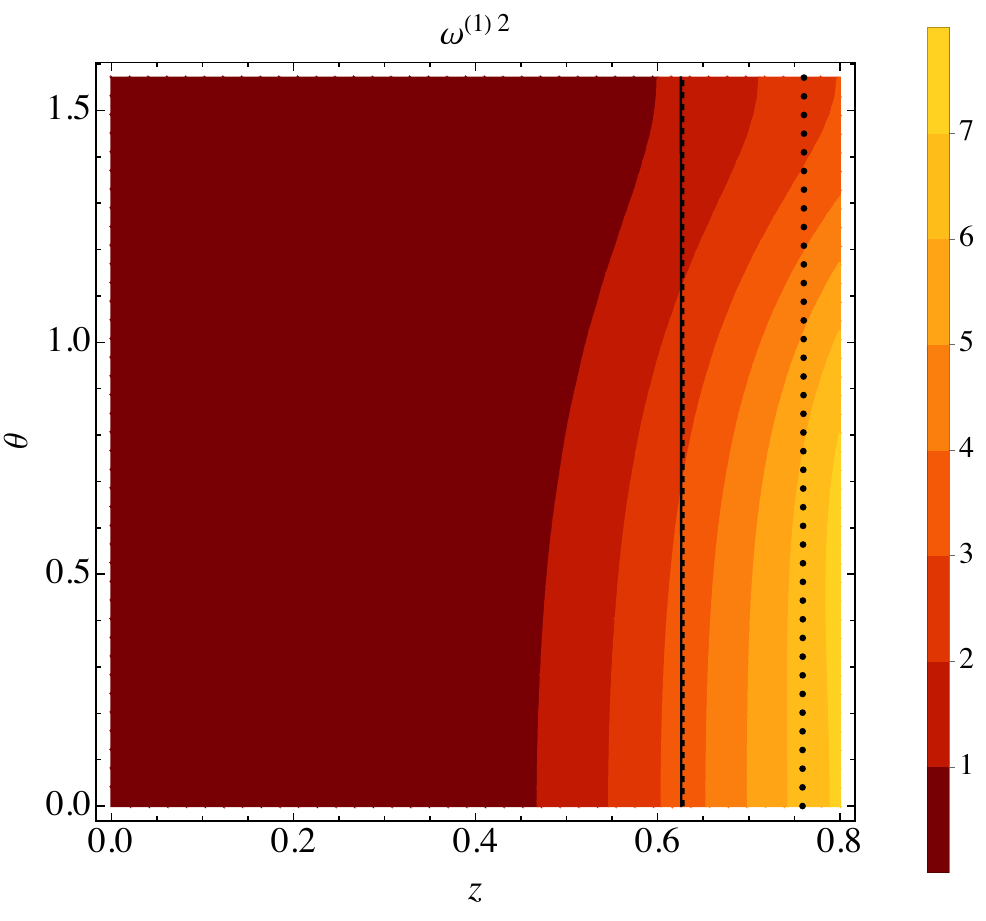}
  \includegraphics[scale=0.5]{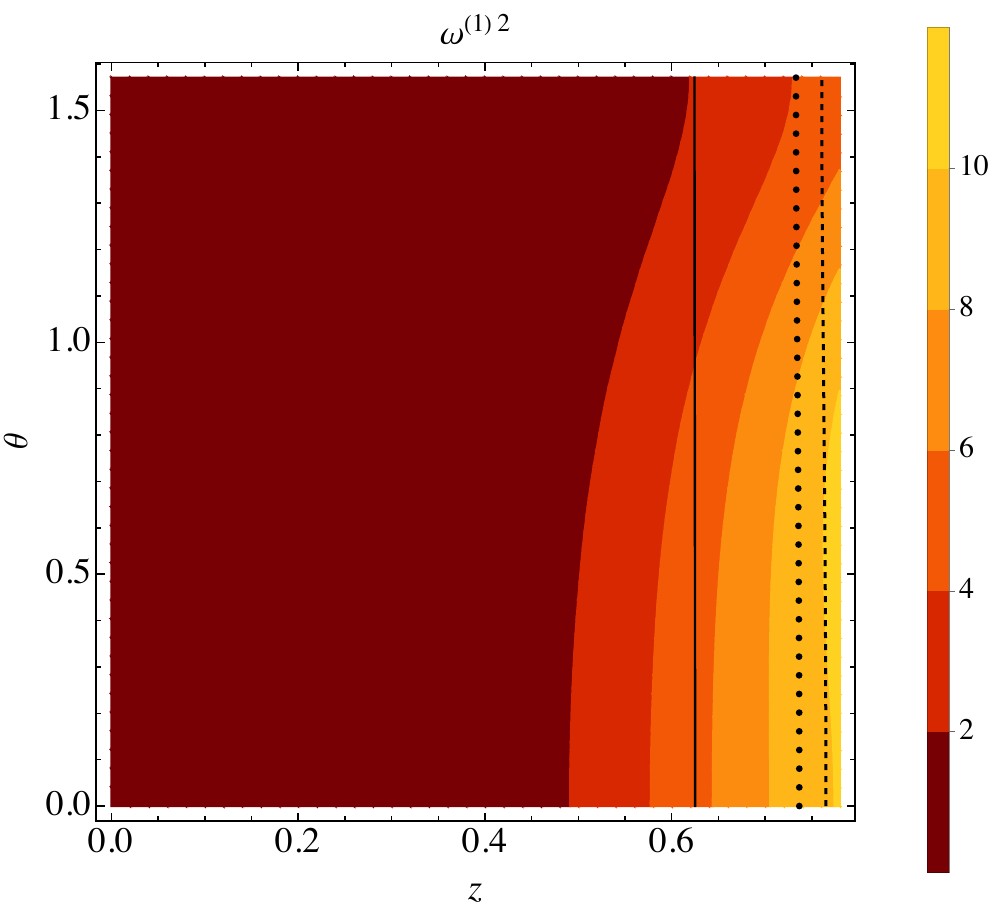}
  }
  \caption{\label{fig:leadingtwist}
  Figure showing $\omega^{(1)2}$ in our coordinate domain for black holes with spin $j = 0.8$ for the Family IA (left) and IIB (right), extrapolated from our small $\epsilon$ solutions. The dashed, dotted and solid black lines are as for the previous figures.
  }
\end{figure}

Recall that for the `painted' on aether the twist vanishes as $\epsilon \to 0$, going as, $\omega_{\mu\nu} = \epsilon \omega^{(1)}_{\mu\nu} + \ldots$. In figure~\ref{fig:leadingtwist} 
is plotted the squared norm of the leading twist $\omega^{(1)}_{\mu\nu}$
for the region I and region II families,
extrapolated 
from our small $\epsilon$ solutions. This quantity takes a slightly different form for each family of solutions, but common to all of them
is firstly that it does not vanish in the interior of our coordinate domain, and secondly this leading twist
grows larger at points deeper in the black hole interior. 
If it continues to grow, there can be
no universal horizon. 
In the limit $c_a = \epsilon \to 0$,
on the other hand, the aether becomes twist free. We 
expect that 
in that limit a universal horizon
arises, because the spin-1 
(and spin-0 for region II)
wave speed diverges,
so the horizon for that wave mode 
becomes a universal horizon.

\section{Discussion}
\label{sec:discussion}

In this work we have numerically constructed 
families of rotating black hole solutions in Einstein-aether theory, focusing on 
on the case of coupling parameters that are
consistent with the many observational and theoretical constraints.
Constraining the spin-2 wavespeed to be that of light leaves the two regions of parameter space, regions I and II, which have one and two large couplings respectively,
but still obey all existing constraints.
In the strong field regime one would have naively expected quantitatively different behaviour from that of GR, for example for stationary black holes.
However we have found 
that while the aether action has terms in with large couplings, the aether may be approximately twist free, and for region II also expansion free, such that these terms are small. The aether then has little backreaction on the spacetime geometry which is then approximately that of a GR solution. Furthermore the leading aether behaviour then does not depend on the  $O(1)$ 
aether couplings, taking an approximately universal form that depends only on the GR solution it is `painted' on to, and whether the couplings lie in
region I or II.

We expect that this approximate GR behaviour applies not just for stationary black hole spacetimes, but more generally for dynamical strong field solutions including  matter. It requires asymptotic behaviour that is compatible with the twist and expansion free conditions, 
such as 
asymptotic flatness, but a cosmological setting would not allow an expansion free aether compatible with cosmological symmetry. 
While we have argued such approximately GR behaviour may occur, we are unable to analytically prove existence of such solutions beyond weak field settings, as the aether twist potential obeys complicated equations of motion for both region I and II. 
For region II the question is whether 
there exists a foliation by maximal slices, whose unit normal
congruence would serve as an aether with vanishing 
twist and expansion. As mentioned earlier, existence has been proved for vacuum spacetimes ``close'' to Minkowski spacetime, but we know of no result that could apply in strong field regimes such as for black hole spacetimes.
It is also possible there may  exist solutions with a non-GR behaviour where the aether does not take such a nearly twist (and expansion) free form. 

We have used
a new numerical approach to construct rotating stationary black holes in the Einstein-aether theory, and using these we show
 this approximate GR behaviour persists to the strong field regime. 
The main challenge to numerical construction is that there are multiple horizons corresponding to the spin-2, -1 and -0 degrees of freedom. 
 A fascinating property of these rotating black hole solutions
 is that they do not possess the familiar $t$-$\phi$ orthogonality property usually enjoyed by asymptotically flat black holes. Furthermore the metric horizon and the wave mode horizons defined by the effective propagation metrics for the various degrees of freedom are not Killing horizons.
 Indeed we believe these are the first examples of black holes without $t$-$\phi$ orthogonality and with non-Killing horizons 
 which are asymptotically flat.
That they are not Killing horizons is presumably of paramount importance
to whether and how the laws of black hole mechanics might or might not apply to them. 
Since no linear combination of the Killing vectors is tangent to the horizon generators, none of the available definitions of surface gravity is suitable,\footnote{This is unlike for the non-Killing horizons considered in \cite{Cropp:2013zxi}, which have well-defined surface gravity.} so the zeroth law fails not only because the surface gravity is not constant, but because it is undefined. The ``first law", if one exists, might at best involve an integral of some local variation over the horizon, but 
that could not have the integrated form $T\delta S$. In fact, 
it seems unlikely that even a meaningful {\it local} notion of temperature or surface entropy density exists.
Previous attempts to make sense of these laws sometimes assumed
that the horizon is a Killing horizon,
and floundered even in spherical symmetry~\cite{Foster:2005fr,Pacilio:2017emh,Ding:2020bwa}. 
A different approach to Lorentz violating 
black hole thermodynamics has focused on the universal horizon, and has found 
partial success in limited cases~\cite{Berglund:2012bu,Mohd:2013zca,Pacilio:2017swi}. 
This program, too, is called into question
for the case of Einstein-aether theory
(but not for Ho\v{r}ava gravity), since
our results strongly suggest that rotating Einstein-aether black holes possess no universal horizon.

Perhaps  the most interesting feature of these solutions is that in the allowed parameter regions I and II these solutions do indeed take the approximate form of Kerr, with the nearly twist free aether `painted' on top and taking a universal form which is approximately independent of the $O(1)$ aether parameters within these regions. The existence of smooth horizons for all degrees of freedom then seems sufficient to determine a unique aether profile that depends only on the mass and angular momentum parameters.
We have found no non-GR behaviours in the families of solutions we have been able to construct for these phenomenologically allowed parameters.

This new numerical method is  based on the harmonic Einstein equation \eqref{eq:harmonic} with a suitable choice of reference metric, and
solved using an iteration scheme starting from an initial guess for the solution. It is suitable for studying black holes in theories with multiple wave mode speeds
leading to distinct horizons at all of which the spacetime should be regular.
Rather than solving for the metric exterior to a black hole Killing horizon on a spacelike surface extending from spatial infinity to the bifurcation surface, which yields an elliptic pde problem that may be solved as a boundary value problem with the horizon forming one such boundary, instead the equation is solved on an ingoing slice that pierces the future metric horizon and all future horizons associated to the different wave modes. The resulting pde system is only elliptic outside all wave mode horizons. Inside all wave mode horizons we expect it is hyperbolic, and correspondingly we do not impose a boundary condition on the innermost boundary of the coordinate domain.
Instead the solution is specified by requiring asymptotic flatness and regularity at all wave mode horizons, which in turn is achieved by having suitably regular metric functions in the ingoing coordinate chart, 
and 
fixing
the physical data for the solution, which for our rotating Einstein-aether black holes is their mass and angular momentum.
The pde system is solved by the Newton method. Starting with an initial guess metric with smooth metric functions, we expect iterations of the method to maintain this smoothness. If the method converges---which we expect and indeed find happens only when all horizons are captured in the coordinate domain---then smoothness of the metric functions ensures these horizons are smooth.

The method works reasonably well in practice, although we have found that reasonably good initial guesses are required to achieve convergence to a solution.
One interesting question that remains from our work is how far the coordinate domain may be extended inside the wave mode horizons. 
If the domain did not extend far enough into the spacetime to include all horizons, then smoothness could not be imposed on these, and no unique solution to the p.d.e. problem would exist for the Newton method to find.
However we also find that we are unable to find solutions if we extend the coordinate domain too far into the interior of the the spacetime. Perhaps this is because the domain then approaches closer to a singularity which destabilizes the numerical system. Alternatively perhaps in the region inside all horizons the expected hyperbolic character leads to behaviour of metric or aether functions that is hard to resolve, such as oscillations~\cite{Eling:2006ec,Barausse:2011pu}.
It would be interesting to better understand more formally 
how the method works, and how best to choose the reference metric and coordinate domain for a given problem.

Unfortunately our method gives no obvious information about the dynamical stability of the black holes we find. Even for the previously found static spherically symmetric black holes of~\cite{Eling:2006ec,Barausse:2011pu}, this remains an open question for general perturbations.  For perturbations respecting the spherical symmetry the  dynamical simulations of~\cite{Garfinkle:2007bk} show certain black holes  are stable since they form the end state of collapse, although these are not for phenomenologically allowed parameters. However, the recent~\cite{Tsujikawa:2021typ} reports some static spherically symmetric solutions are unstable to perturbations that break the spherical symmetry. Constructing dynamical linearized Einstein-aether perturbations about our numerical stationary solutions seems like a difficult task. Since for the phenomenological regimes the spacetime is close to Kerr, an interesting and simpler question might be to ask whether the aether on a fixed Kerr background is stable.

Let us return to the Einstein-aether theory and its phenomenology.
In Appendix B we have shown
that for regions I and II in the weak field regime the aether is approximately twist free and is `painted on' to a metric that is approximately that of usual GR. 
If weak field initial data later enters the strong field regime, such as in gravitational collapse, then the painting picture
dictates that the metric and matter will evolve to remain approximately that of GR, with the aether continuing to be painted on top during this strong field dynamics. 
The only caveat to this is if at some point in the evolution there fails to be a smooth solution to the aether twist potential equation (so for region II the zero expansion condition, and for I equation~\eqref{eq:aetherRegion1}). In this case one might expect that either a singularity forms in the aether, or the subleading terms in $c_a$ become large and the aether departs from the painting picture in the future of where this singular behaviour develops.
In particular using this argument we deduce that the Kerr-like black hole solutions found here numerically are presumably the ones that originate from weak field gravitational collapse.

An important question is whether 
for coupling parameters
in regions I and II one may construct initial data for a Cauchy evolution with large twist. If such data did exist then its evolution would give rise to  non-GR behaviours, in contrast to that of the `painted on' dynamics. The question would then be does the large twist quickly dissipate, or can it persist so that the non-GR behaviour is not just localized near this initial data surface?
For example, if strong field initial data with large twist collapses to a black hole, can the
twist fall into the black hole and produce 
an exotic black hole, different from the approximately Kerr black holes found here? 
While it is possible that there exist such exotic solutions for region I and II which do not have a 4d GR-like behaviour, it seems likely that in phenomenological settings the relevant behaviours would closely follow GR, as in practice, apart from cosmological solutions, strong field gravitational dynamics in our universe originate from the evolution of weak field initial conditions. 
For example, we expect astrophysical black holes will have formed from a prior gravitational collapse of weak field initial data.

Assuming this picture is correct, our conclusion is that in these phenomenological parameter regions, where some aether couplings are $O(1)$, approximate  GR behaviour for black holes 
is nonetheless recovered, with the metric deviating from a GR solution by $\sim O(c_a)$. Recall that the largest this can be is for region I, where $c_a = O(10^{-5})$. If deviations of the quadrupole moment from the Kerr value are of that same order, they would lie around the lower limit of what is projected to be measurable using LISA observations of EMRIs~\cite{Babak:2017tow}. It would be very interesting to understand precisely what is the potential for observing small deviations in physical observables such as this and others, using LISA and other high precision observations of gravitational dynamics.

We note that any twist free solution to Einstein-aether can be embedded as a solution of the low energy limit of Ho\v{r}ava gravity~\cite{Jacobson:2010mx}.
The limiting solutions, taking $\epsilon \to 0$ for the various region I and II families in section~\ref{sec:phenobh}, have a twist free aether on an exact Kerr background. 
The limiting couplings all vanish in Ho\v{r}ava gravity for the region I families, but
for the region II families a nonzero $c_\theta$ remains. Hence this limiting universal solution we have found for the region II families is also a nontrivial solution in Ho\v{r}ava gravity.
Other solutions identical to GR in Ho\v{r}ava gravity with only nonvanishing coupling $c_\theta$ are the slowly moving spherical black hole found in \cite{Ramos:2018oku}
and the spherical collapse and quasinormal mode 
solutions found in \cite{Franchini:2021bpt}. 
These are all examples of the general fact,
argued in \cite{Bellorin:2010je}, that the asymptotically flat solutions of (the IR limit of) Ho\v{r}ava gravity with only nonzero $c_\theta$ coupling are {\it all} identical to GR solutions, with the aether orthogonal to an asymptotically static maximal foliation.\footnote{It was not shown in \cite{Bellorin:2010je} that the foliation {\it must} be asymptotically static, but this has been explicitly established for the spherical solutions in \cite{Franchini:2021bpt}}

Finally, we conclude 
by considering why Einstein-aether admits this `painted on' aether, and hence approximate GR behaviour, in the phenomenological parameter regions I and II where the twist coupling and, for region II, also the 
expansion coupling, are large.
Such a painting picture does not work generally in the parameter space of the theory. Why in the regimes where the theory satisfies the tight weak field Solar system constraints does it allow for such approximate non-linear GR behaviour?

We have shown that for weak fields, at leading order in the PPN expansion, a vanishing shear coupling $c_\sigma$ implies the aether is in fact twist free. Exactly for this reason the weak field Solar system observations do not constrain the twist coupling $c_\omega$. Taking this to be $O(1)$ then results in the allowed regions I and II. In the weak field regime the aether theory passes the phenomenological tests since the metric behaves 
as for GR up to corrections suppressed by the small coupling parameters
precisely because the aether takes a twist free, and for region II also an approximately expansion free, configuration painted over a GR weak field solution. 

Thus the recovery of GR at a non-linear level 
can
be viewed as an extension of the recovery of GR in the weak field regime. It is the same mechanism behind both -- a near twist free aether, allowing little backreaction given a large twist coupling. And for region II the key point is that it is compatible to have an aether that is both twist and expansion free. The twist condition reduces the aether to one scalar function, the twist potential, and then an expansion free condition gives a local equation of motion for that scalar function. 
Hence it is no coincidence that the large parameter regions I and II determined by weak field constraints happen to allow non-linear near-GR behaviour.

\section*{Acknowledgements}
We thank Yosuke Misonoh for discussion and collaboration at early stages of this project. We are grateful to Enrico Barausse, Alessandra Buonanno,
Claudia de Rham, David Mattingly, Harvey Reall, Thomas Sotiriou, Andrew Tolley and Robert Wald 
for useful discussions.
PF was supported by supported by the European Research Council grant ERC-2014- StG 639022-NewNGR and by a Royal Society University Research Fellowship Grant No. UF140319 and URF\textbackslash R\textbackslash 201026.
TJ was supported in part by 
NSF Grants No. PHY-1708139 and PHY-2012139.
TW acknowledges support from the STFC grants ST/P000762/1 and ST/T000791/1.

\newpage
\appendix

\section{Relation between irreducible and original parameterization of Einstein-aether theory}
\label{app:aetherTheory}

In the original form of the Einstein-aether in~\cite{Jacobson:2004ts} the action is written as,
\begin{equation}
I=\int d^4x\sqrt{-g}\bigg[R-K^{abcd}(\nabla_au_c)(\nabla_bu_d)+\tilde{\lambda}(g^{ab}u_a\,u_b+1)\bigg]
\end{equation}
where,
\begin{equation}
K^{abcd}=c_1\,g^{ab}g^{cd}+c_2\,g^{ac}g^{bd}+c_3\,g^{ad}g^{bc} - c_4\,u^a\,u^b\,g^{cd}
\end{equation}
with $c_{1,2,3,4}$ the dimensionless parameters that govern the aether dynamics. The translation to the couplings we use in the main text is,
\begin{eqnarray}
\frac{c_\theta}{3} & = & \frac{1}{3} c_{13} +  c_2 \\
c_\sigma & = & c_{13} \\
c_\omega & = & c_1 - c_3 \\
c_a  & = &  c_{14} \; .
\end{eqnarray}
A subtlety is that the two forms of action only agree when $u^2 = -1$, and as a consequence the Lagrange multiplier in the main text $\lambda$, and $\tilde{\lambda}$ here are not equal.
Following~\cite{Jacobson:2004ts} the various degrees of freedom have speeds,
\begin{eqnarray}
\mathrm{spin-0}: && s_{(0)}^2 = \frac{ c_{123} \left( 2 - c_{14} \right) }{ c_{14} \left( 1 - c_{13} \right) \left( 2 + c_{13} + 3 c_2 \right)} \\ 
\mathrm{spin-1}: && s_{(1)}^2 = \frac{2 c_1 - c_1^2 + c_3^2}{2 c_{14} \left( 1 - c_{13} \right)} \\ 
\mathrm{spin-2}: && s_{(2)}^2 = \frac{1}{1 - c_{13}}
\end{eqnarray}
where we use the notation $c_{13} = c_1 + c_3$, $c_{14} = c_1 + c_4$, $c_{123} = c_1 + c_2 + c_3$, and the PPN parameters were determined in~\cite{Foster:2005dk} to be,
\begin{eqnarray}
\alpha_1 & = & - \frac{8 \left( c_3^2 + c_1 c_4 \right) }{2 c_1 - c_1^2 + c_3^2} \\
\alpha_2 & = & \frac{\alpha_1}{2} - \frac{\left( c_1 + 2 c_3 - c_4 \right) \left( 2 c_1 + 3 c_2 + c_3 + c_4 \right)}{c_{123} \left( 2 - c_{14} \right) } 
\end{eqnarray} 
with Solar system tests and binary pulsar constraints requiring $|\alpha_1| \lesssim 10^{-5}$ and $|\alpha_2| \lesssim 10^{-7}$.

\section{Weak field calculation of the aether}
\label{app:PPN}

In order to give a brief discussion we will follow exactly the conventions of~\cite{Foster:2005dk} so that we can use results from that paper. Note in particular the opposite metric sign convention in that work, which we will employ within this section of the appendix.

Let us collect the relevant results. At PPN order $O(1)$ we see from~\cite{Foster:2005dk} that the metric perturbation to Minkowski spacetime is, 
\begin{eqnarray}
h_{00} = - 2 U \; , \quad h_{ij} = - 2 U \delta_{ij} \; .
\end{eqnarray}
The leading order behaviour for $h_{0i}$ is at PPN order $O(1.5)$, and we will not need the explicit form although we will use the gauge condition for $h_{0i}$,
\begin{eqnarray}\label{B2}
h_{0i,i} = -3 U_{,0} + \theta n_{i,i} \; , \quad n_i = u_i - h_{0i} \; , \quad \theta = - \frac{c_1 + 2 c_3 - c_4}{2 - c_{14}} \; .
\end{eqnarray}
The potential $U$ is derived from the matter energy density $\rho$ as,
\begin{eqnarray}
U(t,\vec{x}) = G_N \int d\vec{y} \frac{\rho(t,\vec{y})}{| \vec{x} - \vec{y} |}
\end{eqnarray}
where $G_N = G (1 - \frac{c_{14}}{2} )^{-1}$.
From~\cite{Foster:2005dk} the leading PPN behaviour at $O(1.5)$ for the aether $u_\mu$ is,
\begin{eqnarray}
u_0 = 1 - U
\; , \quad
u_i =  h_{0i} \left( 1 -  \frac{c_-}{2 c_1} \right)  + \frac{1}{2 c_1}  \left( 2 c_1 A + c_- \left( \frac{3}{2} + A \theta \right) \right) \chi_{,0i}  
\end{eqnarray}
where the constant $A$ and potential $\chi$ are,
\begin{eqnarray}
A = - \frac{2 c_1 + 3 c_2 + c_3 + c_4}{2 c_{123}} \; , \quad \chi(t,\vec{x}) = - G_N \int d\vec{y} \rho(t,\vec{y}) | \vec{x} - \vec{y} |
\end{eqnarray}
First we reexpress these formulae using the couplings of the irreducible parameterization, $c_{\omega,\sigma,\theta,a}$: 
\[\theta = \frac{c_a-2c_\sigma}{2-c_a},\qquad
A= -\frac32\,\frac{c_\theta + c_a}{c_\theta +2c_\sigma}
\]
\[
u_i =  \frac{c_\sigma}{c_\omega+c_\sigma}
h_{0i}  + \left(A+\frac{c_\omega}{c_\omega+
c_\sigma}\left(\frac32+ A\theta\right)\right) \chi_{,0i} \; .
\] 
Now imposing the constraint that
$c_\sigma=0$ yields for the spatial 
aether components,
\begin{eqnarray}
\label{ueq}
u_i &=& - \frac{3 c_a}{2 c_\theta} \left(  \frac{  2 + c_\theta   }{ 2 - c_{a}}   \right) \chi_{,0i}
\end{eqnarray}
and then we see that to $O(1.5)$ the aether takes the twist free form,
\begin{eqnarray}
\label{eq:PPNtwistfree}
u_\mu = k^{(PPN)} \partial_\mu f^{(PPN)} \; , \quad f^{(PPN)} = t - \frac{3 c_a}{2 c_\theta} \left(  \frac{  2 + c_\theta   }{ 2 - c_{a}}   \right) \chi_{,0} \; , \quad k^{(PPN)} = 1 - U
\end{eqnarray}
Thus we see that simply the condition $c_\sigma = 0$ implies the weak field aether at leading PPN order is twist free.

In region I we have $| c_a | \ll 1$ and $c_\theta \simeq 3 c_a$ so that the twist potential takes the form,
\begin{eqnarray}
f^{(PPN)} = t - \frac{1}{2} \chi_{,0} + O(c_a)
\end{eqnarray}
which determines the twist potential for $\bar{u}_\mu = k \partial_\mu f$ in our 'painting' picture -- it is just the expression above setting $c_a \to 0$, so
\begin{eqnarray}
f = \left. f^{(PPN)} \right|_{c_a = 0} \; , \quad k = \left. k^{(PPN)} \right|_{c_a = 0}
\end{eqnarray}
We see this is indeed universal, being independent of the aether parameters within region I.
For region II we have $| c_a | \ll 1$ and $c_\theta = O(1)$ so the twist potential is approximately trivial,
\begin{eqnarray}
f^{(PPN)} = t + O(c_a) \; .
\end{eqnarray}
Hence this simply yields the leading twist free aether $\bar{u} = (1 - U) dt$.
Thus indeed the aether at leading order in PPN is compatible with the painting picture for a twist free leading aether $\bar{u}$ as given above for region I and II. Furthermore since at this PPN order equation~\eqref{eq:PPNtwistfree} gives a twist free aether for any $c_a$, the correction $u^{(1)}_\mu$ to the leading twist free aether in the painting picture
is in fact also twist free, as are all subleading corrections in $c_a$.

The leading order expansion of the aether at $O(1.5)$ is,
\begin{eqnarray}
\nabla \cdot u = - \nabla_i u_i &=& - u_{i,i} + h_{0i,i} - \frac{1}{2} h_{ii,0}  \nonumber \\
& = &  - \frac{ 1}{ 1 + \theta } u_{i,i} + \frac{3 \theta}{1 + \theta}  U_{,0}, 
\end{eqnarray}
using $h_{0i,i} =  (\theta u_{i,i} -3 U_{,0})/(1 + \theta )$ 
from \eqref{B2}. Then,
setting $c_\sigma = 0$ so that 
$\theta = c_a/(2 - c_{a})$,
we find 
\begin{eqnarray}\label{B12}
\nabla \cdot u & = &   \frac{3 c_a}{4 c_\theta} \left(   2 + c_\theta     \right) \chi_{,0ii} + \frac{3 }{2 }c_a U_{,0}
\end{eqnarray}
In region I this yields an expansion,
\begin{eqnarray}
\nabla \cdot u & = &   \frac{1}{2}\chi_{,0ii} + O(c_a)
\end{eqnarray}
and hence the painted leading twist free aether $\bar{u}_\mu$ has expansion $\bar{\nabla} \cdot \bar{u} = \frac{1}{2}\chi_{,0ii}$. 
For region II the painted leading aether must be twist and expansion free, and indeed \eqref{B12} yields $\nabla \cdot u = O(c_a)$, so that the expansion vanishes in the limit $c_a \to 0$.

\section{Convergence and accuracy for Kerr and for Einstein-aether black holes}
\label{app:conv}

In this section we provide convergence tests for both our example construction of the Kerr black hole in Section~\ref{sec:KerrExample} and some representative rotating Einstein-aether black holes from Section~\ref{sec:results}. Our numerical code uses sixth order finite differencing on uniform grids in both the $z$ and $\theta$ directions of our rectangular coordinate domain. We take $N_z$ grid points in the $z$ direction and $N_\theta$ in the $\theta$ direction. We wish to investigate the order at which our numerical solutions converge to a continuum limit. We note it is possible to have lower than sixth order convergence if the functions one solves for are less differentiable than $C^6$. At the least we must have second order convergence as the Einstein-aether equations involve second derivatives. As we will show, we generally find convergence consistent with our sixth order differencing.

\subsection{Convergence tests for Kerr}

In Section~\ref{sec:KerrExample} we demonstrated our in-going method using the toy problem of `finding' the Kerr solution in vacuum GR. Here we present several convergence tests that provide evidence that our numerical solutions indeed converge to Kerr. 

\begin{figure}[t]
\centering
\includegraphics[scale=0.8]{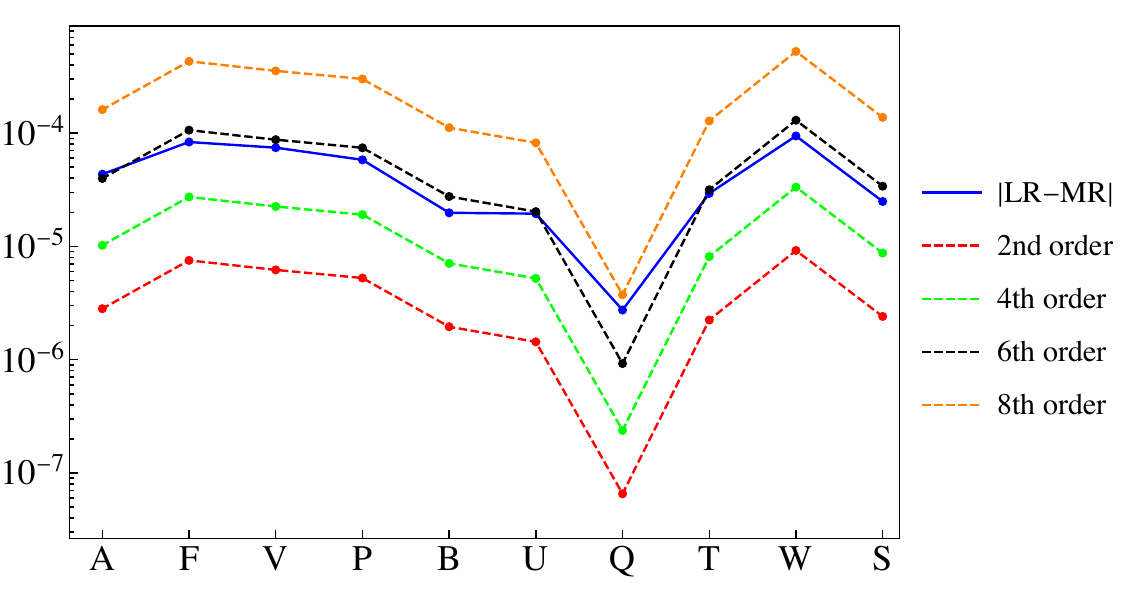}
\caption{Convergence order for all the variables in our numerical construction of a Kerr black hole with $j=0.70$. We have considered the resolutions $(N_z,N_\theta)=(20,20),(40,40),(60,60)$. For all variables, we see that the convergence order is very close to 6, in accordance with our differencing scheme.}
\label{fig:pointwise_kerr}
\end{figure}

In the first test, we show that each of the functions that we solve for in our code converges, pointwise, to the continuum limit with a convergence order which is close to 6, as it should be given our differencing stencils. To see this, for each of the metric functions, for example $A$, we consider the differences between solutions computed at some resolution $(N_z, N_\theta) = (N,N)$, at $(2 N, 2N)$ and at $(3 N, 3N)$. 
We denote these solutions $A_N$, $A_{2 N}$ and $A_{3 N}$ respectively.  Let us define,
\begin{eqnarray}
L^{(A)}_N = ||A_{N}-A_{2 N}|| \; , \quad R^{(A)}_N = ||A_{2 N}-A_{3 N}|| \times \frac{h_{N}^Q-h_{2 N}^Q}{h_{2 N}^Q-h_{3 N}^Q}
\end{eqnarray}
where the norm above is taken to be the $L_1$-norm of the function values at the grid points at the resolution $(N,N)$ (which are common to the higher resolutions since the grids are uniform), and where $h_{N}$, $h_{2 N}$, $h_{3 N}$ are the grid spacing for those resolutions.
Then if the numerical scheme converges with an order of convergence $Q$, asymptotically for increasing resolution we would expect,
\begin{eqnarray}
\lim_{N\to\infty}\left( L^{(A)}_N - R^{(A)}_N \right) \to 0
\end{eqnarray}
and similarly for the other 9 metric functions $\mathcal{F} = \{ T, V, W, \ldots\}$. To estimate the order of convergence from our solutions in figure~\ref{fig:pointwise_kerr} we show $L_{20}$ and $R_{20}$ for all the metric functions for convergence orders $Q=2,4,6,8$ for the solution with spin $j = 0.70$ and $z_{max} = 0.65$. This compares the resolutions $(N_z, N_\theta) = (20,20)$, $(40,40)$ and $(60,60)$. 
As this figure shows, the order of convergence for all variables is very close to 6. We see similar behaviours for other spins. 

\begin{figure}[t]
\centering
\includegraphics[scale=0.5]{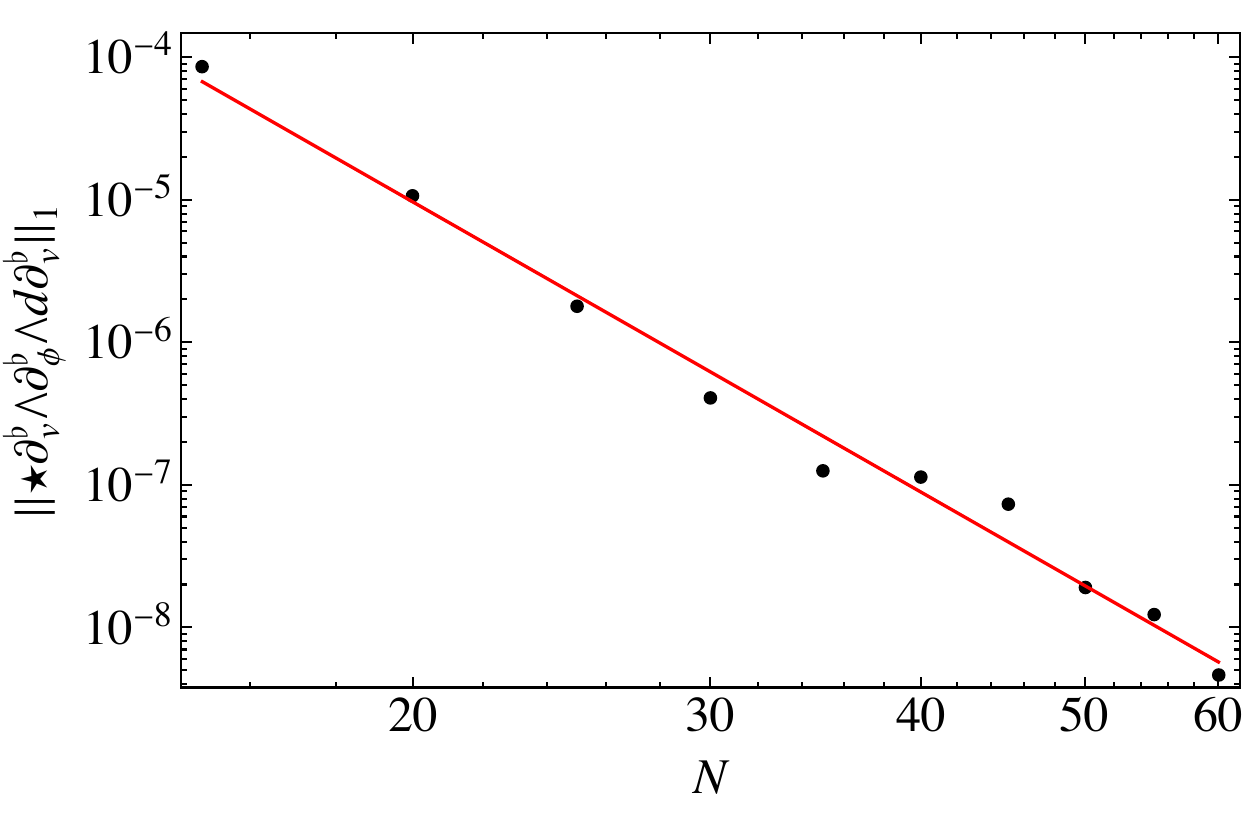}\\
\centering
\includegraphics[scale=0.5]{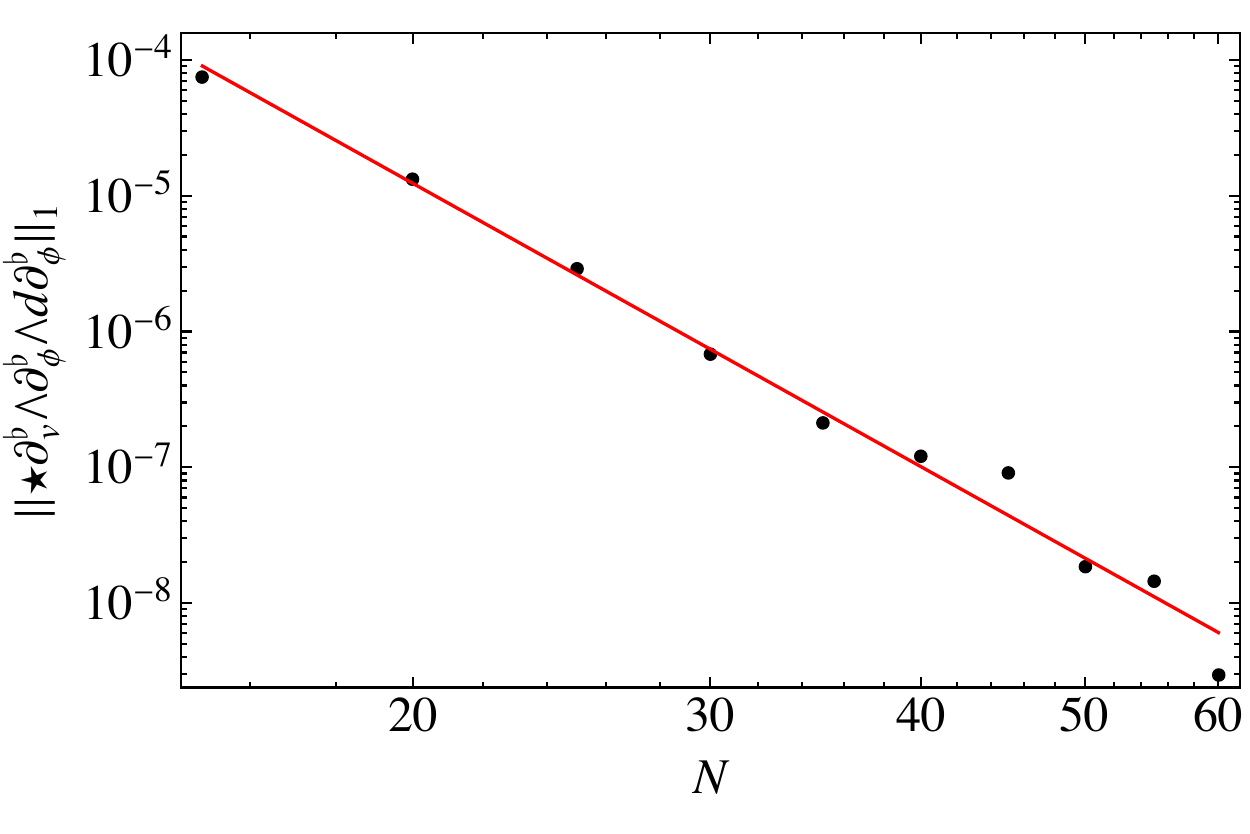}
\caption{$L_1$ norm of $\star \partial_v^\flat\wedge \partial_\phi^\flat\wedge d\partial_v^\flat$ and $\star \partial_v^\flat\wedge \partial_\phi^\flat\wedge d\partial_\phi^\flat$ for our numerical construction of a Kerr black hole with $j=0.70$ and $z_{max}=0.65$. This quantity should vanish for a spacetime with $v$-$\phi$ orthogonality property such as Kerr. The slopes of the fits are $-6.76$ (top) and $-6.93$ (bottom), indicating better than sixth order convergence. Here $N$ denotes the number of grid points along the $z$ and $\theta$ directions.}
\label{fig:vphidv_Kerr}
\end{figure}

Another relevant test that we can carry out to check that the geometry constructed with our method approaches the Kerr spacetime is to verify that it enjoys the $v$-$\phi$ orthogonality property, i.e., that the planes spanned by the Killing directions $v$-$\phi$ are integrable. Recall that our metric ansatz does not have this property built in so recovering it in the continuum limit is non-trivial.  Following our discussion in section~\ref{sec:nonKillinghorizons}, in figure~\ref{fig:vphidv_Kerr} we show that $\star (\partial_v^\flat\wedge \partial_\phi^\flat\wedge d\partial_v^\flat)$ approaches zero in the continuum limit with a slope that is consistent with our differencing scheme. We observe a similar behaviour for $\star (\partial_v^\flat\wedge \partial_\phi^\flat\wedge d\partial_\phi^\flat)$. The vanishing of these two quantities implies that indeed the continuum spacetime has the $v$-$\phi$ orthogonality property.

\begin{figure}[t]
\centering
\includegraphics[scale=0.6]{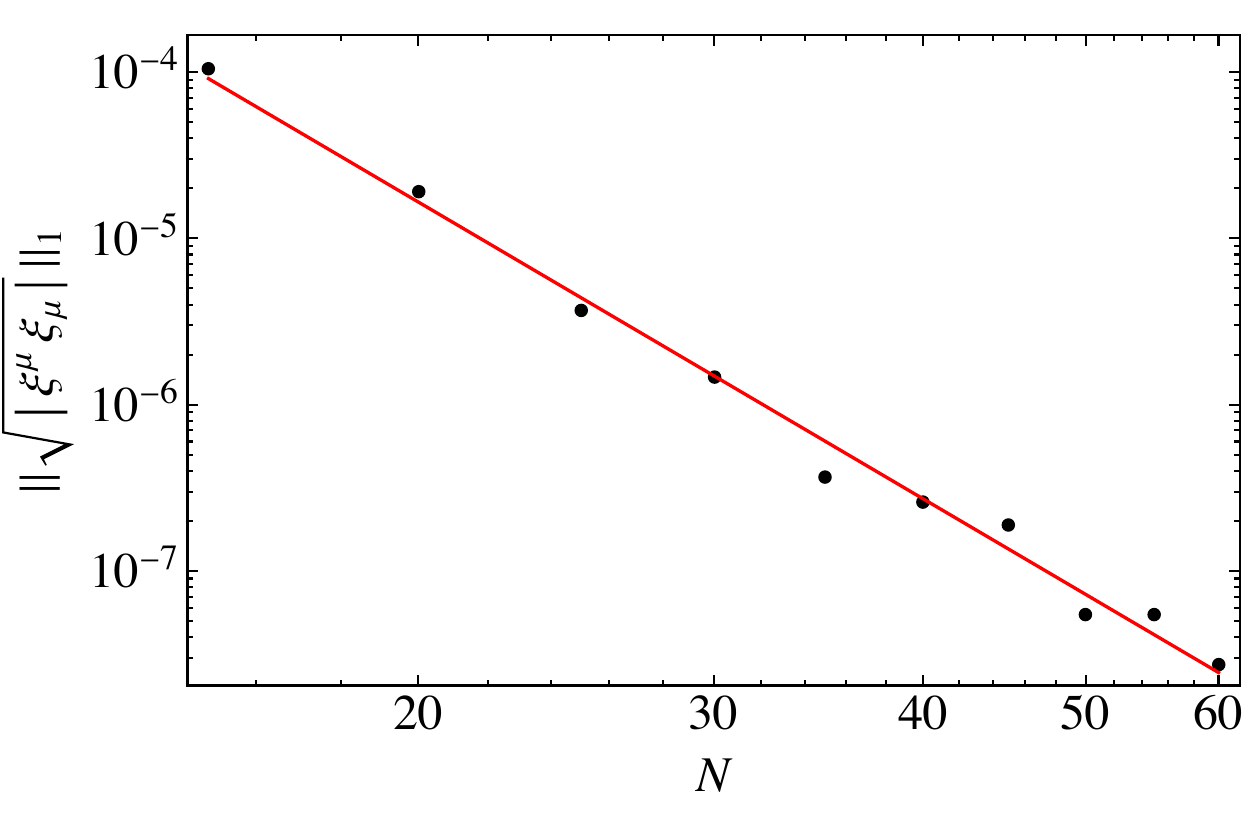}
\caption{$L_1$ norm of $\sqrt{|\xi^\mu\xi_\mu|}$ for a Kerr black hole with $j=0.70$ and $z_{max}=0.65$. The slope of the fit is $-5.93$, indicating near sixth order convergence, as one would expect. Here $N$ denotes the number of grid points along the $z$ and $\theta$ directions.}
\label{fig:phimax_kerr}
\end{figure}

In figure~\ref{fig:phimax_kerr} we display the $L_1$-norm of $\sqrt{|\xi^\mu\xi_\mu|}$ evaluated in our computational domain for resolutions $(N_z, N_\theta) = (N,N)$ as $N$ is varied. We compute the norm as,
\begin{eqnarray}
|| \sqrt{|\xi^\mu\xi_\mu|} ||_1 = \int_0^{z_{max}} dz \int_0^{\pi/2} d\phi \sqrt{|\xi^\mu\xi_\mu|}
\end{eqnarray}
first constructing the function $\sqrt{|\xi^\mu\xi_\mu|}$ from our numerical solution using sixth order interpolation, and then computing the integral using {\tt Mathematica}'s \texttt{NIntegrate} function. Recall that we are solving the Einstein-DeTurck equations \eqref{eq:harmonic} and we need $\xi^a\to 0$ in the continuum limit to recover an actual solution of the Einstein equations. This figure indicates that indeed $\sqrt{|\xi^\mu\xi_\mu|}\to 0$ in the continuum limit. Moreover, the slope of the fit turns out to be $\sim -5.93$, and hence close to the order of our finite differencing scheme.

\begin{figure}[t]
\centering
\includegraphics[scale=0.9]{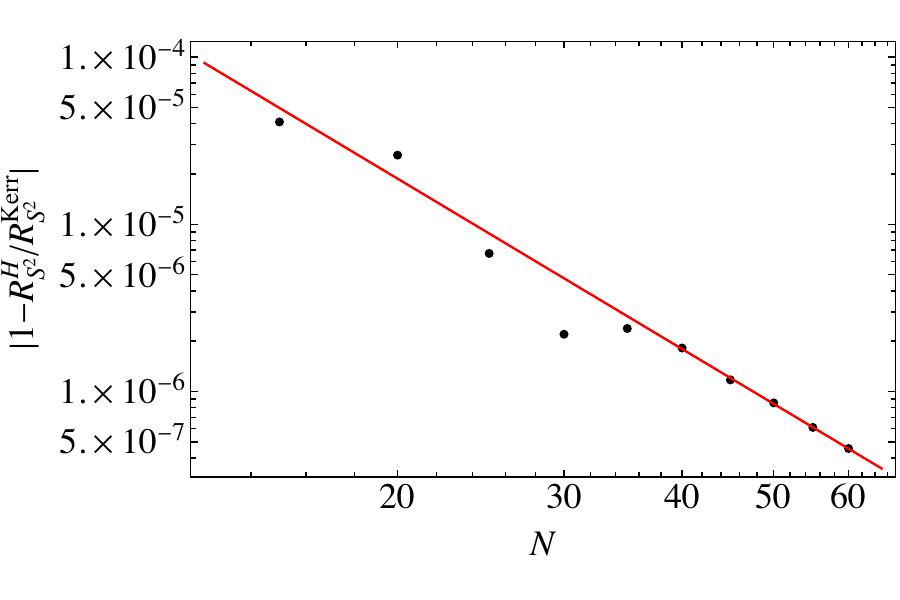}
\caption{Convergence of the equatorial radius of the horizon $S^2$ for a Kerr black hole with $j=0.70$. The slope of the fit is $-3.38$. Here $N$ denotes the number of grid points along the $z$ or $\theta$ directions (we take them to be the same).}
\label{fig:horizon_kerr}
\end{figure}

Finally in figure~\ref{fig:horizon_kerr} since we know the numerical solution should tend to Kerr in the continuum, we study the convergence of the radius of the horizon $S^2$ at its equator to its true value for the Kerr black hole.  This is a geometric quantity which is gauge invariant in the symmetry class of spacetimes that we consider in this paper. For the actual Kerr black hole, this quantity turns out to be $2\,M$, where $M$ is the mass parameter. The figure shows data for spin $j=0.70$, but we observe similar behaviour for other spins. In order to compute this quantity the equatorial coordinate position of the horizon must be found, and we do this by interpolating the metric functions to sixth order (in accordance with our differencing order) in {\tt Mathematica} and then using the {\tt FindRoot} function to locate the horizon. As this plot indicates, our numerical solutions converge to the analytic solution. The fit to our numerical data has slope $\sim -3.38$, which would suggest a convergence order less than four (but still greater than two, which is the highest derivative order in the Einstein equations). Note however that this convergence reflects not only that of the underlying numerical solution of the harmonic Einstein equation, but also the subsequent post-processing to locate the horizon. It is therefore not surprising that the convergence order is less than that of the sixth order differencing of the p.d.e.s.

\subsection{Convergence tests for rotating Einstein-Aether black holes}
In this subsection we present results of the convergence tests for some representative rotating Einstein-Aether black holes.  We show results for a black hole with $j=0.8$ in Family IA with $\epsilon=0.10$ and $z_{max} = 0.75$ and also one with $\epsilon=0.61$ and $z_{max} = 0.63$ as two representative examples. The latter is the largest value of $\epsilon$ for which we could construct black holes with this spin in this family. We have considered other families, other values of $\epsilon$, and also different spins and find similar results. 

\begin{figure}[t]
\centering
\includegraphics[scale=0.75]{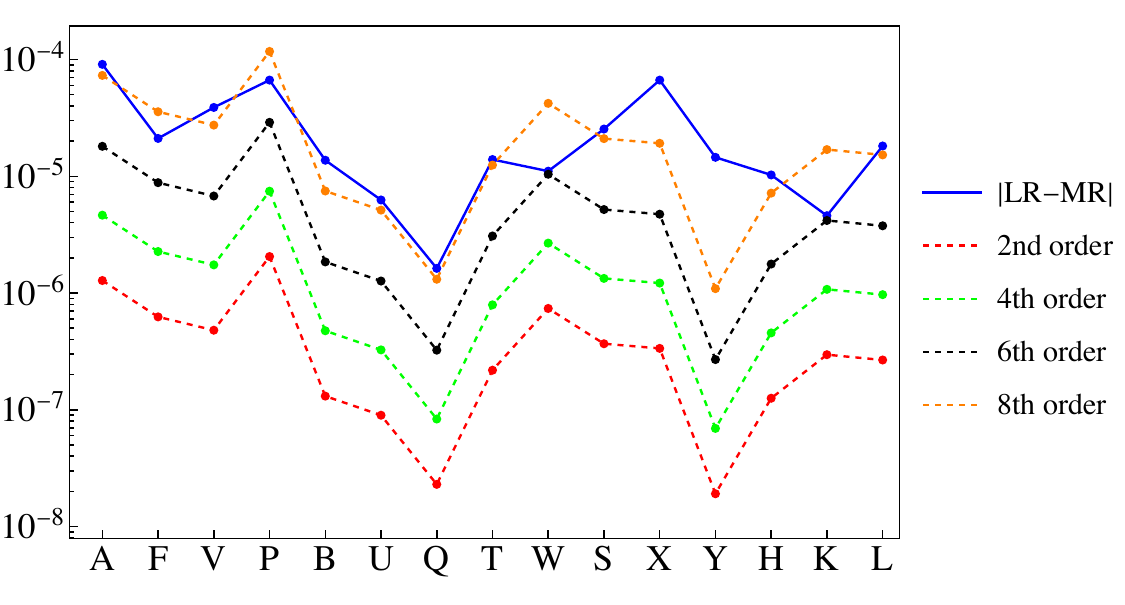}
\includegraphics[scale=0.75]{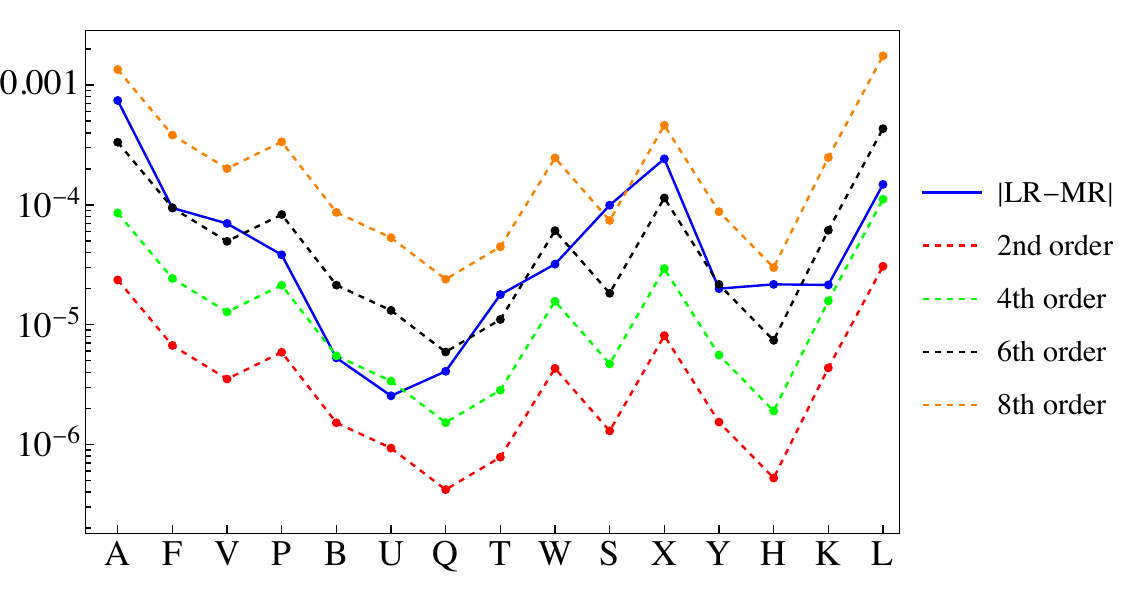}
\caption{Convergence of the $L_1$ norms of the unknown variables for two rotating Einstein-Aether black holes in Family IA with $j=0.8$ and $\epsilon=0.1$ (top) and  $\epsilon=0.612$ (bottom). }
\label{fig:pointwise_ae}
\end{figure}

We first consider the pointwise convergence of the 15 metric and aether functions, $\mathcal{J}$, by computing the $L_1$ norms of their differences for various resolutions as for the Kerr toy example using the resolutions $(N_z,N_\theta) = (20,20)$, $(40,40)$ and $(60,60)$. This is shown in Fig. \ref{fig:pointwise_ae}. 
For the solution with $\epsilon=0.10$ (top plot) we observe a sixth order or better convergence. For the $\epsilon=0.61$ solution we see most metric functions give sixth order convergence, although some seem to display slightly worse fourth order convergence, although we note that since this is the largest $\epsilon$ where solutions were found for this family and spin this is perhaps not surprising.
Also as mentioned in the main text, we observe that the order of convergence is sensitive to the relative location of the various horizons and also the extent of our computational domain in the radial direction, $z_{max}$.

\begin{figure}[t]
\centering
\includegraphics[scale=0.7]{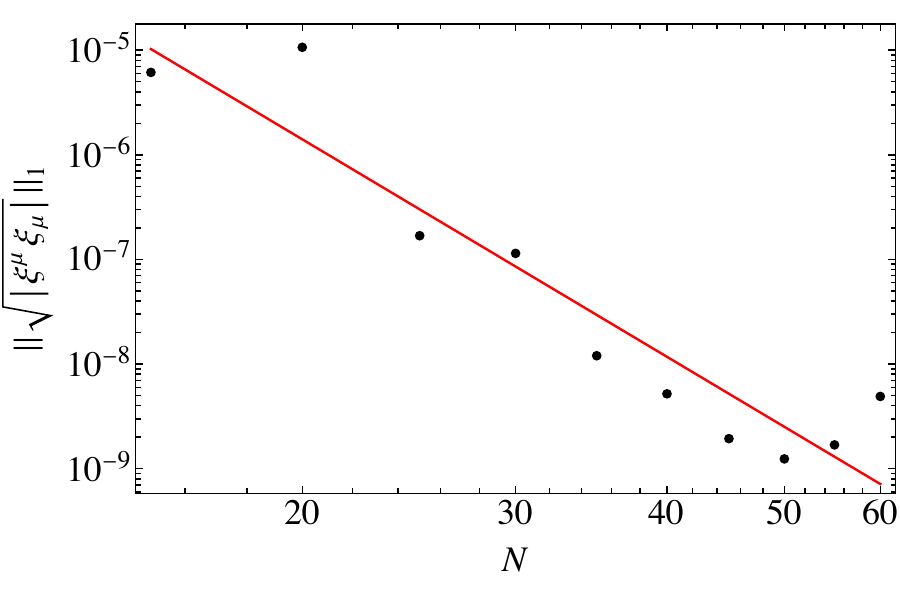}
\includegraphics[scale=0.75]{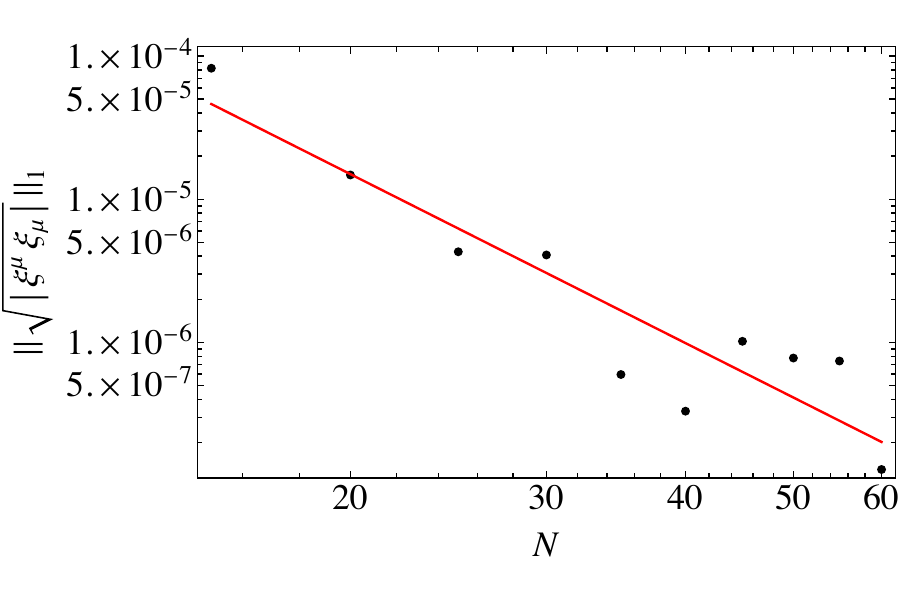}
\caption{Convergence of the $L_1$ norms of $|\xi_\mu\xi^\mu|$ for two rotating Einstein-Aether black holes in Family IA with $j=0.8$ and $\epsilon=0.1$ (top) and  $\epsilon=0.61$ (bottom). For the solutions with $\epsilon=0.1$ the slope is $\sim-6.9$, while for the solutions with $\epsilon=0.61$ the slope is $\sim-3.9$. Here $N$ denotes the number of grid points in both the $z$ and $\theta$ directions.}
\label{fig:pointwise_phi_ae}
\end{figure}

In Fig. \ref{fig:pointwise_phi_ae} we show the $L_1$ norm of $\sqrt{|\xi_\mu\xi^\mu|}$ for different resolutions for the same Einstein-Aether black holes in Family IA. For the solutions with $\epsilon=0.1$, the convergence order of this quantity is better than six, while for the solutions with $\epsilon=0.61$, the convergence order is slightly less than four. We note for the black holes with $\epsilon=0.1$, the quantity $\sqrt{|\xi_\mu\xi^\mu|}$ obtained from solutions with $50\times50$ grid points or more is already affected by finite machine precision (for our C implementation we represent real numbers using double precision).

\section{Details of Einstein-aether black hole constructions}
\label{app:details}

In this appendix we provide more details about the construction of the rotating Einstein-aether black holes presented in the main text.

\begin{figure}
\centering
  \includegraphics[scale=0.55]{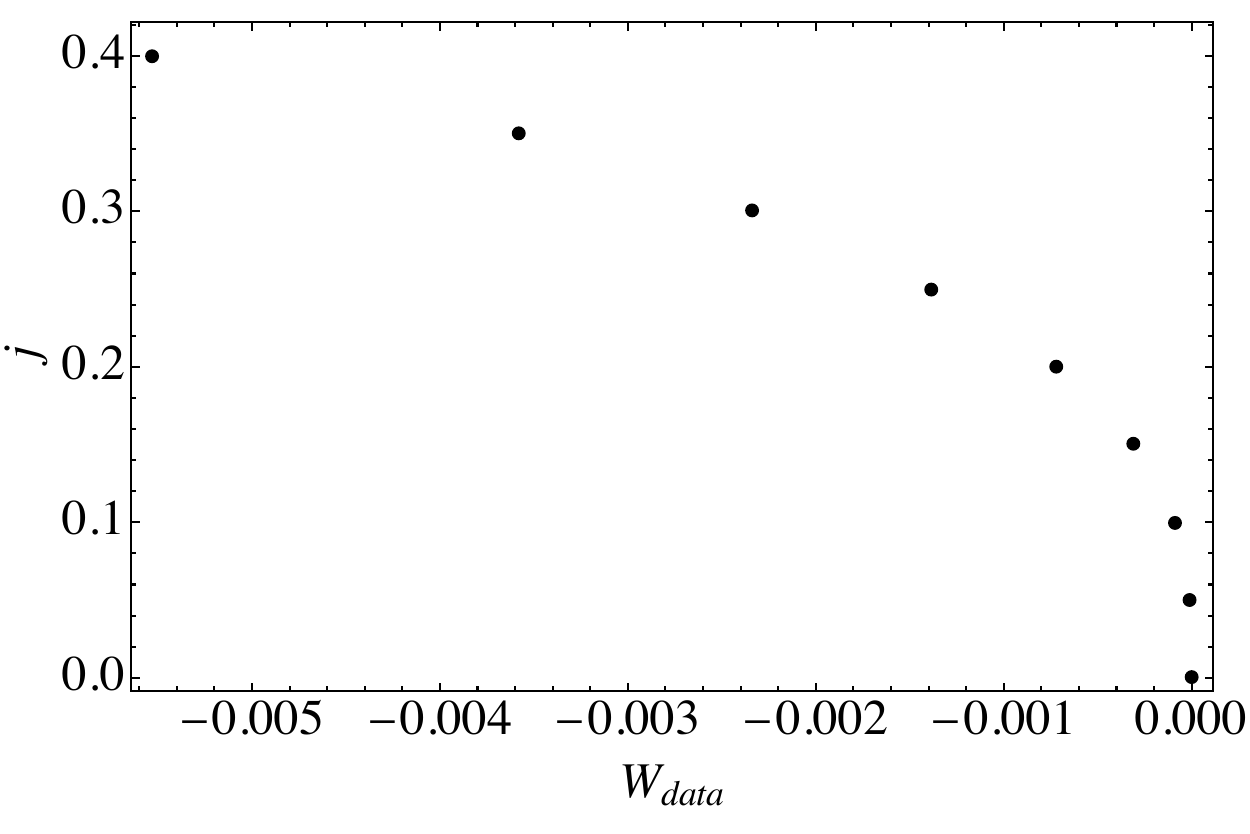}
  \caption{\label{fig:Wmap}
  Example of the map from $W_{data}$ to the dimensionless spin $j$ for Kerr black holes found with our ingoing method (see section~\ref{sec:KerrExample}). In this case we fixed $z_{max}=0.56$
  and took $\mu = 1$, $a = 0$ for the reference metric in equation~\eqref{eq:ExampleRef}. Then varying $W_{data}$ gives solutions with spins $j$ as shown above.
  }
\end{figure}

As discussed in the main text, our solutions have two moduli, which we determine by fixing $T=T_{data}$ and $W=W_{data}$ at $z=z_{max}$ and $\theta=\pi/2$. Because of the scale invariance of the theory, we can set $T_{data}=0$ without loss of generality; then, we vary $W_{data}$ to obtain the desired value of the dimensionless spin $j$. In Fig. \ref{fig:Wmap} we illustrate the map between $W_{data}$ and
the spin $j$ in the case of the toy example of finding Kerr in section~\ref{sec:KerrExample}. Fixing the reference metric given in equation~\eqref{eq:ExampleRef}, here taking $\mu = 1$ and $a = 0$, then varying the data $W_{data}$ gives solutions with different spins. We see a smooth relation between $W_{data}$ and $j$ which then allows one to tune solutions to a desired value of $j$ using a Newton solver wrapped around the harmonic Einstein equation solver to then determine the necessary $W_{data}$. This works in the same way for the rotating Einstein-Aether black holes we find in section~\ref{sec:results}.

\begin{figure}
\centerline{
  \includegraphics[scale=0.8]{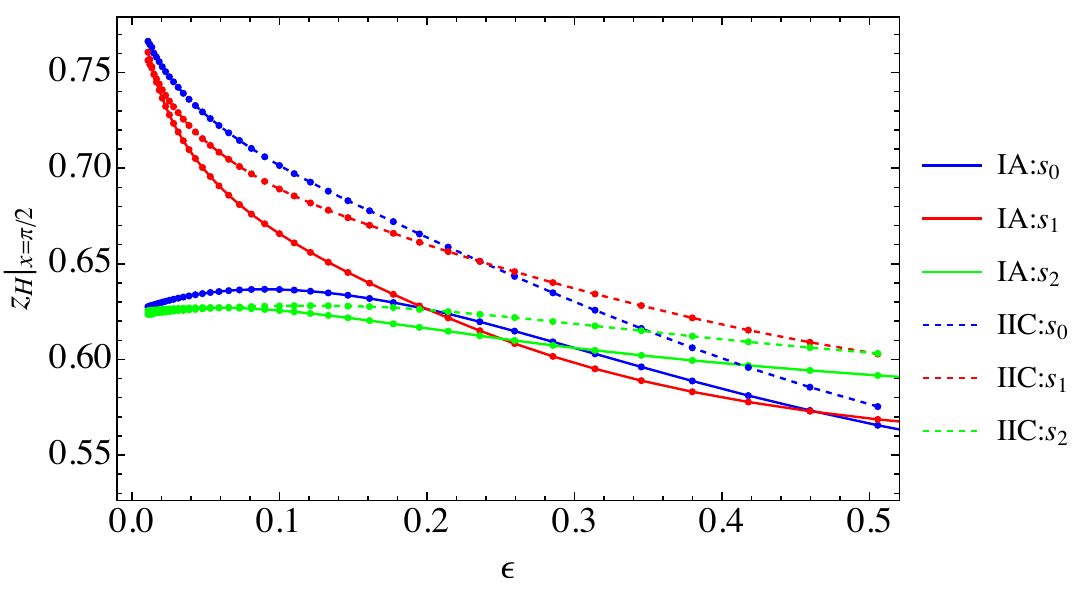}
  }
  \caption{\label{fig:horizonstwo}
  Coordinate positions of various spin horizons for two representative families as a function of $\epsilon$ and fixed $j=0.8$. The spin 2 horizon is always the outermost and it approaches the location of the horizon of a Kerr black hole with the same $j$, i.e., $z_H^\textrm{Kerr}=0.625$, in the $\epsilon\to 0$ limit. For Family I, the spin 1 horizon is the innermost one and it approaches a limiting value as $\epsilon\to 0$, corresponding to $s_1^2\to\infty$. Similarly, for Family II, the spin 0 and the spin 1 horizons are inside the spin 2 horizon, and they approach the same limiting value in the Kerr limit. Note that the coordinate location of the spin-0 and spin-1 horizons in the limit $\epsilon \to 0$  will be different for Families I and II since the limiting aether configuration is different. 
  }
\end{figure}

One of the key points of our method is to consider a computational domain that covers all horizons; in the particular case of the Einstein-aether black holes, these are the spin 0, 1 and 2 horizons. For a given mass and spin, we do not know a priori where the coordinate locations of the various horizons are going to be until we have found the full solution. As we discussed in the main text, the location of each of the horizons depends on the effective metric \eqref{eq:effmetric} which controls the propagation of the corresponding wave mode. This effective metric depends on both the spacetime metric and the aether.
To illustrate this, in figure~\ref{fig:horizonstwo} we show the coordinate position of the horizon on the equatorial plane for the various spin degrees of freedom for a representative $j = 0.8$ black hole in the family IA, and for another in the family IIC, as a function of $\epsilon$. As noted earlier, for our choice of reference metric we find the coordinate position of the various horizons varies only slightly as a function of $\theta$ for the Einstein-aether black holes found here, as we saw already in the earlier figure~\ref{fig:horizons}.

 In practice, to find the rotating Einstein-aether black holes we start by considering a computational domain with extent $z_{max}$ such that it covers the horizon of a Kerr black hole with the same dimensionless spin $j$ as our sought Einstein-aether black hole. Once we have found the black hole solution, we can locate the various horizons. If we do not find a solution we try with slightly larger values for $z_{max}$. Sometimes our method converges to solutions where the innermost horizon is not quite captured in the coordinate domain, but appears to lie just outside it. In such a case one can estimate the slightly larger $z_{max}$ required to capture all the horizons and then try to solve the equations again.

Once we have found an Einstein-aether black hole with dimensionless spin $j$ in a given family, we move along the family by varying the parameter $\epsilon$ while keeping $z_{max}$ fixed. As we vary $\epsilon$, the coordinate locations of the various horizons change and eventually one (or more) horizon(s) will no longer be covered by the computational domain. To  proceed we first guess by extrapolation where the horizons are going to be, and then extend $z_{max}$ accordingly, and finally extrapolate the previous solution to obtain a good initial guess for the new computational domain. 

Note that when we move along a given family by decreasing $\epsilon$, we approach the phenomenological regime discussed in Section \ref{sec:phenobh}. In this region, depending on the family that we consider, the propagation speeds of the spin 1 and spin 0 modes diverge, which implies that the corresponding horizon limits to become a universal horizon, while the spin 2 horizon converges to the horizon of a Kerr black hole with the same $j$. This forces us to increase $z_{max}$ as we decrease $\epsilon$. On the other hand, as $\epsilon\to0$, the backreaction of the aether configuration on the background Kerr black hole becomes less important. In practice, for $\epsilon \lesssim 0.1$, we do not need to tune $W_{data}$ to keep $j$ fixed to a reasonable accuracy; for the spin $j = 0.8$ solutions presented here we choose the parameters of the background metric $\mu=1$ and $a=0.8$ and taking $W_{data} = 0$, we find that the dimensionless spin $j$ of the resulting Einstein-aether black hole satisfies $j=0.8$ to a few percent or better for $\epsilon \le 0.1$. This accuracy increases for smaller $\epsilon$, and for $\epsilon \simeq 0.01$ the black holes satisfy having spin $j = 0.8$ to $\sim 0.01\%$. 
 On the other hand, as we increase $\epsilon$ to $O(1)$ values the propagation speeds of the spin 0 and 1 modes drop below the speed of light 
and hence their respective horizons lie outside the spin 2 horizon. In this case, we have to decrease $z_{max}$ while making sure that the spin 2 horizon lies within our computational domain. For $\epsilon \ge 0.1$ the backreaction of the aether is strong and we need to tune $W_{data}$ so that $j=0.8$ is kept fixed to good accuracy as we move along the family. We do this by solving for $W_{data}$ using the Newton method, and this gives solutions with a spin accurate to better than 1 part in $10^8$.

We require a good initial guess for the Newton-Raphson scheme. We have tried a variety of initial guesses, and importantly precisely the same numerical solutions were found (up to numerical precision) indicating the system of equations is well-posed.
For a spin $j$ Einstein-aether black hole we use a reference metric which is the Kerr black hole with parameters $\mu=1$ and $a=j$ and also take this as the initial guess for the metric; for the aether, an example of the initial guess we use is
\begin{equation}
\begin{aligned}
    X &=\mu(1+0.4\,z)\,,\\
    Y &= 0\,,\\
    K &= -a\,z\,H\,,\\
    L &=-0.01\,z^2\,,
\end{aligned}
\label{eq:ansatz_aether}
\end{equation}
and the function $H$ is fixed by requiring that the aether vector has unit timelike norm.
Finally we take $T_{data} = 0$ and $W_{data} = 0$. If a solution is found, then it will have spin close to, but not equal to the desired value $j$. Then $W_{data}$ can be tuned by a separate Newton solver to accurately obtain this spin $j$.
As mentioned above for the spin $j = 0.8$ black holes of the region I and II families presented here with $\epsilon \le 0.1$ the spin is sufficiently close to the desired spin (to better than a few percent) that we do not tune $W_{data}$. However we must tune it for the $\epsilon > 0.1$ solutions.
Often the above initial guess for the metric and aether are not good enough, and 
in practice we often start with the above initial guess for the aether \eqref{eq:ansatz_aether} and solve the aether equations on a fixed Kerr background. Once we have found this aether configuration, we use it as an improved initial guess for the full Einstein-aether system. This procedure works well for the region I and II families of solutions with $\epsilon \simeq 0.1$. Having found a first Einstein-aether black hole for a given family it is then straightforward to  move along that family taking small steps in $\epsilon$ by using the previous solution as the initial guess.    
For each solution found we check that the vector $\xi^\mu$ is consistent with being zero within numerical error to ensure it is a solution of the Einstein equation, rather than a Ricci soliton. In fact none of the solutions we found in this work were Ricci solitons. 

Finally, we implement solution of the  Einstein-aether equations in {\tt C}.
Following \cite{Figueras:2012rb}, we evaluate the full harmonic Einstein and aether equations directly in the {\tt C} code. Similarly, we compute the linearised operator that we need for the Newton solver iterations by numerically differentiating these. The reason is that 
while substituting our metric and aether ansatz into the Einstein-aether equations would in principle give analytic expressions for the equations and their linearization, in practice they result in exceedingly complicated expressions that we have not been able to use.

\newpage
{
\bibliographystyle{jhep}
\footnotesize
\bibliography{refs}

\providecommand{\href}[2]{#2}\begingroup\raggedright\begin{thebibliography}{10}

\bibitem{LIGOScientific:2016aoc}
{\scshape LIGO Scientific, Virgo} collaboration, \emph{{Observation of
  Gravitational Waves from a Binary Black Hole Merger}},
  \href{https://doi.org/10.1103/PhysRevLett.116.061102}{\emph{Phys. Rev. Lett.}
  {\bfseries 116} (2016) 061102}
  [\href{https://arxiv.org/abs/1602.03837}{{\ttfamily 1602.03837}}].

\bibitem{EventHorizonTelescope:2019dse}
{\scshape Event Horizon Telescope} collaboration, \emph{{First M87 Event
  Horizon Telescope Results. I. The Shadow of the Supermassive Black Hole}},
  \href{https://doi.org/10.3847/2041-8213/ab0ec7}{\emph{Astrophys. J. Lett.}
  {\bfseries 875} (2019) L1}
  [\href{https://arxiv.org/abs/1906.11238}{{\ttfamily 1906.11238}}].

\bibitem{Jacobson:2004ts}
T.~Jacobson and D.~Mattingly, \emph{{Einstein-Aether waves}},
  \href{https://doi.org/10.1103/PhysRevD.70.024003}{\emph{Phys. Rev. D}
  {\bfseries 70} (2004) 024003}
  [\href{https://arxiv.org/abs/gr-qc/0402005}{{\ttfamily gr-qc/0402005}}].

\bibitem{TheLIGOScientific:2017qsa}
{\scshape LIGO Scientific, Virgo} collaboration, \emph{{GW170817: Observation
  of Gravitational Waves from a Binary Neutron Star Inspiral}},
  \href{https://doi.org/10.1103/PhysRevLett.119.161101}{\emph{Phys. Rev. Lett.}
  {\bfseries 119} (2017) 161101}
  [\href{https://arxiv.org/abs/1710.05832}{{\ttfamily 1710.05832}}].

\bibitem{Creminelli:2017sry}
P.~Creminelli and F.~Vernizzi, \emph{{Dark Energy after GW170817 and
  GRB170817A}},
  \href{https://doi.org/10.1103/PhysRevLett.119.251302}{\emph{Phys. Rev. Lett.}
  {\bfseries 119} (2017) 251302}
  [\href{https://arxiv.org/abs/1710.05877}{{\ttfamily 1710.05877}}].

\bibitem{deRham:2018red}
C.~de~Rham and S.~Melville, \emph{{Gravitational Rainbows: LIGO and Dark Energy
  at its Cutoff}},
  \href{https://doi.org/10.1103/PhysRevLett.121.221101}{\emph{Phys. Rev. Lett.}
  {\bfseries 121} (2018) 221101}
  [\href{https://arxiv.org/abs/1806.09417}{{\ttfamily 1806.09417}}].

\bibitem{Oost:2018tcv}
J.~Oost, S.~Mukohyama and A.~Wang, \emph{{Constraints on Einstein-aether theory
  after GW170817}},
  \href{https://doi.org/10.1103/PhysRevD.97.124023}{\emph{Phys. Rev. D}
  {\bfseries 97} (2018) 124023}
  [\href{https://arxiv.org/abs/1802.04303}{{\ttfamily 1802.04303}}].

\bibitem{Gupta:2021vdj}
T.~Gupta, M.~Herrero-Valea, D.~Blas, E.~Barausse, N.~Cornish, K.~Yagi et~al.,
  \emph{{Updated Binary Pulsar Constraints on Einstein-${\ae}$ther Theory in
  Light of Gravitational Wave Constraints on the Speed of Gravity}},
  \href{https://arxiv.org/abs/2104.04596}{{\ttfamily 2104.04596}}.

\bibitem{Kleihaus:2000kg}
B.~Kleihaus and J.~Kunz, \emph{{Rotating hairy black holes}},
  \href{https://doi.org/10.1103/PhysRevLett.86.3704}{\emph{Phys. Rev. Lett.}
  {\bfseries 86} (2001) 3704}
  [\href{https://arxiv.org/abs/gr-qc/0012081}{{\ttfamily gr-qc/0012081}}].

\bibitem{Headrick:2009pv}
M.~Headrick, S.~Kitchen and T.~Wiseman, \emph{{A New approach to static
  numerical relativity, and its application to Kaluza-Klein black holes}},
  \href{https://doi.org/10.1088/0264-9381/27/3/035002}{\emph{Class. Quant.
  Grav.} {\bfseries 27} (2010) 035002}
  [\href{https://arxiv.org/abs/0905.1822}{{\ttfamily 0905.1822}}].

\bibitem{Eling:2006df}
C.~Eling and T.~Jacobson, \emph{{Spherical solutions in Einstein-aether theory:
  Static aether and stars}},
  \href{https://doi.org/10.1088/0264-9381/23/18/008}{\emph{Class. Quant. Grav.}
  {\bfseries 23} (2006) 5625}
  [\href{https://arxiv.org/abs/gr-qc/0603058}{{\ttfamily gr-qc/0603058}}].

\bibitem{Oost:2021tqi}
J.~Oost, S.~Mukohyama and A.~Wang, \emph{{Spherically symmetric exact vacuum
  solutions in Einstein-aether theory}},
  \href{https://arxiv.org/abs/2106.09044}{{\ttfamily 2106.09044}}.

\bibitem{Garfinkle:2007bk}
D.~Garfinkle, C.~Eling and T.~Jacobson, \emph{{Numerical simulations of
  gravitational collapse in Einstein-aether theory}},
  \href{https://doi.org/10.1103/PhysRevD.76.024003}{\emph{Phys. Rev. D}
  {\bfseries 76} (2007) 024003}
  [\href{https://arxiv.org/abs/gr-qc/0703093}{{\ttfamily gr-qc/0703093}}].

\bibitem{Eling:2006ec}
C.~Eling and T.~Jacobson, \emph{{Black Holes in Einstein-Aether Theory}},
  \href{https://doi.org/10.1088/0264-9381/23/18/009}{\emph{Class. Quant. Grav.}
  {\bfseries 23} (2006) 5643}
  [\href{https://arxiv.org/abs/gr-qc/0604088}{{\ttfamily gr-qc/0604088}}].

\bibitem{Barausse:2011pu}
E.~Barausse, T.~Jacobson and T.~P. Sotiriou, \emph{{Black holes in
  Einstein-aether and Horava-Lifshitz gravity}},
  \href{https://doi.org/10.1103/PhysRevD.83.124043}{\emph{Phys. Rev. D}
  {\bfseries 83} (2011) 124043}
  [\href{https://arxiv.org/abs/1104.2889}{{\ttfamily 1104.2889}}].

\bibitem{Berglund:2012bu}
P.~Berglund, J.~Bhattacharyya and D.~Mattingly, \emph{{Mechanics of universal
  horizons}}, \href{https://doi.org/10.1103/PhysRevD.85.124019}{\emph{Phys.
  Rev. D} {\bfseries 85} (2012) 124019}
  [\href{https://arxiv.org/abs/1202.4497}{{\ttfamily 1202.4497}}].

\bibitem{Zhang:2020too}
C.~Zhang, X.~Zhao, K.~Lin, S.~Zhang, W.~Zhao and A.~Wang, \emph{{Spherically
  symmetric static black holes in Einstein-aether theory}},
  \href{https://doi.org/10.1103/PhysRevD.102.064043}{\emph{Phys. Rev. D}
  {\bfseries 102} (2020) 064043}
  [\href{https://arxiv.org/abs/2004.06155}{{\ttfamily 2004.06155}}].

\bibitem{Eling:2003rd}
C.~Eling and T.~Jacobson, \emph{{Static postNewtonian equivalence of GR and
  gravity with a dynamical preferred frame}},
  \href{https://doi.org/10.1103/PhysRevD.69.064005}{\emph{Phys. Rev. D}
  {\bfseries 69} (2004) 064005}
  [\href{https://arxiv.org/abs/gr-qc/0310044}{{\ttfamily gr-qc/0310044}}].

\bibitem{Barausse:2015frm}
E.~Barausse, T.~P. Sotiriou and I.~Vega, \emph{{Slowly rotating black holes in
  Einstein-\ae{}ther theory}},
  \href{https://doi.org/10.1103/PhysRevD.93.044044}{\emph{Phys. Rev. D}
  {\bfseries 93} (2016) 044044}
  [\href{https://arxiv.org/abs/1512.05894}{{\ttfamily 1512.05894}}].

\bibitem{Jacobson:2013xta}
T.~Jacobson, \emph{{Undoing the twist: The Ho\v{r}ava limit of Einstein-aether
  theory}}, \href{https://doi.org/10.1103/PhysRevD.89.081501}{\emph{Phys. Rev.
  D} {\bfseries 89} (2014) 081501}
  [\href{https://arxiv.org/abs/1310.5115}{{\ttfamily 1310.5115}}].

\bibitem{Sarbach:2019yso}
O.~Sarbach, E.~Barausse and J.~A. Preciado-L\'opez, \emph{{Well-posed Cauchy
  formulation for Einstein-\ae{}ther theory}},
  \href{https://doi.org/10.1088/1361-6382/ab2e13}{\emph{Class. Quant. Grav.}
  {\bfseries 36} (2019) 165007}
  [\href{https://arxiv.org/abs/1902.05130}{{\ttfamily 1902.05130}}].

\bibitem{Withers:2009qg}
B.~Withers, \emph{{Einstein-aether as a quantum effective field theory}},
  \href{https://doi.org/10.1088/0264-9381/26/22/225009}{\emph{Class. Quant.
  Grav.} {\bfseries 26} (2009) 225009}
  [\href{https://arxiv.org/abs/0905.2446}{{\ttfamily 0905.2446}}].

\bibitem{Foster:2005dk}
B.~Z. Foster and T.~Jacobson, \emph{{Post-Newtonian parameters and constraints
  on Einstein-aether theory}},
  \href{https://doi.org/10.1103/PhysRevD.73.064015}{\emph{Phys. Rev. D}
  {\bfseries 73} (2006) 064015}
  [\href{https://arxiv.org/abs/gr-qc/0509083}{{\ttfamily gr-qc/0509083}}].

\bibitem{Frusciante:2020gkx}
N.~Frusciante and M.~Benetti, \emph{{Cosmological constraints on Ho\v{r}ava
  gravity revised in light of GW170817 and GRB170817A and the degeneracy with
  massive neutrinos}},
  \href{https://doi.org/10.1103/PhysRevD.103.104060}{\emph{Phys. Rev. D}
  {\bfseries 103} (2021) 104060}
  [\href{https://arxiv.org/abs/2005.14705}{{\ttfamily 2005.14705}}].

\bibitem{Shao:2013wga}
L.~Shao, R.~N. Caballero, M.~Kramer, N.~Wex, D.~J. Champion and A.~Jessner,
  \emph{{A new limit on local Lorentz invariance violation of gravity from
  solitary pulsars}},
  \href{https://doi.org/10.1088/0264-9381/30/16/165019}{\emph{Class. Quant.
  Grav.} {\bfseries 30} (2013) 165019}
  [\href{https://arxiv.org/abs/1307.2552}{{\ttfamily 1307.2552}}].

\bibitem{Will:2014kxa}
C.~M. Will, \emph{{The Confrontation between General Relativity and
  Experiment}}, \href{https://doi.org/10.12942/lrr-2014-4}{\emph{Living Rev.
  Rel.} {\bfseries 17} (2014) 4}
  [\href{https://arxiv.org/abs/1403.7377}{{\ttfamily 1403.7377}}].

\bibitem{Eardley:1978tr}
D.~M. Eardley and L.~Smarr, \emph{{Time function in numerical relativity.
  Marginally bound dust collapse}},
  \href{https://doi.org/10.1103/PhysRevD.19.2239}{\emph{Phys. Rev. D}
  {\bfseries 19} (1979) 2239}.

\bibitem{PhysRevD.31.1267}
M.~J. Duncan, \emph{Maximally slicing a black hole with minimal distortion},
  \href{https://doi.org/10.1103/PhysRevD.31.1267}{\emph{Phys. Rev. D}
  {\bfseries 31} (1985) 1267}.

\bibitem{Gomes:2013bbl}
H.~Gomes and G.~Herczeg, \emph{{A Rotating Black Hole Solution for Shape
  Dynamics}},
  \href{https://doi.org/10.1088/0264-9381/31/17/175014}{\emph{Class. Quant.
  Grav.} {\bfseries 31} (2014) 175014}
  [\href{https://arxiv.org/abs/1310.6095}{{\ttfamily 1310.6095}}].

\bibitem{Bergamini:2003ch}
R.~Bergamini and S.~Viaggiu, \emph{{A Novel derivation for Kerr metric in
  Papapetrou gauge}},
  \href{https://doi.org/10.1088/0264-9381/21/19/006}{\emph{Class. Quant. Grav.}
  {\bfseries 21} (2004) 4567}
  [\href{https://arxiv.org/abs/gr-qc/0305035}{{\ttfamily gr-qc/0305035}}].

\bibitem{Estabrook:1973ue}
F.~Estabrook, H.~Wahlquist, S.~Christensen, B.~DeWitt, L.~Smarr and E.~Tsiang,
  \emph{{Maximally slicing a black hole}},
  \href{https://doi.org/10.1103/PhysRevD.7.2814}{\emph{Phys. Rev. D} {\bfseries
  7} (1973) 2814}.

\bibitem{Beig:1997fp}
R.~Beig and N.~O'Murchadha, \emph{{Late time behavior of the maximal slicing of
  the Schwarzschild black hole}},
  \href{https://doi.org/10.1103/PhysRevD.57.4728}{\emph{Phys. Rev. D}
  {\bfseries 57} (1998) 4728}
  [\href{https://arxiv.org/abs/gr-qc/9706046}{{\ttfamily gr-qc/9706046}}].

\bibitem{Jacobson:1993vj}
T.~Jacobson, G.~Kang and R.~C. Myers, \emph{{On black hole entropy}},
  \href{https://doi.org/10.1103/PhysRevD.49.6587}{\emph{Phys. Rev. D}
  {\bfseries 49} (1994) 6587}
  [\href{https://arxiv.org/abs/gr-qc/9312023}{{\ttfamily gr-qc/9312023}}].

\bibitem{Racz:1995nh}
I.~Racz and R.~M. Wald, \emph{{Global extensions of space-times describing
  asymptotic final states of black holes}},
  \href{https://doi.org/10.1088/0264-9381/13/3/017}{\emph{Class. Quant. Grav.}
  {\bfseries 13} (1996) 539}
  [\href{https://arxiv.org/abs/gr-qc/9507055}{{\ttfamily gr-qc/9507055}}].

\bibitem{Reall:2021voz}
H.~S. Reall, \emph{{Causality in gravitational theories with second order
  equations of motion}},
  \href{https://doi.org/10.1103/PhysRevD.103.084027}{\emph{Phys. Rev. D}
  {\bfseries 103} (2021) 084027}
  [\href{https://arxiv.org/abs/2101.11623}{{\ttfamily 2101.11623}}].

\bibitem{Adam:2011dn}
A.~Adam, S.~Kitchen and T.~Wiseman, \emph{{A numerical approach to finding
  general stationary vacuum black holes}},
  \href{https://doi.org/10.1088/0264-9381/29/16/165002}{\emph{Class. Quant.
  Grav.} {\bfseries 29} (2012) 165002}
  [\href{https://arxiv.org/abs/1105.6347}{{\ttfamily 1105.6347}}].

\bibitem{Figueras:2012rb}
P.~Figueras and T.~Wiseman, \emph{{Stationary holographic plasma quenches and
  numerical methods for non-Killing horizons}},
  \href{https://doi.org/10.1103/PhysRevLett.110.171602}{\emph{Phys. Rev. Lett.}
  {\bfseries 110} (2013) 171602}
  [\href{https://arxiv.org/abs/1212.4498}{{\ttfamily 1212.4498}}].

\bibitem{DeTurck}
D.~DeTurck, \emph{{Deforming metrics in the direction of their Ricci tensors}},
  {\emph{J. Diff. Geom.} {\bfseries 18} (1983) 157}.

\bibitem{Friedrich}
H.~Friedrich, \emph{{On the hyperbolicity of Einstein's and other gauge field
  equations}}, {\emph{Commun. Math. Phys.} {\bfseries 100} (1985) 525}.

\bibitem{Garfinkle:2001ni}
D.~Garfinkle, \emph{{Harmonic coordinate method for simulating generic
  singularities}},
  \href{https://doi.org/10.1103/PhysRevD.65.044029}{\emph{Phys. Rev. D}
  {\bfseries 65} (2002) 044029}
  [\href{https://arxiv.org/abs/gr-qc/0110013}{{\ttfamily gr-qc/0110013}}].

\bibitem{Figueras:2011va}
P.~Figueras, J.~Lucietti and T.~Wiseman, \emph{{Ricci solitons, Ricci flow, and
  strongly coupled CFT in the Schwarzschild Unruh or Boulware vacua}},
  \href{https://doi.org/10.1088/0264-9381/28/21/215018}{\emph{Class. Quant.
  Grav.} {\bfseries 28} (2011) 215018}
  [\href{https://arxiv.org/abs/1104.4489}{{\ttfamily 1104.4489}}].

\bibitem{Figueras:2016nmo}
P.~Figueras and T.~Wiseman, \emph{{On the existence of stationary Ricci
  solitons}}, \href{https://doi.org/10.1088/1361-6382/aa764a}{\emph{Class.
  Quant. Grav.} {\bfseries 34} (2017) 145007}
  [\href{https://arxiv.org/abs/1610.06178}{{\ttfamily 1610.06178}}].

\bibitem{Sonner:2017jcf}
J.~Sonner and B.~Withers, \emph{{Universal spatial structure of nonequilibrium
  steady states}},
  \href{https://doi.org/10.1103/PhysRevLett.119.161603}{\emph{Phys. Rev. Lett.}
  {\bfseries 119} (2017) 161603}
  [\href{https://arxiv.org/abs/1705.01950}{{\ttfamily 1705.01950}}].

\bibitem{Adam:2013aa}
A.~Adam, \emph{Numerical General Relativity in Exotic Settings}, Ph.D. thesis,
  Imperial College London, 2013.

\bibitem{Hawking:1971vc}
S.~W. Hawking, \emph{{Black holes in general relativity}},
  \href{https://doi.org/10.1007/BF01877517}{\emph{Commun. Math. Phys.}
  {\bfseries 25} (1972) 152}.

\bibitem{Carter}
B.~Carter, \emph{Black Holes}, ch.~Black Hole Equilibrium States, pp.~57--214.
\newblock Gordon and Breach, 1973.

\bibitem{Fischetti:2012vt}
S.~Fischetti, D.~Marolf and J.~E. Santos, \emph{{AdS flowing black funnels:
  Stationary AdS black holes with non-Killing horizons and heat transport in
  the dual CFT}},
  \href{https://doi.org/10.1088/0264-9381/30/7/075001}{\emph{Class. Quant.
  Grav.} {\bfseries 30} (2013) 075001}
  [\href{https://arxiv.org/abs/1212.4820}{{\ttfamily 1212.4820}}].

\bibitem{Cropp:2013zxi}
B.~Cropp, S.~Liberati and M.~Visser, \emph{{Surface gravities for non-Killing
  horizons}},
  \href{https://doi.org/10.1088/0264-9381/30/12/125001}{\emph{Class. Quant.
  Grav.} {\bfseries 30} (2013) 125001}
  [\href{https://arxiv.org/abs/1302.2383}{{\ttfamily 1302.2383}}].

\bibitem{Foster:2005fr}
B.~Z. Foster, \emph{{Noether charges and black hole mechanics in
  Einstein-aether theory}},
  \href{https://doi.org/10.1103/PhysRevD.73.024005}{\emph{Phys. Rev. D}
  {\bfseries 73} (2006) 024005}
  [\href{https://arxiv.org/abs/gr-qc/0509121}{{\ttfamily gr-qc/0509121}}].

\bibitem{Pacilio:2017emh}
C.~Pacilio and S.~Liberati, \emph{{Improved derivation of the Smarr formula for
  Lorentz-breaking gravity}},
  \href{https://doi.org/10.1103/PhysRevD.95.124010}{\emph{Phys. Rev. D}
  {\bfseries 95} (2017) 124010}
  [\href{https://arxiv.org/abs/1701.04992}{{\ttfamily 1701.04992}}].

\bibitem{Ding:2020bwa}
H.-F. Ding and X.-H. Zhai, \emph{{Entropies and The First Laws of Black Hole
  Thermodynamics in Einstein-aether-Maxwell Theory}},
  \href{https://doi.org/10.1088/1361-6382/aba31d}{\emph{Class. Quant. Grav.}
  {\bfseries 37} (2020) 185015}
  [\href{https://arxiv.org/abs/2001.06261}{{\ttfamily 2001.06261}}].

\bibitem{Mohd:2013zca}
A.~Mohd, \emph{{On the thermodynamics of universal horizons in
  Einstein-{\textbackslash{}AE}ther theory}},
  \href{https://arxiv.org/abs/1309.0907}{{\ttfamily 1309.0907}}.

\bibitem{Pacilio:2017swi}
C.~Pacilio and S.~Liberati, \emph{{First law of black holes with a universal
  horizon}}, \href{https://doi.org/10.1103/PhysRevD.96.104060}{\emph{Phys. Rev.
  D} {\bfseries 96} (2017) 104060}
  [\href{https://arxiv.org/abs/1709.05802}{{\ttfamily 1709.05802}}].

\bibitem{Tsujikawa:2021typ}
S.~Tsujikawa, C.~Zhang, X.~Zhao and A.~Wang, \emph{{Odd-parity stability of
  black holes in Einstein-Aether gravity}},
  \href{https://arxiv.org/abs/2107.08061}{{\ttfamily 2107.08061}}.

\bibitem{Babak:2017tow}
S.~Babak, J.~Gair, A.~Sesana, E.~Barausse, C.~F. Sopuerta, C.~P.~L. Berry
  et~al., \emph{{Science with the space-based interferometer LISA. V: Extreme
  mass-ratio inspirals}},
  \href{https://doi.org/10.1103/PhysRevD.95.103012}{\emph{Phys. Rev. D}
  {\bfseries 95} (2017) 103012}
  [\href{https://arxiv.org/abs/1703.09722}{{\ttfamily 1703.09722}}].

\bibitem{Jacobson:2010mx}
T.~Jacobson, \emph{{Extended Horava gravity and Einstein-aether theory}},
  \href{https://doi.org/10.1103/PhysRevD.81.101502}{\emph{Phys. Rev. D}
  {\bfseries 81} (2010) 101502}
  [\href{https://arxiv.org/abs/1001.4823}{{\ttfamily 1001.4823}}].

\bibitem{Ramos:2018oku}
O.~Ramos and E.~Barausse, \emph{{Constraints on Ho\v{r}ava gravity from binary
  black hole observations}},
  \href{https://doi.org/10.1103/PhysRevD.99.024034}{\emph{Phys. Rev. D}
  {\bfseries 99} (2019) 024034}
  [\href{https://arxiv.org/abs/1811.07786}{{\ttfamily 1811.07786}}].

\bibitem{Franchini:2021bpt}
N.~Franchini, M.~Herrero-Valea and E.~Barausse, \emph{{Relation between general
  relativity and a class of Ho\v{r}ava gravity theories}},
  \href{https://doi.org/10.1103/PhysRevD.103.084012}{\emph{Phys. Rev. D}
  {\bfseries 103} (2021) 084012}
  [\href{https://arxiv.org/abs/2103.00929}{{\ttfamily 2103.00929}}].

\bibitem{Bellorin:2010je}
J.~Bellorin and A.~Restuccia, \emph{{On the consistency of the Horava Theory}},
  \href{https://doi.org/10.1142/S021827182500290}{\emph{Int. J. Mod. Phys. D}
  {\bfseries 21} (2012) 1250029}
  [\href{https://arxiv.org/abs/1004.0055}{{\ttfamily 1004.0055}}].

\end{thebibliography}\endgroup
}


\end{document}